\documentclass[12pt]{article}

\usepackage{newtxtext,newtxmath}

\usepackage{graphicx}

\usepackage[letterpaper,margin=1in]{geometry}

\renewenvironment{abstract}
	{\quotation}
	{\endquotation}

\date{}

\makeatletter
\renewcommand{\fnum@figure}{\textbf{Figure \thefigure}}
\renewcommand{\fnum@table}{\textbf{Table \thetable}}
\makeatother

\usepackage{scicite}

\usepackage{url}

\def\scititle{
	Predicting solid-state reaction onset temperatures from high-throughput pairwise synthesis
}
\title{\bfseries \boldmath \scititle}

\author{
	Lauren~N.~Walters$^{1,2\ast\dagger}$,
	Matthew~J.~McDermott$^{2\dagger}$,
	Bernardus~Rendy$^{3,2}$,\and
	Yuxing~Fei$^{3,2}$,
	Gerbrand~Ceder$^{2,3}$\and
	\small$^{1}$Bakar Institute of Digital Materials for the Planet, University of California Berkeley,\and
	\small 373 Cory Hall, Berkeley, CA 94720, U.S.A.\and
	\small$^{2}$Division of Materials Science, Lawrence Berkeley National Laboratory,\and
	\small 1 Cyclotron Rd, Berkeley, CA 94720, U.S.A.\and
	\small$^{3}$Department of Materials Science and Engineering, University of California Berkeley,\and
	\small 210 Hearst Memorial Mining Building, Berkeley, CA 94720, U.S.A.\and
	\small$^\ast$Corresponding author. Email: laurenwalters@berkeley.edu (L.N.W.); gceder@berkeley.edu (G.C.)\and
	\small$^\dagger$These authors contributed equally to this work.
}

\begin{document}

\maketitle

\begin{abstract}

Solid-state synthesis still relies on heuristics to choose reaction temperatures.
The most widely used, Tammann's rule, places reaction onset at two-thirds of the lowest precursor melting point, but has never been systematically tested.
Using an automated laboratory, we heated 1,035 pairwise combinations of 46 common precursors at their Tammann temperatures and identified the products by x-ray diffraction.
Only 25\% of pairs formed products containing both precursors. Combining these outcomes with text-mined literature syntheses, we trained log-normal accelerated failure time models that treat each fixed-temperature experiment as a bound on the unknown onset temperature.
On 26 reactions with onset temperatures measured by \textit{in situ} x-ray diffraction, a three-descriptor model reduced the mean absolute error from 131\,$^\circ$C (Tammann's rule) to 88\,$^\circ$C; adding 29 further physicochemical descriptors did not improve the model.
Reaction onset can therefore be predicted from routine pass/fail synthesis outcomes, and predictions will improve as such data accumulates.

\end{abstract}

\noindent\textbf{Teaser:} Automated tests of 1,035 precursor pairs show
Tammann's rule misses most onsets; a three-descriptor model cuts the error by a third.

\section*{Introduction}
Computational and artificial intelligence models now routinely generate thousands of candidate inorganic compounds, yet their physical realization remains limited and a bottleneck for the deployment of novel predicted materials. 
Synthesis science relies on isolated case studies and qualitative heuristics, producing neither the scale nor the quantitative rigor needed to match digital discovery.
Solid-state experimental methods are mature, but their predictive theory remains underdeveloped: a chemist's intuition and rules of thumb, like Tammann's rule, while useful, need to be replaced by a fully systematic, data-driven understanding of reactivity.
Closing that gap requires moving beyond one-off observations toward reproducible, predictive platforms that can be deployed to deliver the next generation of materials.

The Tammann--rule estimates the onset reaction temperature of a precursor mixture as two-thirds the melting point of the lowest melting point precursor \cite{Tammann_rule_1, Tammann_rule_2, Tammann_rule_3}.
Thus, Tammann's rule has been used as a guideline for temperature selection within solid state synthesis development \cite{Huo2022_RevisedTammanRule, Tammann_rule_2, Tammann_rule_3}, as well as sintering \cite{Tammann_rule_use_1}, diffusion \cite{Tammann_rule_use_2}, operating temperature \cite{Tammann_rule_use_3}, and catalysis \cite{Tammann_rule_use_4}.
A traditional justification for the Tammann--rule connects the onset of reactivity as a consequence of defect formation in transport controlled systems \cite{Tammann_rule_1}.
A positive feedback loop between defect interactions at high temperature leads to a critical temperature ($T_T$) at which the defect concentration increases by multiple orders of magnitude, \textit{i.e.} a defect avalanche, and triggering a chemical reaction \cite{Tammann_rule_1}.
While the onset reaction temperature for chemical reactions has been verified for select systems \cite{Tammann_rule_1, Tammann_rule_2, Tammann_rule_use_1}, the generalized application of Tammann's rule has yet to be experimentally verified on a wide range of solid state syntheses.

The recent emergence of autonomous and self-driving laboratories (SDLs) has transitioned materials discovery from singular manual experiments to high-throughput, standardized experimentation \cite{Alab, Alab_os}.
This shift is exemplified by platforms such as PASCAL, Ada, and the Mobile Robot Chemist, which navigate high-dimensional experimental parameter spaces for thin-film and molecular discovery with minimal human intervention \cite{Cakan2024_SDL2, MacLeod2020_SDL1, Burger2020_SDL3}.
Unlike conventional reporting in synthesis, which is biased towards successful outcomes, these automated platforms can systematically collect datasets that include varied reaction results.
Such machine-readable records are increasingly recognized as a prerequisite for the development of predictive synthesis theory, and are now central to major community efforts, including the Materials Project and the Open Reaction Database \cite{MaterialsProject, Kearnes2021_openreactiondatabase}.
By increasing experimental throughput, automated and autonomous laboratories enable the systematic testing of long-standing heuristics, such as Tammann's rule, across diverse chemical spaces.

In this study, we performed 1,035 pairwise chemical reactions spanning 46 distinct precursors and 39 elements, with each trial executed at the temperature prescribed by Tammann's rule.
The scale of experimental synthesis, characterization, and analysis was made possible through the integrated autonomous framework of the A-Lab, AlabOS, and Dara \cite{Alab, Alab_os, Dara}.
While the approximately 75\% non-reactivity rate observed at the Tammann temperature challenges the rule's general utility, we demonstrate that alternative models incorporating empirical reaction outcomes, thermodynamics, and melting temperatures achieve high predictive accuracy.
The performance of these models would likely scale with data volume, underscoring the critical need for the generation of reliable, unbiased experimental datasets to advance the field of predictive synthesis science.

\subsection*{Results}

\subsubsection*{Pairwise Syntheses with A-Lab}

\begin{figure}
    \centering
    \includegraphics[width=0.9\linewidth]{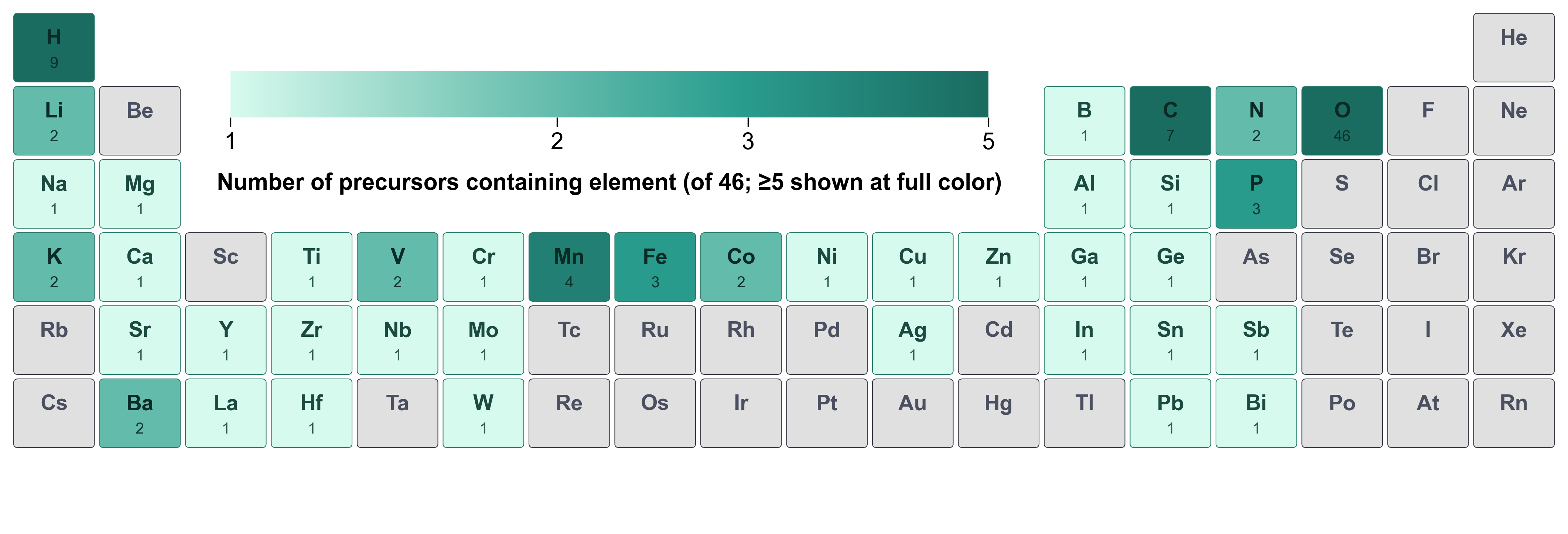}
    \caption{\textbf{The setup of the Precursor Genome framework.} Schematic of the periodic table. Elements shaded in teal represent elements present within our precursor data set. A color bar in the middle depicts the number of different precursors that contain that element.
    %
    }
    \label{fig:intro-figure}
\end{figure}

We curated a precursor library designed to span a diverse chemical space, prioritizing materials common to solid-state synthesis with an emphasis on precursors commonly used for electrochemical energy storage. Selection criteria included cost, stability in air, hygroscopicity, and safety.
Figure~\ref{fig:intro-figure} shows the extent that the chosen precursors span the periodic table, where precursors represented within our project are shaded teal. In total, 39 elements are represented across 46 precursors in the genome.
To maintain high data integrity, every precursor underwent rigorous pre-synthesis characterization by powder XRD, SEM, and EDS, and hygroscopicity testing. Detailed information on procurement, processing, and characterization profiles are provided in Supplementary Text, section~S1.

Using the autonomous A-Lab platform \cite{Alab}, we performed solid-state synthesis on 1,035 enumerated precursor pairs (Supplementary Text, section~S1).
Following precursor selection and reaction enumeration, we calculated the Tammann temperature and generated automated submission files for all pairwise samples.
The reference temperature for each precursor is its first thermally accessible transition, which depending on the precursor is a melting ($T_{melt}$), decomposition, reduction, sublimation, or phase-transformation temperature, as listed in the Supplementary Materials (table~\ref{tab:precursor-melting-points}).
After synthesis, all samples underwent automated powder XRD characterization and analysis, with the resulting refinements subjected to a two-step manual validation to ensure phase identification accuracy (see Materials and Methods).

Prior to experimental execution, we calculated reaction energies for each precursor pair in the genome.
Fig.~\ref{fig:staircase-energies-plot} presents a staircase plot of these energies, utilizing enthalpies approximated from 0\,K DFT data (GGA) from the Materials Project \cite{MaterialsProject} and adjusted for the calculated Tammann-rule reaction temperature via the SISSO method \cite{Gibbs_Bartel}.
69.3\% of the reactions exhibit negative (thermodynamically favorable) reaction energies, typically ranging from -0.2\,eV/atom to 0.0\,eV/atom. While the distribution of reaction energies varies across the chemical space, certain precursors demonstrate consistently high reactivity; for example, iron oxalate (FeC$_2$O$_4\cdot$2H$_2$O) maintains a negative reaction energy across all pairings. Conversely, HfO$_2$ and SiO$_2$ have no favorable predicted reaction ($\Delta G_rxn = 0$) for most of their pairings (25 and 27 of 45 pairings). To prevent selection bias, we executed all 1,035 reactions regardless of their calculated thermodynamic reaction energy; 1,000 yielded powders that could be characterized (see Materials and Methods ~\ref{sec:mat_method}, table ~\ref{tab:errored-samples}).

\begin{figure}
    \centering
    \includegraphics[width=1.0\linewidth]{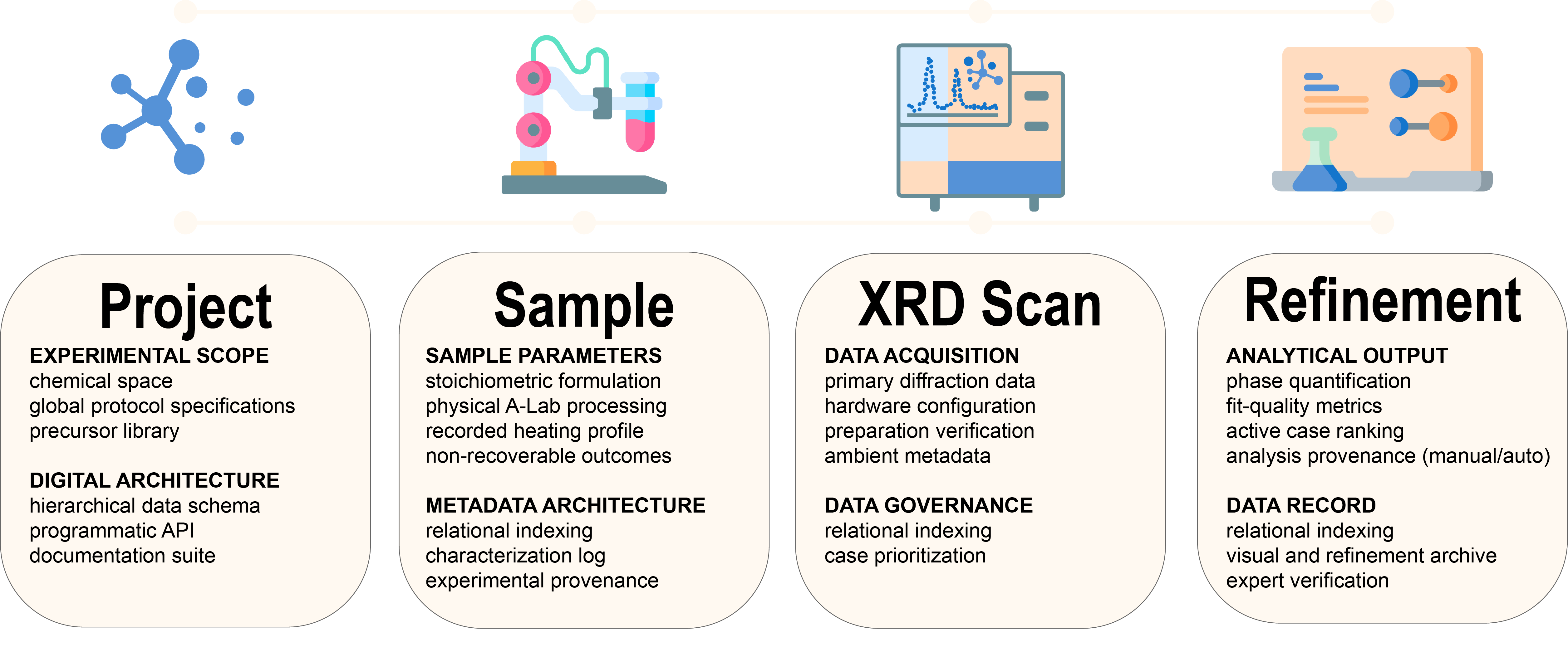}
    \caption{\textbf{The Precursor Genome nested data hierarchy.}
    Project: Defines chemical space boundaries and global metadata schemas.
    Sample: Houses experimental synthesis parameters, measured thermal profile, and relational indexing.
    Scan: Primary characterization record container for raw diffraction files (.xrdml), validated preparation flags, and hardware configuration.
    Refinement: Converged analytical results including phase weight fractions, quality metrics (e.g., $R_{wp}$), and expert verification of refinement analysis.
    }
    \label{fig:database-map}
\end{figure}

Due to the number of reactions, all raw diffractograms, visualized Rietveld refinements, identified products, and associated weight fractions are available via the Precursor Genome repository only (see Data and materials availability and ~\cite{PG_SciData}).
The Precursor Genome is a structured, machine-readable database of solid-state reactivity designed for automated analysis and AI/statistical modeling. The database follows a nested relational hierarchy (Figure~\ref{fig:database-map}) linking synthesis conditions to phase identification: Project containers define the chemical space; Sample objects store experimental metadata (e.g., precursor masses and dwell temperatures); and Scan objects hold raw diffractograms with associated Refinement objects containing CIF structure files, refinement parameters, and fit metrics (e.g., $R_{wp}$). The repository offers notebooks to provide easy access custom extraction and re-refinement.

\subsubsection*{Pairwise Solid-State Reaction Outcomes}

We categorize experimental outcomes into four distinct classes: \textit{unreacted}, \textit{transformed}, \textit{complete}, and \textit{partial}. The \textit{unreacted} classification indicates that only the original precursors were recovered after heating, while the \textit{transformed} designation is assigned when refinement of XRD characterization identifies reduction, oxidation, or decomposition products of individual precursors (e.g., Fe$_3$O$_4$ + O$_2$ $\rightarrow$ Fe$_2$O$_3$). A reaction is considered \textit{complete} if $\geq$90wt\% of the recovered product consists of phases incorporating ions from both precursors (e.g., Fe$_2$O$_3$ + Na$_2$CO$_3$ $\rightarrow$ NaFeO$_2$) or if one precursor is entirely consumed, suggesting a complete reaction to a stoichiometry deviating from the initial 1:1 ratio. A \textit{partial} reaction outcome refers to instances where Rietveld refinement identifies at least one product phase containing ions from both precursors, yet the weight fraction remains below 90wt\% with residual precursors or transformed phases in the recovered sample.

\begin{figure}
    \centering
    \includegraphics[width=0.5\linewidth]{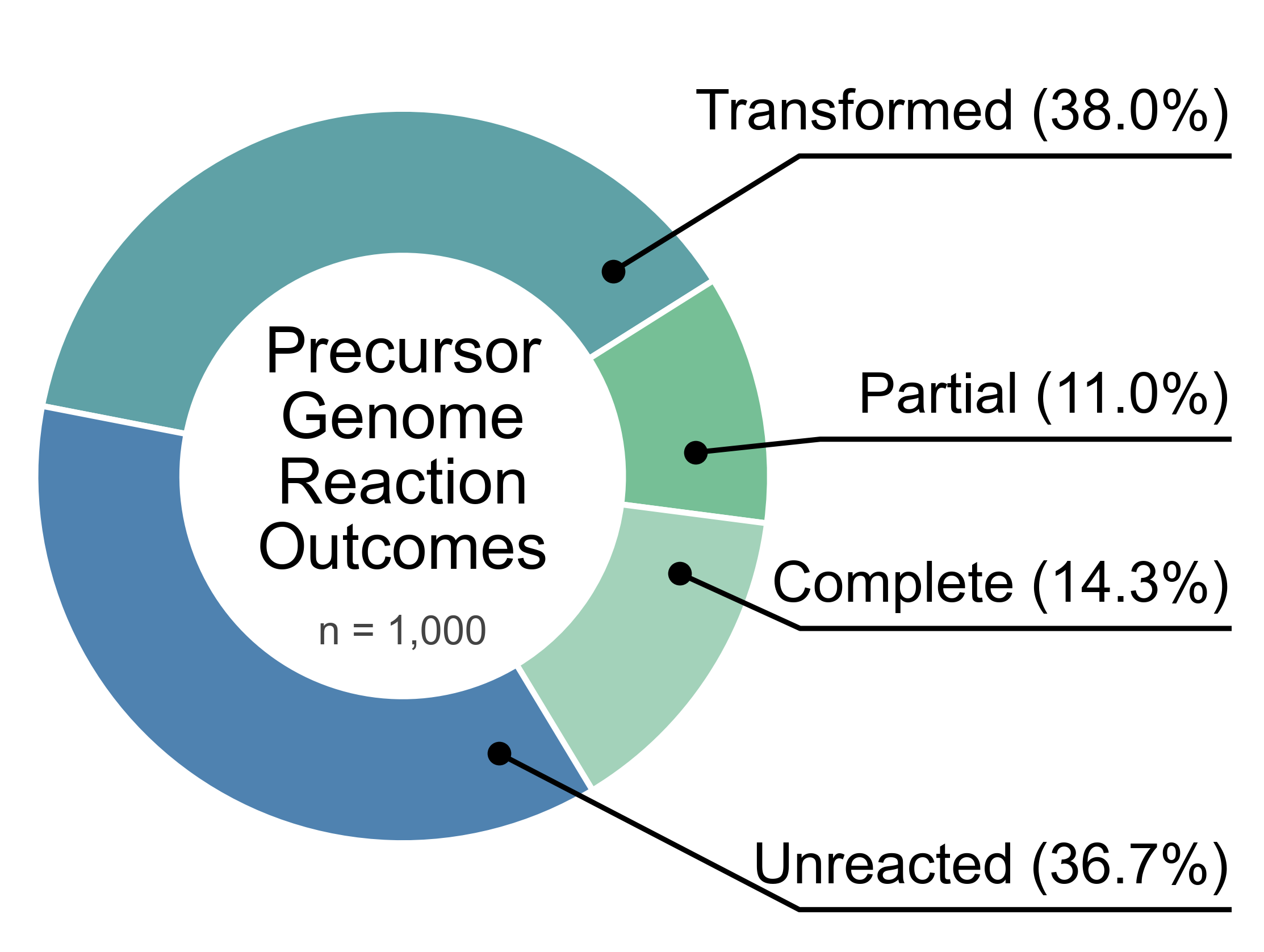}
    \caption{Ring chart visualizing all n\,=\,1,000 reaction outcomes: unreacted (36.7\%), transformed (38.0\%), partial (11.0\%), and complete (14.3\%).}
    \label{fig:all-results-pie}
\end{figure}

Figure~\ref{fig:all-results-pie} shows the distribution for all reactions in the Precursor Genome: unreacted (36.7\%), transformed (38.0\%), partial (11.0\%), and complete (14.3\%).
Notably, over 35\% of the precursor pairs showed no change at the Tammann temperature, a figure that rises to 74.7\% when including transformed reactions.
With complete and partial reactions accounting for only 25.3\% of the total outcomes, these results suggest that for a vast majority of bulk, powdered solid-state systems, the Tammann-rule heuristic significantly underpredicts the temperatures necessary for detectable reaction onset.

\begin{figure}
    \centering
    \includegraphics[width=1.0\linewidth]{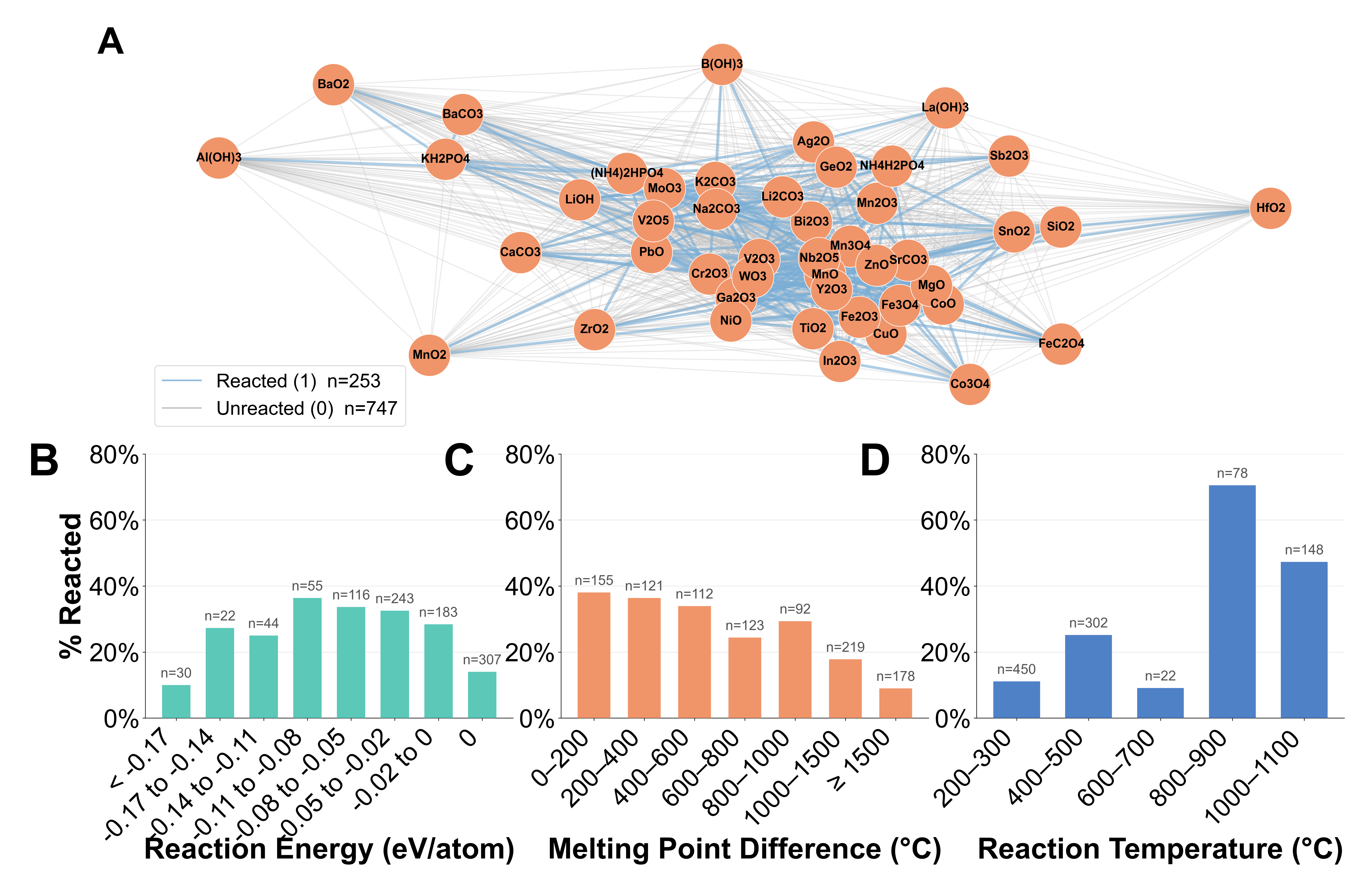}
    \caption{\textbf{Reaction outcomes of precursor genome.}
    (\textbf{A}) Network graph of precursor reaction outcomes constructed using force-directed layout (Fruchterman-Reingold algorithm implemented in NetworkX \cite{Fruchterman1991_NetworkGraph, Hagberg2008_NetworkX}). Edge weights are proportional to the fraction of reacted experiments per pair. Reacted pairs (n\,=\,253) are shown as blue, thicker edges and unreacted pairs (n\,=\,747) as gray edges.
    %
    %
    Histograms showing the percentage of precursor pairs that reacted as a function of (\textbf{B}) reaction energy, (\textbf{C}) melting point difference between precursor pairs, and (\textbf{D}) Tammann-rule predicted reaction temperature (200--1,100$^\circ$C).
    }
    \label{fig:reaction-outcome-categorization}
\end{figure}

We further analyze Precursor Genome outcomes by precursor, chemical series, reaction energy, and melting point difference. For this analysis, \textit{complete} and \textit{partial} outcomes are grouped as \textit{reactive}, while \textit{unreacted} and \textit{transformed} outcomes are grouped as non-reactive, as neither produces inter-precursor phase formation.
Figure~\ref{fig:reaction-outcome-categorization}A shows a force-directed network graph where nodes represent precursors and edges represent reaction outcomes; thicker blue lines indicate reactive pairs, thinner, gray lines non-reactive. Precursors with higher connectivity are pulled toward the center, providing a qualitative reactivity map at the Tammann temperature. Transition metal oxides such as manganese and iron, occupy the network core due to their broad reactivity, while peripheral precursors such as SnO$_2$, Sb$_2$O$_3$, and Al(OH)$_3$ are largely inert, reacting with at most three partners each. Notably, precursors with volatile groups (e.g., carbonates, hydroxides, ammonia) react less frequently than many transition metal oxides, despite the high thermodynamic driving force associated with gas release. This suggests that other features, such as melting point and reaction temperature, are also strong determinants of reactivity, and are explored in the following sections.

Fig.~\ref{fig:heatmap-element-type-only} shows a heatmap of reaction outcomes across three cation series: (\textit{i}) Group 1/2 alkali and alkaline earth metals, (\textit{ii}) transition metals, and (\textit{iii}) post-transition \textit{p}-block elements. Transition metal self-pairs are the most reactive at 40.5\%, followed by Group 1/2 self-pairs (29.4\%) and transition metal--Group 1/2 cross-pairs (21.8\%). Precursors with post-transition metals are consistently the least reactive, reacting with themselves only 9.1\% of the time and only modestly more when paired with other series (9.1\% with Group 1/2, 17.4\% with transition metals).

Figures~\ref{fig:reaction-outcome-categorization}B-D compare the fraction of reacted pairs across three descriptors: DFT-calculated reaction energy (B), melting point difference (C), and Tammann-rule reaction temperature (D); bar labels indicate the number of precursor pairs per bin. Reaction energy shows no strong trend with reactivity at the Tammann temperature. If anything, reactivity is slightly elevated at moderate energies (-0.11 to -0.08\,eV/atom), though the high-energy bin ($<$\,-0.170\,eV/atom) contains only 30 pairs and should be interpreted cautiously. Pairs with no predicted driving force ($ \Delta G_{rxn} = 0$, n\,=\,307) react least often (14.0\%).
Melting point difference and reaction temperature show clearer, opposing trends. As melting point difference increases, reactivity declines markedly: pairs with melting temperatures within 0–200\,$^\circ$C react 38\% of the time (n = 155), dropping to 18\% at 1000–1500\,$^\circ$C (n = 219) and 9\% above 1500\,$^\circ$C (n = 178). Reaction temperature shows the reverse: only 11\% of pairs react at 200–300\,$^\circ$C (n = 450), rising to 70\% at 800–900\,$^\circ$C (n = 78) and 47\% at 1000–1100\,$^\circ$C (n = 148).

Several inverse trends elucidate the limitations of these heuristics. For instance, fig.~\ref{fig:supp_positive_energy_bins} shows a binned analysis of reactions with non-negative reaction energies ($\Delta G_{rxn} \geq 0.0$ eV). While n = 264 samples remained unreacted, n = 43 precursor pairs successfully reacted either partially or completely despite the lack of a predicted driving force. We can similarly evaluate the normalized reaction temperatures ($T/T_{melt}$) for reacted samples to identify precursor-specific reactivity thresholds. As shown in fig.~\ref{fig:supp_precursor_onset_boxwhisker}, the effective reaction onset for many precursors occurs below the two-thirds melting point threshold, with an average of 0.60 of the reference temperature $T_{melt}$ per precursor across all reactions. Notably, the refractory oxides ZrO$_2$, MgO, HfO$_2$, Cr$_2$O$_3$, and Y$_2$O$_3$ all reacted near 0.4$T_{melt}$ when paired with precursors able to react at these lower temperatures, despite chemical intuition that they are relatively inert. In contrast, other precursors required temperatures at or above their Tammann threshold to initiate; CaCO$_3$ and MnO$_2$ reacted near 0.7$T_{melt}$, and Ag$_2$O only above 0.85$T_{melt}$.

Taken together, these trends suggest that melting point disparity and reaction temperature correlate with reactivity. However, the two are physically coupled to chemistry and cannot be fully disentangled from this analysis alone. The following sections use machine learning and statistical methods to more rigorously isolate the features driving reaction outcome.

\subsubsection*{Multivariate Models for Predicting Reaction Onset}

To move beyond Tammann's rule, we develop models to predict the onset temperature using discrete, single-temperature dwell experiments. This experiment format is more common than continuous measurements that explicitly track onset (e.g., thermogravimetric analysis, \textit{in situ} XRD). By leveraging existing synthesis outcomes, we demonstrate how reaction thresholds can be predicted using fundamental descriptors such as precursor identity, melting point, and reaction energy, augmented by a broader set of 
physicochemical descriptors derived with large-language model (LLM) assistance.

\begin{figure}
    \centering
    \includegraphics[width=1.0\linewidth]{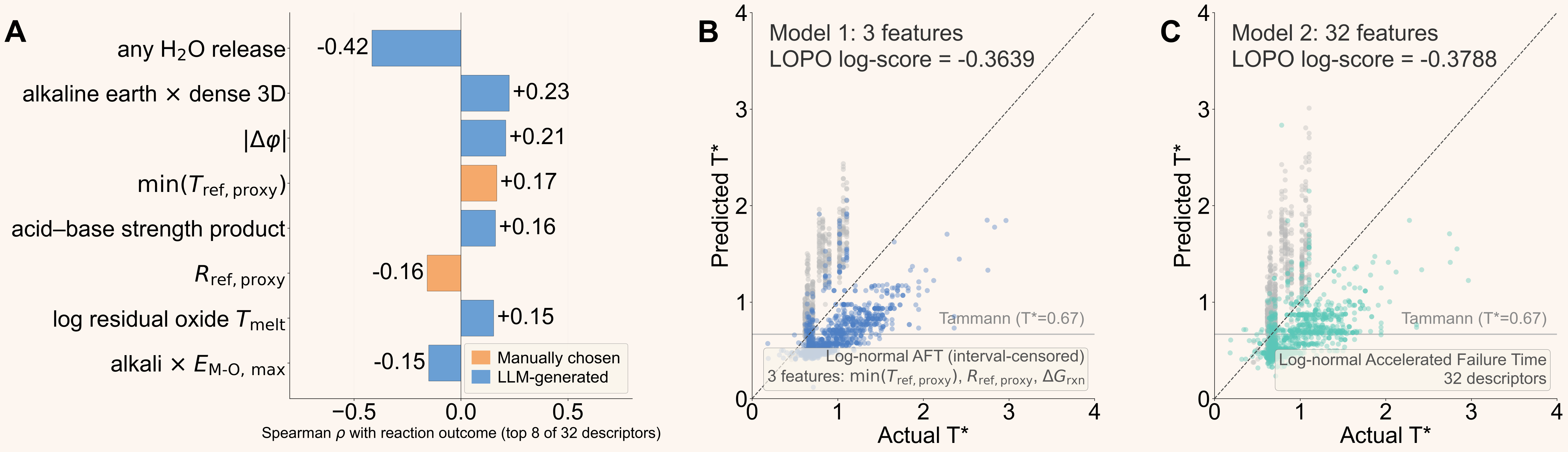}
    \caption{\textbf{Machine learning models to predict reaction onset temperature trained on data from this work and the dataset from Kononova et al. \cite{Kononova2019_OlgaDataSet}.}
    (\textbf{A}) Spearman correlation coefficients \cite{Spearman1904_CorrelationCoefficients} between physicochemical features and the binary reaction outcome (reacted vs unreacted), across the full PG\,+\,PG-LIT dataset.
    (\textbf{B}-\textbf{C}) Predicted versus actual $T^*$ ($T^* = T/T_\mathrm{ref,min}$) for two machine learning models, with reacted observations in color and unreacted in gray. LOPO log-scores are: (\textbf{B}) Model 1 (3 features): $-0.364$; (\textbf{C}) Model 2 (32 features): $-0.379$. The dashed line is drawn at 1:1, where reacted, colored points are expected at or below the parity line. The horizontal line marks the Tammann reference ($T^*=0.67$). 
    }
    \label{fig:model-building}
\end{figure}

We train on the combined Precursor Genome (PG) + Literature (PG-LIT) dataset: the 1,000 in-house pairwise trials of this work (PG) available for analysis, supplemented with solid-state reactions text-mined from the literature by Kononova et al.\ (PG-LIT) \cite{Kononova2019_OlgaDataSet}, totaling 1,866 experiments across 46 precursors.
Prior to model construction, we systematically evaluated pair-level descriptors to identify the primary drivers of reaction onset temperature. Spearman correlation coefficients ($\rho$) between the binary reaction outcome (reacted vs unreacted) and each descriptor were calculated across the full PG\,+\,PG-LIT training set; the top eight are shown in Figure~\ref{fig:model-building}A, with the complete 32-descriptor list provided in table~\ref{tab:spearman-full-reacted}.
The strongest correlation by a wide margin is the propensity to release H$_2$O (any H$_2$O release, $\rho = -0.42$). Beyond it no single descriptor dominates: the next strongest are a coupled alkaline-earth $\times$ dense-3D-host indicator ($\rho = +0.23$), the ionic-potential difference between ions ($|\Delta \phi|$, $\rho = +0.21$), the minimum proxy reference temperature (min($T_{ref,proxy}$), $\rho = +0.17$), the acid–base strength product ($\rho = +0.16$), the proxy reference-temperature ratio ($R_{ref,proxy}$, $\rho = -0.16$), the log residual oxide melting point ($\rho = +0.15$), and an alkali $\times$ maximum M-O bond energy term ($\rho = -0.15$).
%
This negative association, however, is most likely a consequence of the low \emph{verified} reference temperature of these hydroxide and hydrate precursors, which decompose at low temperature (mean $\min(T_\mathrm{ref}) = 252^\circ$C versus $1088^\circ$C for the remaining pairs without any H$_2$O release; Spearman $\rho = -0.63$ between H$_2$O-release propensity and the verified $\min(T_\mathrm{ref})$): such pairs are tested at correspondingly low Tammann temperatures, where reaction is intrinsically less probable, rather than indicating an intrinsic suppression of reactivity by water release itself. These trends are consistent with the reaction outcome categorization (Figure~\ref{fig:reaction-outcome-categorization}), where the fraction of reacting pairs declined with increasing melting-point difference and rose with the Tammann reaction temperature. Hence, absolute temperature appears to dominate over the Tammann-rule temperature.

We constructed two models at increasing levels of complexity, both log-normal accelerated failure time (AFT) regressions in which the latent onset temperature is treated as log-normal and each single-temperature trial as a one-sided, interval-censored bound on it (Materials and Methods, Eqs.~\ref{eq:aft-onset}--\ref{eq:aft-react}). Model 1 uses three features, minimum precursor thermal transition temperature, melting-point ratio, and DFT reaction energy; these three features are a subset of Model 2's 32 descriptors, whose remaining 29 are LLM-generated. We favored these interpretable AFT models over black-box alternatives such as neural networks, which require substantially larger training sets to generalize reliably.

In Figure~\ref{fig:model-building}B--C, we show predicted versus actual normalized temperature $T^*$ ($T^* = T/T_\mathrm{ref,min}$, where $T_\mathrm{ref,min}$ is the minimum per-precursor melting or decomposition temperature of the pair) for both models. Each plotted point corresponds to a single reaction experiment, colored by outcome (reacted in color, unreacted in gray).
Figure~\ref{fig:model-building}B depicts Model 1, the three-feature AFT trained on minimum precursor thermal transition temperature, melting point ratio, and DFT reaction energy. Figure~\ref{fig:model-building}C shows Model 2, the 32-feature accelerated failure time model. In both panels, the horizontal line marks the Tammann reference ($T^* = 0.67$), and the dashed line marks the 1:1 parity line. Both models are log-normal AFTs trained on the same PG\,+\,PG-LIT dataset with the same decomposition-product proxy melting points; they differ in feature count (3 vs 32) and in that Model 2 carries a weak ridge penalty ($\lambda$ = 0.1) chosen by cross-validation (Materials and Methods). The LOPO log-scores are -0.364 (Model 1) and -0.379 (Model 2): the compact three-feature model attains the better held-out censored log-probability.

The expected behavior under the censored likelihood framework is visible in both panels: reacted observations (colored) cluster predominantly below the parity line, indicating that the experimental dwell temperature exceeded the predicted onset, consistent with reaction having occurred. Unreacted observations (gray) cluster predominantly above the parity line, indicating that the experimental dwell temperature fell below the predicted onset, consistent with no reaction. In both models, the primary failure mode is reacted observations appearing above the parity line, corresponding to pairs where the model overpredicted the onset temperature. These cases represent reactions that occurred at lower temperatures than the model anticipated, and are likely underrepresented in the training data at low $T^*$ due to the higher percentage of reactions completed well above their onset in the combined data set ($\approx 90$\%).

\begin{figure}
    \centering
    \includegraphics[width=1.0\linewidth]{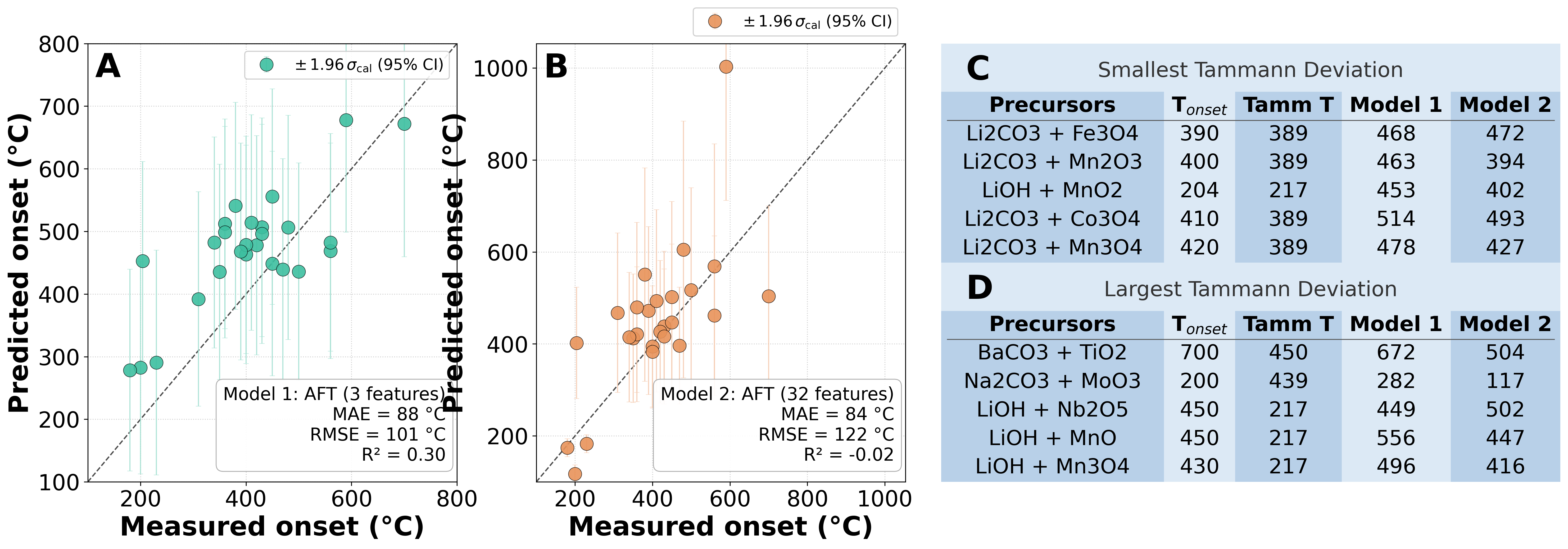}
    \caption{\textbf{Comparison of predictive models for measured solid-state reaction onset temperature.}
    (\textbf{A}) Measured versus predicted onset temperatures ($T_{onset}$) for the three-feature AFT model (Model 1), with error bars showing $\pm1.96\,\sigma_\mathrm{cal}$ (95\% CI), with a calculated $R^2$ of 0.30.
    (\textbf{B}) Measured versus predicted onset temperatures ($T_{onset}$) for the 32-feature AFT model (Model 2), with $\pm1.96\,\sigma_\mathrm{cal}$ (95\% CI) error bars, with a calculated $R^2$ of -0.02.
    (\textbf{C}) Tabulated subset of reactions exhibiting the smallest deviation from the Tammann-rule temperature.
    (\textbf{D}) Tabulated subset of reactions exhibiting the largest deviation from the Tammann-rule temperature.
    Tables in C and D provide the precursor pairs, measured experimental onset temperatures, Tammann-rule temperatures, and predicted onset temperatures for both Model 1 (3-feature AFT) and Model 2 (32-feature AFT).
    }
    \label{fig:validation-insitu}
\end{figure}

To assess model dependence on literature data, we retrained both models utilizing only experimental data from our in-house dataset of 1,000 experiments (fig.~\ref{fig:models_comparison_PGonly}). Both models show a higher fraction of reacted experiments above the parity line and unreacted experiments below the parity line compared to the models trained on the combined precursor genome and text-mined dataset, indicating more frequent overprediction and underprediction of onset temperature respectively. This is reflected in the more negative LOPO log-scores of $-0.492$ for Model 1 and $-0.484$ for Model 2. The reacted/unreacted Spearman structure on PG-only data remains dominated by melting point (fig.~\ref{fig:models_comparison_PGonly}A), but the smaller, less chemically diverse dataset yields weaker models that overpredict onset more strongly. Taken together, this demonstrates the importance of training on large, chemically diverse experimental datasets, and the necessity for this work to supplement in-house data with the Kononova et al.\ dataset \cite{Kononova2019_OlgaDataSet}.

We consolidated all models into the Precursor Genome Onset (PGO) Calculator that can be used and actively trained on real experimental data. This calculator is freely available and open source (see Data and materials availability).

To demonstrate model efficacy, we validated our onset temperature predictions against experimental onset temperatures measured by \textit{in situ} X-ray diffraction \cite{McDermott2023_rxnnetwork2, Szymanski2024_thermoregime}. This validation set comprises 26 reactions whose precursors are all within our precursor genome. For this analysis the 26 corresponding Precursor Genome samples and the 218 literature entries involving these pairs were removed from training, so the models are evaluated on pairs they have not seen; retaining those rows lowers the apparent MAE to 75 °C (Model 1) and 78 °C (Model 2). The complete list of reactions and input parameters is provided in table~\ref{tab:reaction-energies-onset}.

In Figure~\ref{fig:validation-insitu}, we compare predicted onset temperatures using the three-feature AFT (Model 1, Panel A) and the 32-feature AFT (Model 2, Panel B).
The Tammann rule achieves MAE = 131\,$^\circ$C [95\% CI 100, 160], RMSE = 152\,$^\circ$C, and R$^2 = -0.59$ on this held-out set. Model 1 achieves MAE = 88\,$^\circ$C [70, 108], RMSE = 101\,$^\circ$C, and R$^2$ = 0.30, a paired improvement of 43\,$^\circ$C [2, 81] over the Tammann rule. Model 2 achieves MAE = 84\,$^\circ$C [53, 121], RMSE = 122\,$^\circ$C, and R$^2 = -0.02$; the paired difference between the models is -4\,$^\circ$C [-35, +33], so the 29 additional descriptors do not improve held-out accuracy.

Visual inspection of Panel A shows that Model 1, fit raw and unregularized on three features, tracks the measured onset across the full range rather than collapsing toward the dataset mean: it spreads correctly from the lowest-onset carbonate pairs to refractory pairs such as BaCO$_3$ + TiO$_2$. Trained on the combined PG + PG-LIT dataset, Model 1 already outperforms the Tammann rule, so the three physics features alone carry useful onset information.

Model 2 predictions are shown in Figure~\ref{fig:validation-insitu}B. Its MAE matches Model 1's, but its RMSE is larger and its R$^2$ is negative, indicating that the richer 32-descriptor feature set does not capture pair-specific onset behavior beyond what the three physics features already provide. When trained on only the in-house PG data, both models degrade substantially: Model 1 reaches MAE = 180\,$^\circ$C (vs 88\,$^\circ$C on the combined dataset) and Model 2 MAE = 189\,$^\circ$C (vs 84\,$^\circ$C) (fig.~\ref{fig:insitu_validation_PGonly}), confirming that the models benefit from the larger size and broader chemical diversity of the combined PG\,+\,PG-LIT dataset (fig.~\ref{fig:models_comparison_PGonly}).

Tables shown in the final panel as Figures~\ref{fig:validation-insitu}C and D highlight some reactions for which the measured onset temperature deviates most from the Tammann-rule temperature. The table containing the largest deviations (D) is dominated by LiOH paired with high-melting transition metal oxides (Nb$_2$O$_5$, MnO, Mn$_3$O$_4$) and BaCO$_3$ + TiO$_2$, where measured onsets of 430-700 $^\circ$C fall well above the Tammann-rule temperature prediction. For instance, for Na$_2$CO$_3$ + MoO$_3$ (measured onset 200 $^\circ$C), the Tammann-rule predicts 439 $^\circ$C, Model 1 predicts 282 $^\circ$C, and Model 2 predicts 117 $^\circ$C; both models move sharply below the Tammann estimate toward the measured onset, with Model 2 capturing it most closely.

The smallest deviations between the Tammann-rule temperature and measured onset (shown in Figures~\ref{fig:validation-insitu}C) involve Li$_2$CO$_3$ paired with transition metal oxides (TiO$_2$, MnO$_2$, Fe$_3$O$_4$, Co$_3$O$_4$) and LiOH + MnO$_2$. The closest
agreement overall is Li$_2$CO$_3$ + Fe$_3$O$_4$ (measured 390 $^\circ$C), where the Tammann rule deviates by only 1 $^\circ$C. Here both models overpredict (468\,$^\circ$C and 472\,$^\circ$C).

Comparing the high- and low-deviation sets shows that precursor identity alone does not explain the errors: LiOH appears among both the smallest and the largest Tammann deviations (Figures~\ref{fig:validation-insitu}C,D), so the lithium source is not intrinsically early- or late-reacting.
The discriminating information sits instead in the fitted coefficients. After z-scoring, Model 2's largest weights fall on the melting-point inputs ($\beta_{min}(T_{ref,proxy}) = +0.13, \beta\_R_{ref,proxy} = +0.09$), followed by chemical-reactivity descriptors: the ionic-potential mismatch $|\Delta \phi|$, CO$_2$ and H$_2$O release, the M–O bond energies, and the DFT-calculated reaction energy. The reaction-energy term is what lets both models flag the carbonate–acidic-oxide pairs that react hundreds of degrees below their Tammann estimate; for Na$_2$CO$_3$ + MoO$_3$ the measured onset is 200\,$^\circ$C against a Tammann estimate of 439\,$^\circ$C, which Model 1 recovers at 282\,$^\circ$C and Model 2 at 117\,$^\circ$C, and the same pattern holds for Li$_2$CO$_3$ + V$_2$O$_5$.

Model 1's calibrated uncertainty accounts for most of its validation associated errors: 25 of the 26 in situ pairs fall within the $\pm1.96\,\sigma_\mathrm{cal}$ (95\%) prediction interval (Figure~\ref{fig:validation-insitu}A). The one exception is LiOH + MnO$_2$ (measured 204\,$^\circ$C, predicted 453\,$^\circ$C, interval $\pm159\,^\circ$C), where the prediction misses the measurement by more than the interval half-width. 
Model 2's intervals cover 21 of 26 pairs (Figure~\ref{fig:validation-insitu}B), missing LiOH + MnO$_2$ (predicted 402$^\circ$C, $\pm121\,^\circ$C) and BaCO$_3$ + TiO$_2$ (measured 700$^\circ$C, predicted 504$^\circ$C, $\pm196\,^\circ$C) at opposite extremes of the measured onset range, as well as three pairs where its intervals are too narrow ($\pm20\,^\circ$C).
Resolving why
their errors exceed the calibrated uncertainty, whether through additional dwell-temperature measurements, targeted descriptors, or a revised uncertainty model, warrants further investigation.
Our models are trained on binary dwell-temperature outcomes rather than directly measured onset temperatures, with most precursor pairs tested at a single dwell temperature and a subset (199 pairs) carrying outcomes at two or more dwell temperatures after incorporating the PG-LIT literature data, yet Model 1 achieves MAE = 88\,$^\circ$C on held-out \textit{in situ} measurements, comparable to the precision needed for practical synthesis planning. This highlights both the sensitivity of onset temperature to reaction mechanism and the potential value of broadly sampled datasets with diverse precursor chemistries for training onset temperature prediction models.

\subsection*{Discussion}

In this work, we executed 1,035 solid-state reactions across 46 common precursors at a temperature set by the Tammann rule, which predicts reactivity at two-thirds the melting point of the lowest-melting precursor. Of the 1,000 analyzable pairs, a significant majority (74.7\%) of these analyzable pairings fail to react, confirming that the Tammann heuristic fails in both directions, though most consistently underpredicts, the temperature required to initiate chemical reactivity. To address this discrepancy, we identified the governing physical and chemical descriptors of reaction onset and developed the Precursor Genome Onset (PGO) Calculator, which integrates experimental outcomes from binary dwell-temperature observations with \textit{a priori} descriptors to predict onset temperature. 
Validated against held-out \textit{in situ} XRD datasets, the three-descriptor model achieves MAE = 88\,$^\circ$C against 131\,$^\circ$C for the Tammann rule, while a 32-descriptor extension does provide further improvement on temperature prediction \cite{McDermott2023_rxnnetwork2, Szymanski2024_thermoregime}.
The analysis of this dataset and the development of our retrainable framework demonstrate that reaction onset temperature can be predicted from inexpensive binary dwell-temperature observations, and that predictive accuracy improves continuously as new experimental outcomes are contributed.

A central finding from the analysis of reaction outcomes is that no single descriptor is sufficient to predict onset temperature. Utilization of individual descriptors, whether thermodynamic, melting point, or energetic terms, consistently falls short of capturing the complexity of the experimental data.
For instance, reaction energy alone yields essentially no predictive power (R$^2 \approx$0, test RMSE$ = 0.153$, fig.~\ref{fig:supp_energy_only}). Deviations of this magnitude cannot be attributed to standard DFT errors, which typically range around 0.1~eV~atom$^{-1}$ or less \cite{virtual_furnace_TRI1, Gibbs_Bartel}. Melting points (fig.~\ref{fig:supp_meltingpoints_only}, fig.~\ref{fig:models_comparison_PGonly}) or nonlinear interaction terms alone (fig.~\ref{fig:supp_nonlinear_only}) similarly fall short, with test RMSE $= 0.142$ and $0.143$ respectively (see Supplementary Text, section~S5.2).
The performance of Model 1 (3-descriptor) in Figure~\ref{fig:validation-insitu} further confirms that melting point and reaction energy, while necessary, are insufficient descriptors of onset temperature.
Together, these results argue against indiscriminate application of Tammann's rule for synthesis planning, and reinforce that a robust predictive framework must integrate multiple distinct mechanisms, spanning thermodynamics, melting-point, and learned per-precursor reactivity.
%

%

A striking feature of the precursor genome dataset is the significant impact of precursor identity on reaction outcome.
Recorded \textit{reacted} outcomes vary widely across the 46 precursors, from V$_2$O$_3$ (51.2\%) to Al(OH)$_3$ (2.2\%).
Unfortunately, the least reactive precursors, including Al(OH)$_3$ (2.2\%), MnO$_2$ (4.4\%), and Sb$_2$O$_3$ (6.7\%), all decompose or reduce before melting, rendering their listed melting points and chosen temperatures within this work potentially misleading.
Conversely, the most reactive precursors, V$_2$O$_3$ (51.2\%), ZnO (51.2\%), and SrCO$_3$ (48.9\%), are all reactants whose high melting/decomposition temperatures might accurately reflect their reactive behavior, suggesting that for well-behaved materials, melting point remains a reliable reactivity guide.
The manganese oxide series illustrates this sharply: MnO (42.2\%), Mn$_3$O$_4$ 
(45.5\%), Mn$_2$O$_3$ (17.8\%), and MnO$_2$ (4.4\%) span a wide reactivity range 
despite sharing the same cation. Most strikingly, MnO$_2$, despite its high 
oxidation state and the thermodynamic drive to reduce, remains unreacted in the 
majority of experiments, without even reducing to a lower manganese oxide, contrary to expectations. MnO$_2$ decomposes to Mn$_2$O$_3$ at $\sim$535~$^{\circ}$C, meaning most experiments were conducted below its effective reactive threshold, suppressing both reaction and reduction.

Yet, evaluating individual precursor reactivity or energetics cannot fully explain why 64.7\% of unreacted pairs are thermodynamically favorable ($\Delta E < 0$).
%
%
Interestingly, regression analysis illustrates that how far above the lower-melting precursor's minimum reaction temperature an experiment sits is no more predictive than the equivalent quantity for the harder precursor (-0.037 vs. -0.036) (see Supplementary Text, section S5.2).
MgO + WO$_3$ makes this concrete: despite MgO's Tammann temperature of $\approx 1810\,^\circ$C, the pair reacted completely at 900\,$^\circ$C (T$^*$ = 0.38 relative to MgO), driven by WO$_3$ as the lower-melting partner. The intrinsic reactivity of individual precursors and the sensitivity of reaction onset to the partner have a direct practical implication: when selecting a synthesis temperature for a new precursor pair, the lower-melting precursor's thermal floor sets where reaction becomes possible, but the harder precursor's identity determines whether it occurs.

Our work is necessarily limited by its experimental framework: single-temperature dwell experiments conducted \textit{ex situ} provide limited insight into reaction progression, and more granular temperature sampling or longer heating times could modify some observed outcomes. Experimentally, dwell temperatures were rounded to 100\,$^\circ$C and clipped to 200–1,100\,$^\circ$C. Thirty-five samples (3.4\%) could not be characterized automatically because they melted, fused or vitrified, so 25.3\% is a lower bound for ''reacted". Additionally, due to the 1-hour dwell, slow or minor reactions are likely not captured.
Notably absent from our feature set are descriptors known to influence reaction onset including \textit{in situ} reaction kinetics, particle size, and heating rate for which well-documented, consistently measured datasets do not yet exist at the scale required for model training. As these datasets emerge, the AFT framework developed here is well-positioned to incorporate them: each new binary synthesis outcome provides a valid censored observation that can be directly included in retraining, requiring no point measurement of onset temperature and no modification to the model architecture.
Reports of automation for coupled \textit{in situ} synthesis-characterization and \textit{in silico} function-characterization represent exciting opportunities for expanding the experimental basis of data-driven synthesis science.
We ultimately use this work to advocate for data-driven theories of reaction onset grounded in consistently collected, broadly sampled experimental data; frameworks that provide quantified confidence in their predictions while explicitly identifying the conditions under which outliers arise.

Until now, evidence for and application of Tammann's rule has been largely confined to small datasets and case studies \cite{Tammann_rule_1, Tammann_rule_3, Tammann_rule_2}, like many other rules of thumb in synthesis, rather than systematic large-scale evaluation.
Huo et al. proposed an extended Tammann rule trained on text-mined synthesis temperatures, adjustable by materials class and chemistry \cite{Huo2022_RevisedTammanRule}.
However, these models were trained on reported successful synthesis recipes that are neither controlled for methodology nor accompanied by complete experimental metadata, which is a limitation that is particularly consequential given the well-documented biases introduced by trial-and-error reporting of solid-state synthesis conditions \cite{Sun2025_SynthesisReflections}.
This concern extends beyond reaction temperature to other experimental factors such as mixing protocol, particle size, and heating rate, whose influence on reaction onset is real but rarely reported consistently.
The challenge is compounded when moving beyond solid-state methods to hydrothermal, sol-gel, or deposition routes, where additional methodology-dependent and experimentalist-dependent sources of variation further complicate cross-dataset comparison.

Large, self-consistent datasets are essential for formulating and testing theory in a domain as sensitive to experimental variation as solid-state synthesis.
Synthesis science has historically operated in a small-data regime, and even well-reasoned theoretical frameworks such as the Tammann--rule or maximum Gibbs driving force \cite{Ceder_Thermo_Kinetics} can fail to generalize beyond the specific conditions under which they were developed.
Self-driving laboratories represent a promising path toward curating the large, FAIR, and caveat-documented datasets that the synthesis community needs \cite{Alab, Tom2024_SDL-Review}, provided that experimental metadata, such as particle size, heating rate, and mixing protocol, are recorded consistently alongside reaction outcomes.

%
%

\subsection*{Materials and Methods}\label{sec:mat_method}

\subsubsection*{Experimental Synthesis and Characterization}

Reactions were performed via conventional, powdered solid state (ceramic) synthesis. Reagents were purchased from Sigma-Aldrich (Millipore Sigma) at generally a nominal purity of $\geq$ 95\%. Due to the significant number of reagents used, the full list of precursors used, CAS numbers, manufacturer, and purity are included in the Supplementary Materials (table~\ref{tab:precursor-purchase-info}). All precursors were screened for significant hygroscopicity, and characterized by XRD and SEM/EDS prior to use.
%
%
Depending on particle size and mechanical properties, select precursors underwent pre-processing via ball milling or manual grinding with a mortar and pestle to facilitate automatic dispensing and accurate weighing 
(see table~\ref{tab:precursor-prep-info}). 

Solid state reactions were executed using the A-Lab, an autonomous and automated solid state synthesis laboratory with full details of the infrastructure described by N. Szymanski, \textit{et al.} \cite{Alab}. Briefly, the A-Lab performs solid state reactions in three stages that emulate manual procedures: (\textit{i}) a precursor preparation stage, (\textit{ii}) a heating stage, and (\textit{iii}) a sample recovery and characterization stage. Each stage utilizes a 6-axis robot arm that is programmed with high-precision waypoints to transport samples between modular device stations as implemented within AlabOS by Y. Fei, \textit{et al.} \cite{Alab_os}
To prepare samples, powders were dispensed and weighed under ambient conditions using an automatic dispenser balance (Quantos, Mettler Toledo) within a 0.2\% weight threshold into a plastic vial containing ten zirconia balls.  Ethanol was added, and the sample was mixed at 2000\,rpm for 10 minutes to form a slurry. The slurry was pipetted from the vial and dispensed into a crucible. Following evaporation at 80\,$^\circ$C, the precursors formed a densified, mixed mass at the base of the crucible. The samples were then moved to furnaces and heated according to the programmed profile. The default heating profile was 2\,$^\circ$C/min to 300\,$^\circ$C, followed by 5\,$^\circ$C/min up to the maximum dwell temperature, followed by a 1-hour hold, and then uncontrolled cooling in the furnace to room temperature. The samples were recovered autonomously by vertically shaking a crucible covered with a plastic cap with a 10-mm alumina ball inside to dislodge (mill) the sample from the crucible wall and generate powder. The resulting powder was sieved through a steel mesh and pressed into a disc for automated XRD characterization. Remaining powder was stored inside a uniquely-labeled plastic vial.

Synthesis dwell temperatures were calculated following the Tammann's rule formalism established by Tammann \cite{Tammann_rule_3}, and Merkle and Maier \cite{Tammann_rule_1}. In this work, the synthesis temperature is defined as two-thirds of the minimum per-precursor reference temperature ($T_{ref}$) among the precursor set, where $T_{ref}$ is each precursor's first thermally accessible transition: a melting, decomposition, reduction, sublimation, or phase-transformation temperature depending on the precursor (listed in table~\ref{tab:precursor-melting-points}). To accommodate high-throughput batched heating, the dwell temperature ($T_{dwell}$) was determined by rounding the Tammann estimate to the nearest 100\,$^\circ$C. Therefore, the synthesis dwell temperature in Kelvin for a pairwise reaction is set at:
\begin{equation}
    T_{dwell} = 100 \lfloor \frac{2}{3} \left( \mathrm{min}\left( T_{ref, 1}, T_{ref, 2} \right) \right)/100 \rfloor.
\end{equation}
Due to furnace operation limits, the dwell temperature range was constrained to a range of 200\,$^\circ$C to 1,100\,$^\circ$C. Of the 1,035 samples, 26 could not be recovered automatically as powder (table~\ref{tab:errored-samples}) and 9 were excluded for poor XRD refinement analysis, leaving 1,000 for analysis.

Sample characterization was performed primarily by powdered x-ray diffraction (XRD) on an Aeris Minerals diffractometer (Malvern Panalytical) utilizing Cu K$\alpha$ radiation.
Diffractograms were collected over a $2\theta$ range of 10$^\circ$ to 100$^\circ$ for an 8-minute duration, with a step size of $\approx$0.01\,$^\circ$ and an active length of 5.542\,$^\circ$. %
Precursor morphology and elemental composition were evaluated via scanning electron microscopy (SEM) and energy-dispersive X-ray spectroscopy (EDS) using a Thermo Fisher Phenom XL G2 system.

Automatic phase identification and Rietveld refinement were first conducted using Dara, and afterwards evaluated by human experts. Dara \cite{Dara, BGMN} is an automated phase identification and Rietveld refinement algorithm for powder XRD pattern analysis which ranks possible phase sets based on a quality of fit figure of merit. Initial structure searches for phase identification were performed by Dara on known crystal structures from the Inorganic Crystal Structure Database (ICSD) and the Crystallography Open Database (COD) \cite{ICSD, COD_1, COD_2, COD_3, COD_4, COD_5, COD_6, COD_7, COD_8, COD_9}. Lattice parameters were permitted to vary by 1\% when permitted by symmetry. A preferred orientation model with fourth-order spherical harmonics (gewicht=SPHAR4) is used during refinement. Crystalline size parameters (k\textsubscript{1}, B\textsubscript{1}) and microstrain effects (k\textsubscript{2}) were constrained isotropically in the range 0.0 to 0.001 with initial value of 0. Human experts performed final confirmation of phase identification and refinement. In the case of disagreement between Dara and experts, human experts performed manual refinement.

\subsubsection*{Thermodynamic Feasibility}

Reaction energies were calculated by retrieving DFT energies from Materials Project \cite{MaterialsProject} via the pymatgen API \cite{pymatgen, MaterialsProject, Ong2015_mpapi_2}. DFT energies were used to construct phase diagrams \cite{MaterialsProject_PhaseDiagram}, reaction interfaces \cite{MaterialsProject_ReactionInterface1, MaterialsProject_ReactionInterface2}, and reaction networks \cite{McDermott_reactionNetwork_model} at 0\,K and at the Tammann's rule temperature ($T_t$) utilizing physical descriptor fitting to estimate finite-temperature Gibbs free energies \cite{Gibbs_Bartel}.
For the primary analysis, the reported reaction energy ($G_{rxn}$) corresponds to the value calculated for the $T_{dwell}$.

\subsubsection*{Machine Learning Model Development}

Two predictive models for reaction onset temperature were developed as log-normal accelerated failure time (AFT) regressions trained by interval-censored maximum likelihood. These are two distinct ingredients. The AFT is the model: it represents the latent onset temperature as a log-normal function of the descriptors (Eq.~\ref{eq:aft-onset} below). The interval censoring is the training scheme: because the onset temperature is never measured directly in a binary dwell-temperature trial, each reaction outcome is treated as a bracket on the latent onset rather than an exact value, and the coefficients are estimated by maximizing the resulting interval-censored likelihood. For a reaction observed at dwell temperature $T_\mathrm{dwell}$, a positive outcome (reacted) brackets the onset to $T_\mathrm{onset} \leq T_\mathrm{dwell}$, while a negative outcome (unreacted) brackets it to $T_\mathrm{onset} > T_\mathrm{dwell}$. Concretely, the latent onset temperature is modeled as
\begin{equation}\label{eq:aft-onset}
\log T_\mathrm{onset} = \beta^{\top} x + \sigma\varepsilon, \qquad \varepsilon \sim \mathcal{N}(0,1),
\end{equation}
so that $T_\mathrm{onset} = \exp(\beta^{\top} x + \sigma\varepsilon)$ is log-normally distributed with median $\exp(\beta^{\top} x)$. The probability that a pair reacts at dwell temperature $T_\mathrm{dwell}$ is the probability that its onset lies at or below that temperature,
\begin{equation}\label{eq:aft-react}
P(\mathrm{react} \mid x, T_\mathrm{dwell}) = P(T_\mathrm{onset} \leq T_\mathrm{dwell}) = \Phi\!\left(\frac{\log T_\mathrm{dwell} - \beta^{\top} x}{\sigma}\right),
\end{equation}
which links the continuous onset prediction to the binary reaction outcome; the coefficients $\beta$ and scale $\sigma$ are fit by maximizing this censored likelihood \cite{Tobin1958_Survival-Regression}. Here $\Phi$ is the cumulative distribution function of the standard normal distribution, so $\Phi(z)$ gives the probability that a standard normal variable falls at or below $z$; it maps the standardized gap between the log dwell temperature and the predicted log onset, $(\log T_\mathrm{dwell} - \beta^{\top}x)/\sigma$, onto the probability that the reaction has already initiated by $T_\mathrm{dwell}$.
Models were trained on 1,000 analyzable solid-state reaction outcomes collected in this work, supplemented by literature data extracted from Kononova et al. \cite{Kononova2019_OlgaDataSet}. Literature reactions were filtered to include only solid-state syntheses and further restricted to precursor pairs represented in our experimental dataset, yielding a combined training set of 1,866 reactions.

Features predictive of $T^*$ were drawn from three sources: experimentally reported melting points, DFT-calculated reaction energies, and a set of physically motivated descriptors generated with large language model (LLM) assistance (Claude Opus 4.7, Anthropic) and subsequently filtered by expert review. 
Features containing the experimental heating temperature in their construction were explicitly excluded to prevent data leakage. The final descriptor set used in Model 2 comprises 32 features spanning melting point proxies, M-O bond energies, H$_2$O and CO$_2$ release enthalpies, acid-base interaction descriptors, framework topology indicators, and DFT reaction energies (see Supplementary Materials for full list).

Reaction outcomes were classified as binary targets: experiments yielding partially or completely reacted products were assigned a positive label, while unreacted and transformed outcomes were assigned a negative label. The homologous dwell temperature was defined as $T^* = T_\mathrm{exp}/T_\mathrm{ref,min}$, where $T_\mathrm{exp}$ is the experimental dwell temperature and $T_\mathrm{ref,min}$ is the minimum per-precursor reference temperature of the pair. Importantly, this reference temperature ($T_\mathrm{ref,min}$) is not strictly a melting point: it is the first thermally accessible phase-transformation point (PTP) of each precursor, i.e. the lowest temperature at which the solid undergoes a transition under standard atmosphere. Across the precursors used in this work, this PTP is a true melting point for most; for 15 of them it is instead a decomposition or reduction temperature (e.g. CaCO$_3$, BaCO$_3$, Al(OH)$_3$, MnO$_2$). Using the first accessible PTP rather than the melting point matters because many common oxide and salt precursors decompose or reduce well below their nominal melting points, and it is this lower transition that gates the available reaction pathway. Each reference temperature and its transition type were manually verified against standard sources (CRC/Lange's Handbook, American Elements, CommonChemistry) and are tabulated in the Supplementary Materials (see table~\ref{tab:precursor-melting-points}). For the melting-point \emph{features} of both models, the 15 precursors whose first transition is a decomposition or reduction are assigned the melting point of their primary decomposition or reduction product, which we term the \emph{proxy reference temperature} $T_\mathrm{ref,proxy}$ (a "decomposition-product proxy"; e.g. CaCO$_3\to$CaO, BaCO$_3\to$BaO, MnO$_2\to$Mn$_3$O$_4$ and Mn$_2$O$_3\to$Mn$_3$O$_4$, Al(OH)$_3\to$Al$_2$O$_3$), on the basis that the daughter phase governs subsequent reactivity; for precursors that melt without prior decomposition, $T_\mathrm{ref,proxy}$ equals the verified reference temperature. The melting-point features are then the pairwise minimum $\min(T_\mathrm{ref,proxy})$, the ratio $R_\mathrm{ref,proxy} = \max(T_\mathrm{ref,proxy})/\min(T_\mathrm{ref,proxy})$, and the difference $\Delta T_\mathrm{ref,proxy}$. The verified reference temperature ($T_\mathrm{ref}$) is retained for the $T^*$ normalization and the Tammann baseline, so only the model features use the proxy.

\textit{Model 1} is a log-normal AFT on three features: the minimum proxy reference temperature $\min(T_\mathrm{ref,proxy})$, the ratio $R_\mathrm{ref,proxy} = \max(T_\mathrm{ref,proxy})/\min(T_\mathrm{ref,proxy})$, and the DFT reaction energy per atom $\Delta E_\mathrm{rxn}$:
\begin{equation}
\log(T_\mathrm{model\;1}) = \beta_0 + \beta_1 \min(T_\mathrm{ref,proxy}) +
\beta_2 R_\mathrm{ref,proxy} + \beta_3 \Delta E_\mathrm{rxn}
\end{equation}
with fitted coefficients on the raw (unscaled) features (T in K) $\beta_0 = +5.884$, $\beta_1 = 3.66 \time 10^{-4}$\,K$^{-1}$, $\beta_2 = 0.203$, $\beta_3 = 1.21$\,(eV/atom)$^{-1}$ and residual log-scale $\sigma = 0.305$.
Model 1 uses the same interval-censored log-normal AFT loss as Model 2 (implemented in \textit{lifelines}), but with a deliberately minimal fitting procedure: the three features are fed raw, with no z-scoring and no median imputation (the few rows with a non-finite feature are dropped rather than filled), no $L_2$ penalty (penalizer $=0$), and no bootstrap bagging (a single fit). No class reweighting is applied. The melting-point inputs use the decomposition-product proxy described above.

\textit{Model 2} is a log-normal accelerated failure time (AFT) model fit on 32 descriptors using interval censoring implemented via \textit{lifelines} (v0.30.3) \cite{DavidsonPilon2019_lifelines} with a weak L2 regularization (penalizer~$= 0.1$). 
Three manually selected physicochemical features ($\min(T_\mathrm{ref,proxy})$, the ratio $R_\mathrm{ref,proxy}$, and $\Delta E_\mathrm{rxn}$) are combined with 29 descriptors generated with LLM assistance and expert review, covering per-precursor decomposition onset temperatures, atmosphere-release energetics (H$_2$O, O$_2$, CO$_2$ release enthalpies), M--O bond dissociation energies, acid--base character ($\varphi_\mathrm{diff}$), coordination number mismatch, reduction potential, crystal topology indicators, and a residual log-oxide melting point correction (see Supplementary Materials for full feature list).
Model 2 uses a three-step fitting pipeline (Model 1, described above, omits all three steps). First, any missing feature value is filled with the training-set median of that feature (median imputation), so every pair has a complete descriptor vector. Second, each feature is standardized to zero mean and unit variance on the training set (z-scoring), so the regularization penalty acts comparably across descriptors of different physical units. Third, the log-normal AFT is fit by interval-censored maximum likelihood with an $\ell$2 (ridge) penalty of strength $\lambda$ = 0.1 (penalizer), which shrinks the coefficients toward zero to limit overfitting on the modest dataset; $\lambda$ was selected by LOPO log-score over \{0.01, 0.03, 0.1, 0.3, 1, 3\}, with and without bootstrap bagging, and the unbagged $\lambda$ = 0.1 fit was best. No class reweighting was applied.
%
%
The predicted onset temperature is the log-normal median $\exp(\hat{\mu})$ in $^\circ$C. 
Prediction intervals are $\pm 1.96 \sigma_{cal}$, where $\sigma_{cal}$ is the standard deviation of standardized leave-one-pair-out residuals.
Fitted coefficients for both models are listed in table~\ref{tab:spearman-full-reacted}.

Models were validated using leave-one-precursor-out cross-validation (LOPO), in which all experiments involving a given precursor are withheld from training and used exclusively for evaluation, rotated across all 46 precursors. The held-out log-interval-probability (log-score; higher is better) is reported as the primary performance metric, as it is the only valid summary measure for censored observations where the true onset temperature is not directly observed. Both models are log-normal accelerated failure time (AFT) regressions on $T$ ($\log T = \beta^\top x + \sigma\varepsilon$, interval-censored) trained on the same PG\,+\,PG-LIT 46-precursor dataset with the same decomposition-product proxy melting points, so their log-scores are directly comparable; they differ in feature count (3 vs 32) and in regularization (Model 1 is fit raw; Model 2 with imputation, z-scoring and $\lambda$ = 0.1). Model 1 achieves a LOPO log-score of $-0.364$ and Model 2 achieves $-0.379$; the three-feature Model 1 thus attains the better held-out censored log-probability, and the two models give statistically indistinguishable in situ onset errors (below). Additional validation was performed against 26 independently measured in situ pXRD reaction onset temperatures \cite{McDermott2023_rxnnetwork2, Szymanski2024_thermoregime}; for this validation, the 26 in-house samples and 218 literature entries involving these pairs were excluded from training.
Confidence intervals on MAE and on paired MAE differences are from 10,000 bootstrap resamples of the 26 validation pairs.


\clearpage 

\bibliography{precursor_genome_sciadv}
\bibliographystyle{sciencemag}


\section*{Acknowledgments}
\paragraph*{Funding:}
This work was funded and intellectually led by the D2S2 program within the U.S. Department of Energy, Office of Basic Energy Sciences, Materials Sciences and Engineering Division under Contract No. DE-AC02-05-CH11231.
This work acknowledges support by the Genesis Mission as part of the U.S. Department of Energy.
L.N.W. acknowledges funding support from the BIDMaP Postdoctoral Fellowship.
B.R. acknowledges support from the Kavli ENSI Graduate Student Fellowship.
The authors used Claude Opus 4.7 (Anthropic) to generate candidate physicochemical descriptors for the onset-temperature model, as described in Materials and Methods. The authors reviewed, filtered, and take full responsibility for all descriptors and results.

\paragraph*{Author contributions:}

All authors took part in conception of the project; M.J.M, L.N.W., and B.R. planned experimental syntheses. M.J.M, L.N.W., B.R., and Y.F. aided in synthesis execution. L.N.W., M.J.M, and Y.F. performed characterization and analysis. Data preparation was completed by L.N.W., B.R., and Y.F. Data structure creation was completed by L.N.W. and B.R. L.N.W. and M.J.M prepared the manuscript, while all authors aided in manuscript revision. Supervision, funding, and resource acquisition was provided by G.C.

\paragraph*{Competing interests:}
The authors declare that they have no competing interests.

\paragraph*{Data and materials availability:}
The Precursor Genome Onset (PGO) Calculator is open source and freely available at *link*.
All experimental sample information is available at \url{https://doi.org/10.5281/zenodo.21285546} published by Walters, et al. \cite{PG_SciData}, including precursor characterization (pXRD, SEM, EDS) spectra and analysis, pairwise sample \textit{ex situ} pXRD raw spectra and analysis, and some sample metadata (e.g. powder precursor weights, heating profile).
Additional data supporting the findings of this study are available in the Supplementary Materials.



\subsection*{Supplementary materials}
Supplementary Text\\
Figs. S1 to S63\\
Tables S1 to S8


\newpage


\renewcommand{\thefigure}{S\arabic{figure}}
\renewcommand{\thetable}{S\arabic{table}}
\renewcommand{\theequation}{S\arabic{equation}}
\renewcommand{\thepage}{S\arabic{page}}
\setcounter{figure}{0}
\setcounter{table}{0}
\setcounter{equation}{0}
\setcounter{page}{1} 


\begin{center}
\section*{Supplementary Materials for\\ \scititle}

Lauren~N.~Walters$^{\ast\dagger}$,
Matthew~J.~McDermott$^{\dagger}$,
Bernardus~Rendy,
Yuxing~Fei,
Gerbrand~Ceder\\
\small$^\ast$Corresponding author. Email: laurenwalters@berkeley.edu (L.N.W.); gceder@berkeley.edu (G.C.)\\
\small$^\dagger$These authors contributed equally to this work.
\end{center}

\subsubsection*{This PDF file includes:}
Supplementary Text\\
Figures S1 to S63\\
Tables S1 to S8

\newpage


\subsection*{Supplementary Text}

\subsubsection*{Section S1. Precursor Information}

Information for precursors, including precursor identity, vendor, reported purity, additional preparation, and hygroscopicity testing.
Additional characterization for each precursor was completed by powder X-ray diffraction (XRD), scanning electron microscopy (SEM), and energy-dispersive X-ray spectroscopy (EDS).

\clearpage 
\subsubsection*{S1.1. Purchased Precursor Information}

\begin{table}[!htbp]
\linespread{1}\selectfont 
\tiny
\caption{\label{tab:precursor-purchase-info} \textbf{Precursor information, including the CAS identification number, vendor, and reported purity.} Important to note that LiC$_2$H$_3$O$_2\cdot$2H$_2$O, (NH$_4$)$_2$SO$_4$, NH$_4$Cl, NaH$_2$PO$_4\cdot$2H$_2$O, NaNO$_3$, and Si were ultimately removed from the collection due to experimental constraints such as safety and hygroscopicity.}
\begin{tabular}{lccccc}
\hline \hline
\textbf{Index} &\textbf{Precursor Name} & \textbf{Formula} & \textbf{CAS Number} & \textbf{Vendor} & \textbf{Purity} \\
\hline
1	& silver(i) oxide & Ag$_2$O &  20667-12-3 & Sigma Aldrich & 0.99  \\
2	&	aluminium hydroxide & Al(OH)$_3$ & 21645-51-2  & Sigma Aldrich  & - \\
3	&	boric acid & B(OH)$_3$ &  10043-35-3 &  Sigma Aldrich &  0.995 \\
4	&	barium carbonate & BaCO$_3$ &  513-77-9 & Sigma Aldrich  &  0.99 \\
5	&	barium peroxide & BaO$_2$ &  1304-29-6 &  Sigma Aldrich & 0.86  \\
6	&	bismuth(iii) oxide & Bi$_2$O$_3$ &  1304-76-3 &  Sigma Aldrich & 0.999  \\
7	&	calcium carbonate & CaCO$_3$ & 471-34-1  &  Sigma Aldrich & 0.99  \\
8	&	cobalt(ii) oxide & CoO &  1307-96-6 & Sigma Aldrich  & -  \\
9	&	cobalt(ii,iii) oxide & Co$_3$O$_4$ & 1308-06-1  &  Sigma Aldrich &  - \\
10	&	chromium(iii) oxide & Cr$_2$O$_3$ & 1308-38-9  &  Sigma Aldrich &  0.98 \\
11	&	copper(ii) oxide & CuO &  1317-38-0 & Sigma Aldrich  &  0.99 \\
12	&	iron(ii,iii) oxide & Fe$_3$O$_4$ & 1317-61-9  &  Sigma Aldrich &  0.95 \\
13	&	iron(iii) oxide & Fe$_2$O$_3$ &  1309-37-1 &  Sigma Aldrich & 0.96  \\
14	&	iron(ii) oxalate dihydrate & FeC$_2$O$_4\cdot$2H$_2$O &  6047-25-2 &  Sigma Aldrich & 0.99  \\
15	&	gallium(iii) oxide & Ga$_2$O$_3$ & 12024-21-4  &  Sigma Aldrich & 0.9999  \\
16	&	germanium(iv) oxide & GeO$_2$ &  1310-53-8 & Sigma Aldrich  & 0.99998  \\
17	&	hafnium(iv) oxide & HfO$_2$ & 12055-23-1  &  Sigma Aldrich & 0.98  \\
18	&	indium(iii) oxide & In$_2$O$_3$ &  1312-43-2 &  Sigma Aldrich &  0.9999 \\
19	&	potassium carbonate sesquihydrate & K$_2$CO$_3\cdot$1.5H$_2$O &  6381-79-9 & Sigma Aldrich  &  0.99 \\
20	&	potassium phosphate monobasic & KH$_2$PO$_4$ & 7778-77-0  & Sigma Aldrich  & 0.99  \\
21	&	lanthanum(iii) hydroxide & La(OH)$_3$ & 14507-19-8  &  Sigma Aldrich &  0.999 \\
22	&	lithium carbonate & Li$_2$CO$_3$ &  554-13-2 & Sigma Aldrich  &  0.99 \\
23	&	lithium acetate dihydrate & LiC$_2$H$_3$O$_2\cdot$2H$_2$O & 6108-17-4  & Sigma Aldrich  &  $\geq$63 \\
24	&	lithium hydroxide monohydrate & LiOH$\cdot$H$_2$O &  1310-66-3 & Sigma Aldrich  &  0.98 \\
25	&	magnesium oxide  & MgO &  1309-48-4 &  Sigma Aldrich &  0.97 \\
26	&	manganese(ii) oxide & MnO & 1344-43-0  &  Sigma Aldrich & 0.99  \\
27	&	manganese(ii,iii) oxide & Mn$_3$O$_4$ &  1317-35-7 &  Sigma Aldrich &  0.97 \\
28	&	manganese(iii) oxide & Mn$_2$O$_3$ &  1317-34-6 & Sigma Aldrich  &  0.99 \\
29	&	manganese(iv) oxide & MnO$_2$ &  1313-13-9 &  Sigma Aldrich & 0.99  \\
30	&	molybdenum(vi) oxide & MoO$_3$ &  1313-27-5 & Sigma Aldrich  & 0.995  \\
31	&	ammonium phosphate dibasic & (NH$_4$)$_2$HPO$_4$ & 7783-28-0  & Sigma Aldrich  &  0.98 \\
32	&	ammonium sulfate & (NH$_4$)$_2$SO$_4$ & 7783-20-2  & Sigma Aldrich  &  0.99 \\
33	&	ammonium chloride & NH$_4$Cl &  12125-02-9 & Sigma Aldrich  &  0.995 \\
34	&	ammonium phosphate monobasic & NH$_4$H$_2$PO$_4$ & 7722-76-1  & Sigma Aldrich  &  0.98 \\
35	&	sodium carbonate  & Na$_2$CO$_3$ &  497-19-8 &  Sigma Aldrich &  0.995 \\
36	&	sodium phosphate monobasic dihydrate & NaH$_2$PO$_4\cdot$2H$_2$O &  13472-35-0 &  Sigma Aldrich &  0.99 \\
37	&	sodium nitrate & NaNO$_3$ & 7631-99-4  &  Sigma Aldrich &  0.99 \\
38	&	niobium(v) oxide & Nb$_2$O$_5$ &  1313-96-8 & Sigma Aldrich  &  0.999 \\
39	&	nickel(ii) oxide & NiO & 1313-99-1  &  Sigma Aldrich &  0.99 \\
40	&	lead(ii) oxide & PbO &  1317-36-8 & Sigma Aldrich  &  0.999 \\
41	&	antimony(iii) oxide & Sb$_2$O$_3$ & 1309-64-4  &  Sigma Aldrich & 0.99  \\
42	&	silicon & Si & 7440-21-3  &  Sigma Aldrich &  0.99 \\
43	&	silicon dioxide & SiO$_2$ &  60676-86-0 &  Sigma Aldrich &  0.995 \\
44	&	tin(iv) oxide & SnO$_2$ &  18282-10-5 &  Sigma Aldrich &  0.999 \\
45	&	strontium carbonate & SrCO$_3$ &  1633-05-2 &  Sigma Aldrich &  0.999 \\
46	&	titanium(iv) oxide, anatase & TiO$_2$ & 1317-70-0  & Sigma Aldrich  &  0.998 \\
47	&	vanadium(iii) oxide & V$_2$O$_3$ &  1314-34-7 & Sigma Aldrich  &  0.98 \\
48	&	vanadium(v) oxide & V$_2$O$_5$ &  1314-62-1 &  Sigma Aldrich &  0.98 \\
49	&	tungsten(vi) oxide & WO$_3$ & 1314-35-8  & Sigma Aldrich  &  0.999 \\
50	&	yttrium(iii) oxide & Y$_2$O$_3$ &  1314-36-9 & Sigma Aldrich  & 0.9999  \\
51	&	zinc oxide & ZnO &  1314-13-2 &  Sigma Aldrich &  0.99 \\
52	&	zirconium(iv) oxide & ZrO$_2$ &  1314-23-4 &  Sigma Aldrich &   0.99 \\
\hline \hline
\end{tabular}
\end{table}

\clearpage 
\subsubsection*{S1.2. Precursor Hygroscopicity}

Very hygroscopic powders are not easily able to be handled automatically within A-Lab due to clumping. To test how hygroscopic each possible powder is, and if they can be utilized within A-Lab \cite{Alab}, we left a small amount of powder in a container for exactly 48\,hours in open air. XRD patterns and mass change were recorded. The cutoff for significant hygroscopicity was approximately 4\% weight gain, or when a less hygroscopic version was able to be found.

\begin{table}[!htbp]
\linespread{1}\selectfont 
\tiny
\caption{\label{tab:precursor-hygro-info} \textbf{Hygroscopicity information for precursors.} Pre-weight (Pre. Wt.), Post-weight (Post Wt.), and Container weight (Cont. Wt.) are recorded in grams. The mass change $\Delta$Wt. $=$ (Post Wt. $-$ Cont. Wt.) $-$ Pre. Wt. is the mass gained (or lost) by the sample over the exposure period, and \% Change expresses it relative to the pre-exposure mass; positive values indicate moisture uptake.}
\begin{tabular}{llcccccc}
\hline \hline
\textbf{Index} & \textbf{Precursor Name} & \textbf{Formula} & \textbf{Pre. Wt. (g)} & \textbf{Post Wt.} & \textbf{Cont. Wt. (g)} & \textbf{$\Delta$Wt. (g)} & \textbf{\% Change} \\
\hline
1 & aluminium hydroxide & Al(OH)$_3$ & 0.50295 & 2.54955 & 2.0469 & -0.00030 & -0.06 \\
2 & aluminium oxide & Al$_2$O$_3$ & 0.5043 & 2.5844 & 2.0442 & +0.03590 & +7.12 \\
3 & ammonium chloride & NH$_4$Cl & 0.5031 & 2.5607 & 2.05815 & -0.00055 & -0.11 \\
4 & ammonium dihydrogen phosphate & NH$_4$H$_2$PO$_4$ & 0.511 & 2.5982 & 2.08715 & +0.00005 & +0.01 \\
5 & ammonium sulfate & (NH$_4$)$_2$SO$_4$ & 0.5082 & 2.5884 & 2.08035 & -0.00015 & -0.03 \\
6 & antimony(iii) oxide & Sb$_2$O$_3$ & 0.54205 & 2.62275 & 2.0817 & -0.00100 & -0.18 \\
7 & barium oxide & BaO & 0.5373 & 2.71135 & 2.04625 & +0.12780 & +23.79 \\
8 & barium peroxide & BaO$_2$ & 0.5567 & 2.60135 & 2.0421 & +0.00255 & +0.46 \\
9 & bismuth(iii) oxide & Bi$_2$O$_3$ & 0.5203 & 2.5639 & 2.0433 & +0.00030 & +0.06 \\
10 & boric acid & B(OH)$_3$ & 0.5048 & 2.55165 & 2.048 & -0.00115 & -0.23 \\
11 & calcium carbonate & CaCO$_3$ & 0.52415 & 2.5709 & 2.0462 & +0.00055 & +0.10 \\
12 & calcium hydroxide & Ca(OH)$_2$ & 0.5161 & 2.62525 & 2.0884 & +0.02075 & +4.02 \\
13 & chromium(iii) oxide & Cr$_2$O$_3$ & 0.4971 & 2.5833 & 2.08645 & -0.00025 & -0.05 \\
14 & cobalt(ii) oxide & CoO & 0.5238 & 2.58325 & 2.05905 & +0.00040 & +0.08 \\
15 & cobalt(ii,iii) oxide & Co$_3$O$_4$ & 0.5533 & 2.61695 & 2.06725 & -0.00360 & -0.65 \\
16 & copper(ii) oxide & CuO & 0.52545 & 2.57745 & 2.0522 & -0.00020 & -0.04 \\
17 & diammonium hydrogen phosphate & (NH$_4$)$_2$HPO$_4$ & 0.51065 & 2.5654 & 2.0553 & -0.00055 & -0.11 \\
18 & gallium(iii) oxide & Ga$_2$O$_3$ & 0.5244 & 2.58695 & 2.0629 & -0.00035 & -0.07 \\
19 & germanium oxide & GeO$_2$ & 0.5182 & 2.6093 & 2.09135 & -0.00025 & -0.05 \\
20 & hafnium oxide & HfO$_2$ & 0.5037 & 2.58955 & 2.0862 & -0.00035 & -0.07 \\
21 & indium(iii) oxide & In$_2$O$_3$ & 0.5106 & 2.5501 & 2.0398 & -0.00030 & -0.06 \\
22 & iron oxalate & FeC$_2$O$_4$ & 0.49905 & 2.5831 & 2.0855 & -0.00145 & -0.29 \\
23 & iron oxide & Fe$_2$O$_3$ & 0.5371 & 2.58315 & 2.04357 & +0.00248 & +0.46 \\
24 & iron(ii,iii) oxide & Fe$_3$O$_4$ & 0.51885 & 2.5758 & 2.0583 & -0.00135 & -0.26 \\
25 & lanthanum(iii) oxide & La$_2$O$_3$ & 0.52335 & 2.6399 & 2.05426 & +0.06229 & +11.90 \\
26 & lead(ii) oxide & PbO & 0.50915 & 2.5474 & 2.0365 & +0.00175 & +0.34 \\
27 & lithium acetate & LiH$_3$(CO)$_2$ & 0.50625 & 2.68075 & 2.04182 & +0.13268 & +26.21 \\
28 & lithium carbonate & Li$_2$CO$_3$ & 0.50805 & 2.56115 & 2.05335 & -0.00025 & -0.05 \\
29 & lithium hydroxide & LiOH & 0.50785 & 2.6911 & 2.03835 & +0.14490 & +28.53 \\
30 & magnesium carbonate & MgCO$_3$ & 0.5121 & 2.57885 & 2.06645 & +0.00030 & +0.06 \\
31 & magnesium oxide & MgO & 0.5289 & 2.56855 & 2.02398 & +0.01567 & +2.96 \\
32 & manganese dioxide & MnO$_2$ & 0.5158 & 2.56455 & 2.049 & -0.00025 & -0.05 \\
33 & manganese(ii) oxide & MnO & 0.5051 & 2.5748 & 2.06985 & -0.00015 & -0.03 \\
34 & manganese(ii,iii) oxide & Mn$_3$O$_4$ & 0.5003 & 2.5508 & 2.05025 & +0.00025 & +0.05 \\
35 & manganese(iii) oxide & Mn$_2$O$_3$ & 0.52135 & 2.5729 & 2.052 & -0.00045 & -0.09 \\
36 & molybdenum trioxide & MoO$_3$ & 0.50225 & 2.56645 & 2.0644 & -0.00020 & -0.04 \\
37 & monopotassium phosphate & KH$_2$PO$_4$ & 0.5077 & 2.59455 & 2.0868 & +0.00005 & +0.01 \\
38 & monosodium phosphate & NaH$_2$PO$_4$ & 0.5402 & 2.59195 & 2.0373 & +0.01445 & +2.67 \\
39 & nickel(ii) oxide & NiO & 0.53625 & 2.58115 & 2.04475 & +0.00015 & +0.03 \\
40 & niobium(v) oxide & Nb$_2$O$_5$ & 0.51285 & 2.6021 & 2.08945 & -0.00020 & -0.04 \\
41 & potassium carbonate & K$_2$CO$_3$ & 0.5149 & 2.6451 & 2.0435 & +0.08670 & +16.84 \\
42 & silicon & Si & 0.50645 & 2.55515 & 2.0493 & -0.00060 & -0.12 \\
43 & silicon(iv) dioxide & SiO$_2$ & 0.5232 & 2.58105 & 2.0569 & +0.00095 & +0.18 \\
44 & silver(i) oxide & Ag$_2$O & 0.53125 & 2.6019 & 2.05415 & +0.01650 & +3.11 \\
45 & sodium carbonate & Na$_2$CO$_3$ & 0.5276 & 2.5758 & 2.04755 & +0.00065 & +0.12 \\
46 & sodium nitrate & NaNO$_3$ & 0.50875 & 2.5501 & 2.0418 & -0.00045 & -0.09 \\
47 & strontium carbonate & SrCO$_3$ & 0.50185 & 2.55005 & 2.0479 & +0.00030 & +0.06 \\
48 & tin(iv) oxide & SnO$_2$ & 0.5133 & 2.6 & 2.0866 & +0.00010 & +0.02 \\
49 & titanium dioxide & TiO$_2$ & 0.50905 & 2.5801 & 2.07135 & -0.00030 & -0.06 \\
50 & tungsten(vi) oxide & WO$_3$ & 0.5045 & 2.58855 & 2.08475 & -0.00070 & -0.14 \\
51 & vanadium pentoxide & V$_2$O$_5$ & 0.5165 & 2.57255 & 2.05575 & +0.00030 & +0.06 \\
52 & vanadium(iii) oxide & V$_2$O$_3$ & 0.5061 & 2.559 & 2.0527 & +0.00020 & +0.04 \\
53 & yttria & Y$_2$O$_3$ & 0.5112 & 2.5588 & 2.04735 & +0.00025 & +0.05 \\
54 & yttrium chloride & YCl$_3$ & 0.52135 & 2.76745 & 2.08725 & +0.15885 & +30.47 \\
55 & zinc(ii) oxide & ZnO & 0.51475 & 2.5787 & 2.06475 & -0.00080 & -0.16 \\
56 & zirconium dioxide & ZrO$_2$ & 0.50365 & 2.5499 & 2.04555 & +0.00070 & +0.14 \\
57 & barium carbonate & BaCO$_3$ & 0.50955 & 2.5792 & 2.06955 & +0.00010 & +0.02 \\
58 & lithium acetate dihydrate & LiC$_2$H$_7$O$_3$ & NA & NA & NA & NA & NA \\
59 & lithium hydroxide monohydrate & LiH$_3$O$_2$ & 0.4998 & 2.5443 & 2.0412 & +0.00330 & +0.66 \\
60 & potas. carb. sesquihydrate & K$_4$C$_2$O$_9$H$_6$ & 0.50485 & 2.58925 & 2.08205 & +0.00235 & +0.47 \\
61 & lanthanum hydroxide & La(OH)$_3$ & 0.5042 & 2.5521 & 2.04645 & +0.00145 & +0.29 \\
\hline \hline
\end{tabular}
\end{table}

\clearpage 
\subsubsection*{S1.3. Precursor Preparation Prior to A-Lab Reactions}

\begin{table}[!htbp]
\linespread{1}\selectfont 
\tiny
\caption{\label{tab:precursor-prep-info} \textbf{Precursor information on preprocessing before use in A-Lab.} Not Applicable (NA) indicates the precursor was used as supplied. Important to note that LiC$_2$H$_3$O$_2\cdot$2H$_2$O, (NH$_4$)$_2$SO$_4$, NH$_4$Cl, NaH$_2$PO$_4\cdot$2H$_2$O, NaNO$_3$, and Si were ultimately removed from the collection due to experimental constraints such as safety and hygroscopicity.}
\begin{tabular}{lccc}
\hline \hline
\textbf{Index} &\textbf{Precursor Name} & \textbf{Formula} & \textbf{Pre-processing} \\
\hline
1	& silver(i) oxide & Ag$_2$O &  NA \\
2	&	aluminium hydroxide & Al(OH)$_3$ &  NA \\
3	&	boric acid & B(OH)$_3$ &   Dry speedmix 5\,min 3000\,rpm  \\
4	&	barium carbonate & BaCO$_3$ &  NA   \\
5	&	barium peroxide & BaO$_2$ &  NA   \\
6	&	bismuth(iii) oxide & Bi$_2$O$_3$ &  NA   \\
7	&	calcium carbonate & CaCO$_3$ &   NA \\
8	&	cobalt(ii) oxide & CoO &   NA  \\
9	&	cobalt(ii,iii) oxide & Co$_3$O$_4$ &  NA  \\
10	&	chromium(iii) oxide & Cr$_2$O$_3$ &   NA \\
11	&	copper(ii) oxide & CuO &  NA   \\
12	&	iron(ii,iii) oxide & Fe$_3$O$_4$ &  NA  \\
13	&	iron(iii) oxide & Fe$_2$O$_3$ &   NA  \\
14	&	iron(ii) oxalate dihydrate & FeC$_2$O$_4\cdot$2H$_2$O &   NA  \\
15	&	gallium(iii) oxide & Ga$_2$O$_3$ &  NA  \\
16	&	germanium(iv) oxide & GeO$_2$ &  NA  \\
17	&	hafnium(iv) oxide & HfO$_2$ &   NA \\
18	&	indium(iii) oxide & In$_2$O$_3$ &   NA  \\
19	&	potassium carbonate sesquihydrate & K$_2$CO$_3\cdot$1.5H$_2$O &  Mortar \& pestle
5\,min   \\
20	&	potassium phosphate monobasic & KH$_2$PO$_4$ &  Mortar \& pestle
5\,min  \\
21	&	lanthanum(iii) hydroxide & La(OH)$_3$ &  NA  \\
22	&	lithium carbonate & Li$_2$CO$_3$ &  Mortar \& pestle
5\,min  \\
23	&	lithium acetate dihydrate & LiC$_2$H$_3$O$_2\cdot$2H$_2$O &   NA \\
24	&	lithium hydroxide monohydrate & LiOH$\cdot$H$_2$O &  High energy ball mill 5\,min
(WARNING: FUMES)   \\
25	&	magnesium oxide  & MgO &   NA  \\
26	&	manganese(ii) oxide & MnO &   High energy ball mill 10\,min \\
27	&	manganese(ii,iii) oxide & Mn$_3$O$_4$ &   NA  \\
28	&	manganese(iii) oxide & Mn$_2$O$_3$ &   NA  \\
29	&	manganese(iv) oxide & MnO$_2$ &   Dry speedmix 30\,min 3000\,rpm  \\
30	&	molybdenum(vi) oxide & MoO$_3$ &   NA  \\
31	&	ammonium phosphate dibasic & (NH$_4$)$_2$HPO$_4$ &  Mortar \& pestle
5\,min  \\
32	&	ammonium sulfate & (NH$_4$)$_2$SO$_4$ &   NA \\
33	&	ammonium chloride & NH$_4$Cl &   NA  \\
34	&	ammonium phosphate monobasic & NH$_4$H$_2$PO$_4$ &  Dry speedmix 20\,min 2500\,rpm  \\
35	&	sodium carbonate  & Na$_2$CO$_3$ & Mortar \& pestle
5\,min    \\
36	&	sodium phosphate monobasic dihydrate & NaH$_2$PO$_4\cdot$2H$_2$O &  Mortar \& pestle
10\,min (Dry mix leads to liquification)  \\
37	&	sodium nitrate & NaNO$_3$ &   NA \\
38	&	niobium(v) oxide & Nb$_2$O$_5$ &   NA  \\
39	&	nickel(ii) oxide & NiO &   NA \\
40	&	lead(ii) oxide & PbO &  NA   \\
41	&	antimony(iii) oxide & Sb$_2$O$_3$ &  NA  \\
42	&	silicon & Si &  NA  \\
43	&	silicon dioxide & SiO$_2$ &   NA  \\
44	&	tin(iv) oxide & SnO$_2$ &   NA  \\
45	&	strontium carbonate & SrCO$_3$ &   NA  \\
46	&	titanium(iv) oxide, anatase & TiO$_2$ &   NA \\
47	&	vanadium(iii) oxide & V$_2$O$_3$ &    NA \\
48	&	vanadium(v) oxide & V$_2$O$_5$ &   NA  \\
49	&	tungsten(vi) oxide & WO$_3$ &  High energy ball mill 10\,min (Do not dry spin mix)  \\
50	&	yttrium(iii) oxide & Y$_2$O$_3$ &    NA \\
51	&	zinc oxide & ZnO &  NA   \\
52	&	zirconium(iv) oxide & ZrO$_2$ &  NA   \\
\hline \hline
\end{tabular}
\end{table}

\clearpage 
\subsubsection*{S1.4. Precursor Melting Points and Source}

\begin{table}[!htbp]
\linespread{1}\selectfont 
\tiny
\caption{\label{tab:precursor-melting-points} \textbf{Melting points and data sources for A-Lab precursors.}}
\begin{tabular}{llcl}
\hline \hline
\textbf{Index} & \textbf{Precursor Name} & \textbf{Melting Pt. ($^\circ$C)} & \textbf{Source} \\
\hline
1	& silver(i) oxide & 280 & American Elements \\
2	& aluminium hydroxide & 300 & Lange's Handbook (15th Ed.) \\
3	& boric acid & 171 & Lange's Handbook (15th Ed.) \\
4	& barium carbonate & 811 & Fisher Scientific \\
5	& barium peroxide & 450 & Lange's Handbook (15th Ed.) \\
6	& bismuth(iii) oxide & 817 & Lange's Handbook (15th Ed.) \\
7	& calcium carbonate & 825 & Lange's Handbook (15th Ed.) \\
8	& cobalt(ii) oxide & 1935 & Lange's Handbook (15th Ed.) \\
9	& cobalt(ii,iii) oxide & 895 & Sigma-Aldrich \\
10	& chromium(iii) oxide & 2435 & American Elements \\
11	& copper(ii) oxide & 1326 & Weast, CRC Handbook (68th Ed.) \\
12	& iron(ii,iii) oxide & 1597 & Lange's Handbook (15th Ed.) \\
13	& iron(iii) oxide & 1565 & Lange's Handbook (15th Ed.) \\
14	& iron(ii) oxalate dihydrate & 190 & American Elements \\
15	& gallium(iii) oxide & 1900 & American Elements \\
16	& germanium(iv) oxide & 1115 & Lange's Handbook (15th Ed.) \\
17	& hafnium(iv) oxide & 2774 & Lange's Handbook (15th Ed.) \\
18	& indium(iii) oxide & 1910 & CAS Common Chemistry \\
19	& potassium carbonate sesquihydrate & 891 & Fisher Scientific \\
20	& potassium phosphate monobasic & 253 & American Elements \\
21	& lanthanum(iii) hydroxide & 330 & Brown \& Gallagher (2008) \\
22	& lithium carbonate & 720 & Lange's Handbook (15th Ed.) \\
23	& lithium acetate dihydrate & 53 & Sigma-Aldrich \\
24	& lithium hydroxide monohydrate & 462 & American Elements \\
25	& magnesium oxide & 2852 & American Elements \\
26	& manganese(ii) oxide & 1840 & Lange's Handbook (15th Ed.) \\
27	& manganese(ii,iii) oxide & 1567 & Lange's Handbook (15th Ed.) \\
28	& manganese(iii) oxide & 940 & American Elements \\
29	& manganese(iv) oxide & 535 & American Elements \\
30	& molybdenum(vi) oxide & 795 & American Elements \\
31	& ammonium phosphate dibasic & 155 & Lange's Handbook (15th Ed.) \\
32	& ammonium sulfate & 280 & Lange's Handbook (15th Ed.) \\
33	& ammonium chloride & 350 & NIOSH, 2023 \\
34	& ammonium phosphate monobasic & 190 & Lange's Handbook (15th Ed.) \\
35	& sodium carbonate & 856 & Haynes, CRC Handbook (95th Ed.) \\
36	& sodium phosphate monobasic dihydrate & 60 & Fisher Scientific \\
37	& sodium nitrate & 307 & Lange's Handbook (15th Ed.) \\
38	& niobium(v) oxide & 1512 & Lange's Handbook (15th Ed.) \\
39	& nickel(ii) oxide & 1955 & American Elements \\
40	& lead(ii) oxide & 886 & Lange's Handbook (15th Ed.) \\
41	& antimony(iii) oxide & 655 & Lange's Handbook (15th Ed.) \\
42	& silicon & 1412 & Lange's Handbook (15th Ed.) \\
43	& silicon dioxide & 1710 & NIOSH, 2023 \\
44	& tin(iv) oxide & 1630 & Lange's Handbook (15th Ed.) \\
45	& strontium carbonate & 1494 & American Elements \\
46	& titanium(iv) oxide, anatase & 1843 & Lange's Handbook (15th Ed.) \\
47	& vanadium(iii) oxide & 1940 & Lange's Handbook (15th Ed.) \\
48	& vanadium(v) oxide & 690 & American Elements \\
49	& tungsten(vi) oxide & 1473 & American Elements \\
50	& yttrium(iii) oxide & 2440 & Lange's Handbook (15th Ed.) \\
51	& zinc oxide & 1975 & Lange's Handbook (15th Ed.) \\
52	& zirconium(iv) oxide & 2715 & American Elements \\
\hline \hline
\end{tabular}
\end{table}

\clearpage 
\subsubsection*{Section S2. Precursor Genome Reaction Energies}

A staircase chart showing all precursors, with paired cells showing the predicted reaction energy, calculated from the Reaction Interface \cite{MaterialsProject_ReactionInterface1, MaterialsProject_ReactionInterface2} library of pymatgen \cite{pymatgen}.

\begin{figure}[!htbp]
    \centering
    \includegraphics[width=1.0\linewidth]{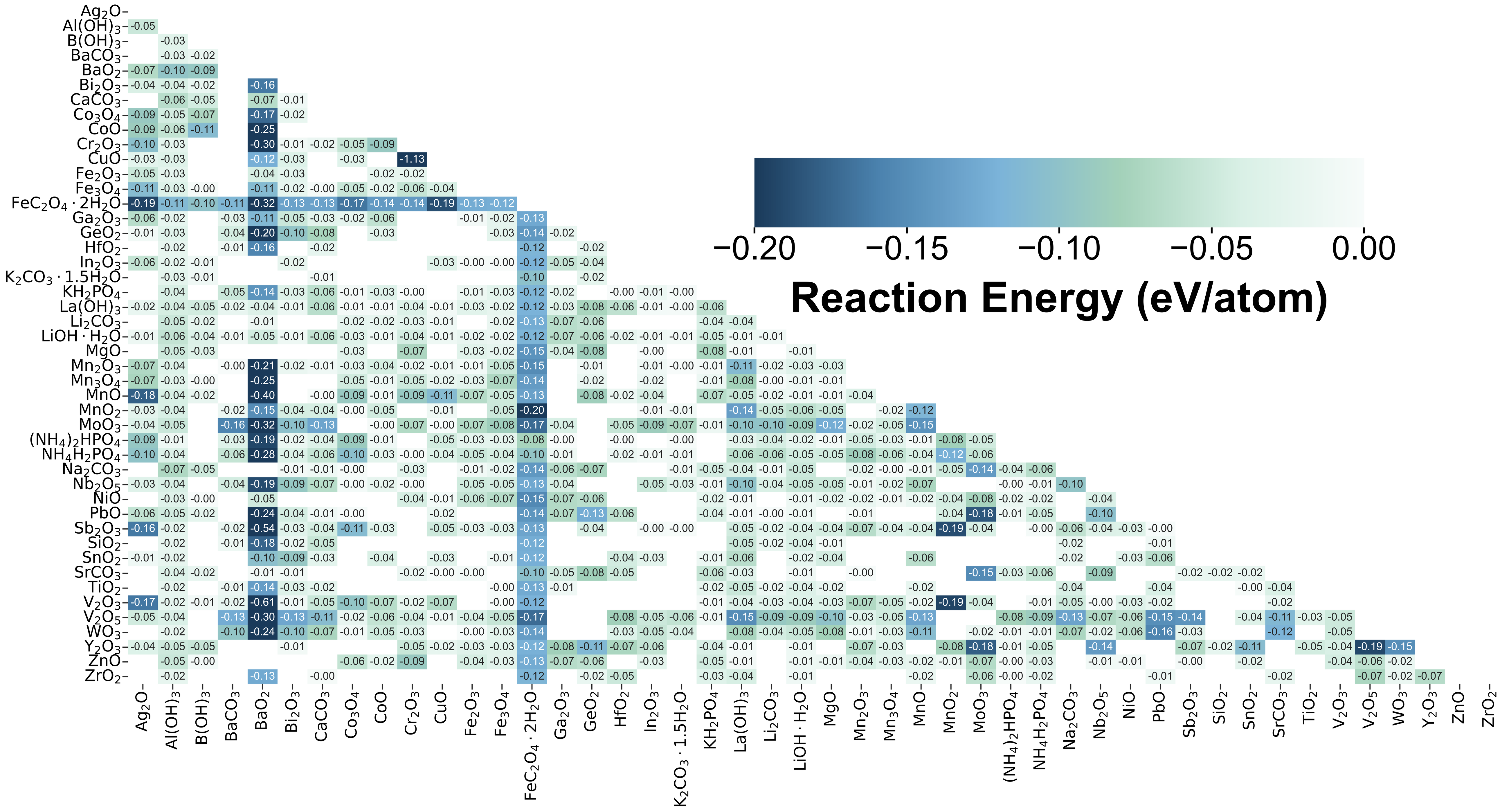}
    \caption{\textbf{Staircase plot depicting precursor genome reaction energies.}}
    \label{fig:staircase-energies-plot}
\end{figure}

\clearpage

\subsubsection*{S2.1. Precursor Characterization}

    \begin{figure}[ht]
        \centering
        \includegraphics[width=0.7\linewidth]{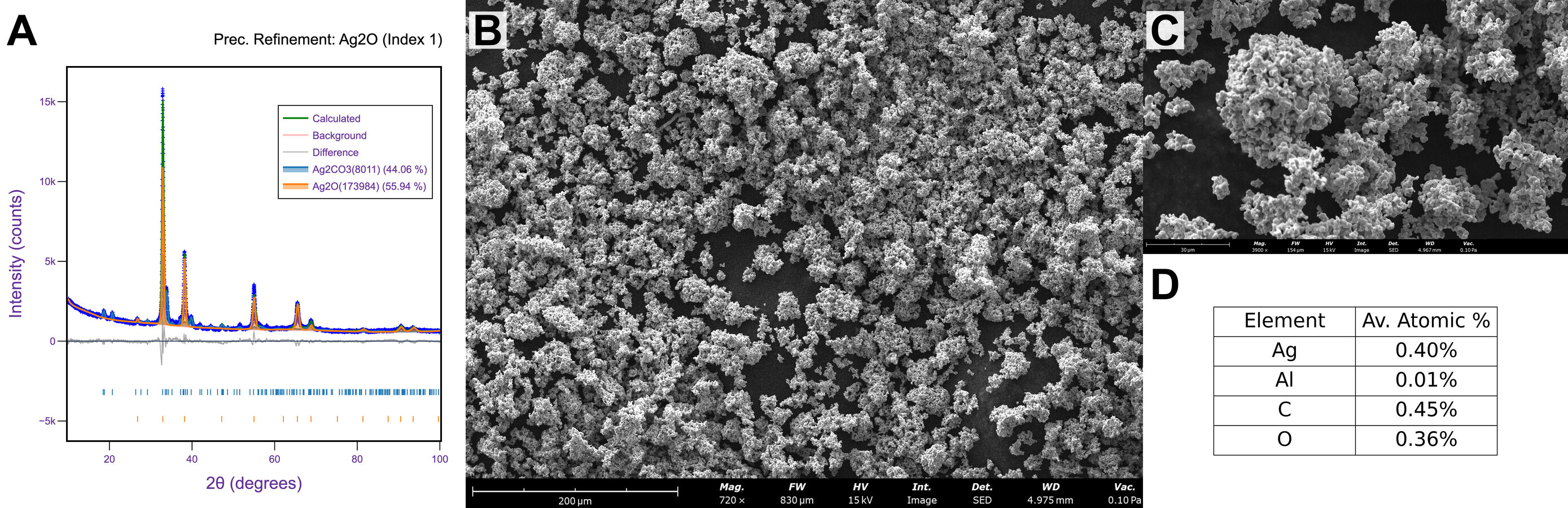}
        \caption{\textbf{Characterization of precursor Ag$_2$O (Project ID: 1).} (A) Powder X-ray diffraction (XRD) pattern with Rietveld refinement.
        (B) Scanning electron microscopy (SEM) showing representative morphology (scale bar around 200\,$\mu$m).
        (C) Scanning electron microscopy (SEM) showing representative morphology (window scaled to particle size).
        (D) Energy-dispersive X-ray spectroscopy (EDS) elemental composition data.}
        \label{fig:SI_1}
        \vspace{-1em}
    \end{figure}
    \begin{figure}[ht]
        \centering
        \includegraphics[width=0.7\linewidth]{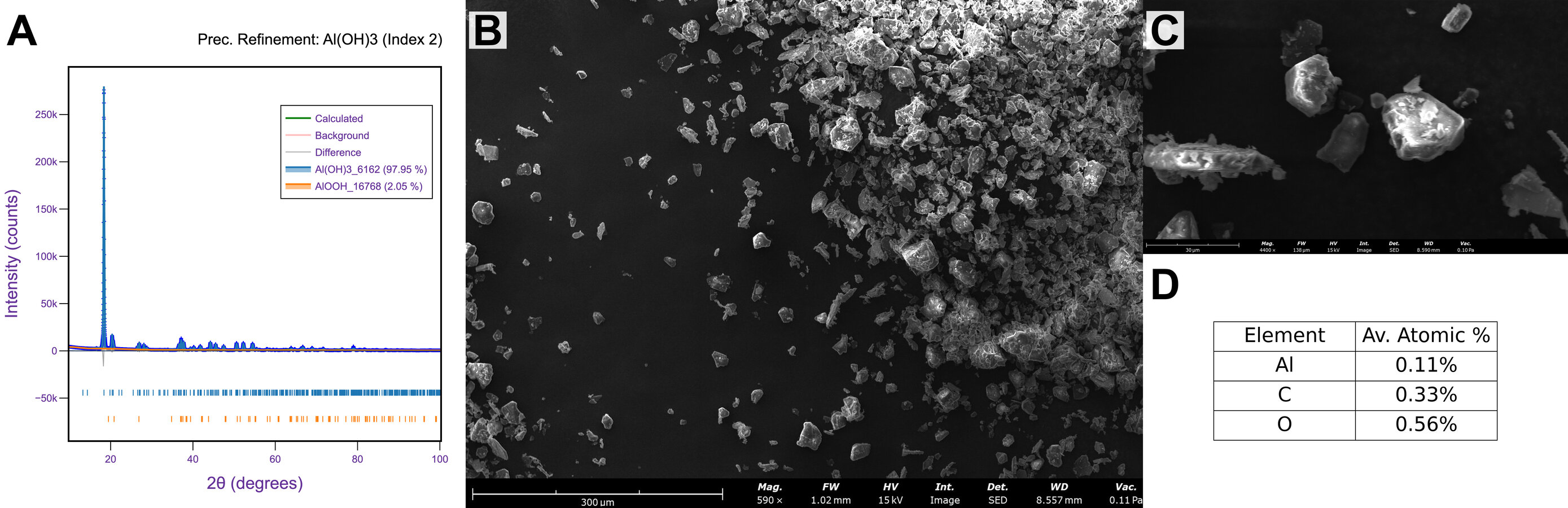}
        \caption{\textbf{Characterization of precursor Al(OH)$_3$ (Project ID: 2).} (A) Powder X-ray diffraction (XRD) pattern with Rietveld refinement.
        (B) Scanning electron microscopy (SEM) showing representative morphology (scale bar around 200\,$\mu$m).
        (C) Scanning electron microscopy (SEM) showing representative morphology (window scaled to particle size).
        (D) Energy-dispersive X-ray spectroscopy (EDS) elemental composition data.}
        \label{fig:SI_2}
        \vspace{-1em}
    \end{figure}
    \begin{figure}[ht]
        \centering
        \includegraphics[width=0.7\linewidth]{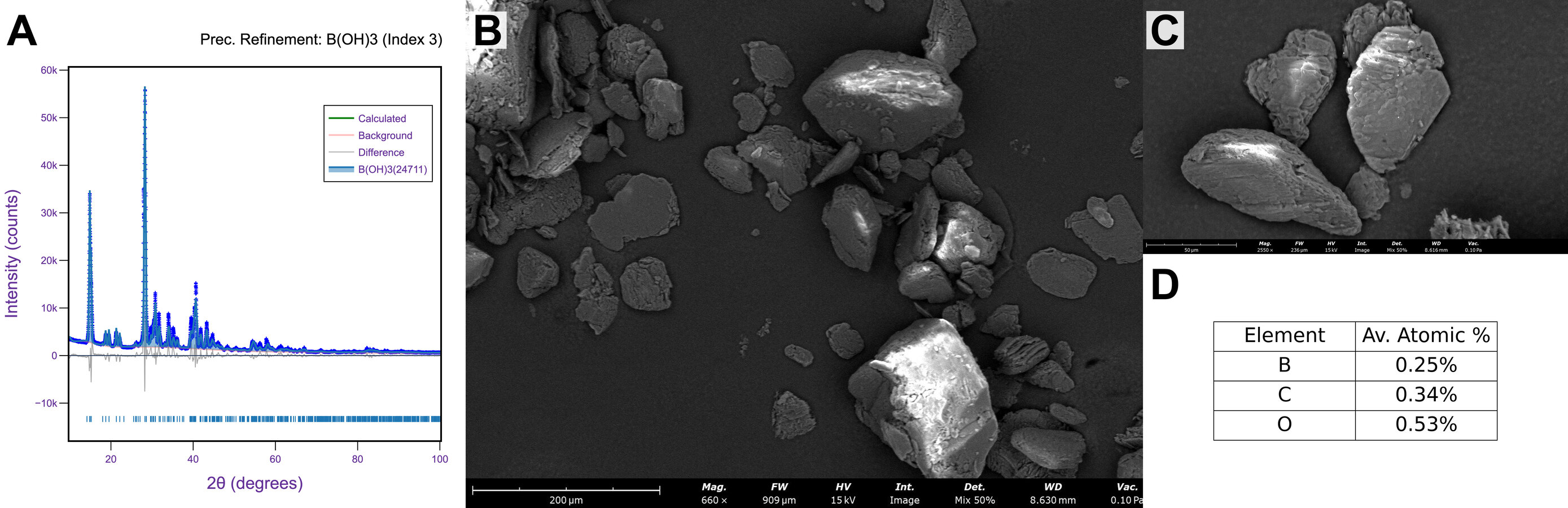}
        \caption{\textbf{Characterization of precursor B(OH)$_3$ (Project ID: 3).} (A) Powder X-ray diffraction (XRD) pattern with Rietveld refinement.
        (B) Scanning electron microscopy (SEM) showing representative morphology (scale bar around 200\,$\mu$m).
        (C) Scanning electron microscopy (SEM) showing representative morphology (window scaled to particle size).
        (D) Energy-dispersive X-ray spectroscopy (EDS) elemental composition data.}
        \label{fig:SI_3}
        \vspace{-1em}
    \end{figure}
    \begin{figure}[ht]
        \centering
        \includegraphics[width=0.7\linewidth]{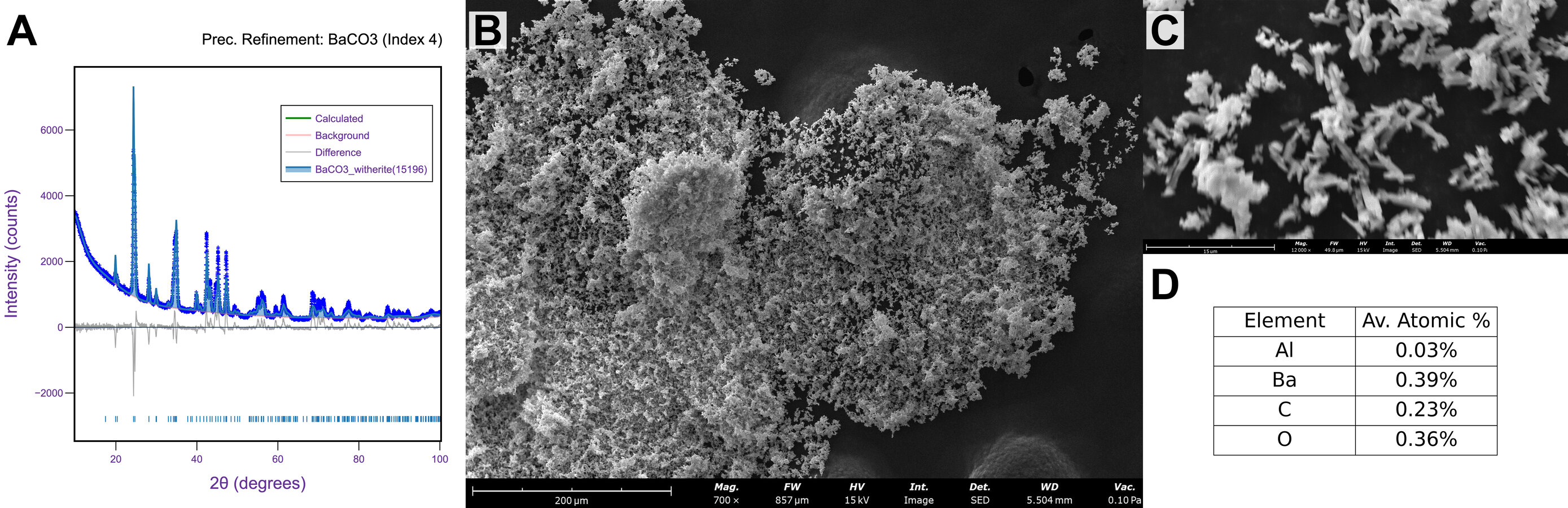}
        \caption{\textbf{Characterization of precursor BaCO$_3$ (Project ID: 4).} (A) Powder X-ray diffraction (XRD) pattern with Rietveld refinement.
        (B) Scanning electron microscopy (SEM) showing representative morphology (scale bar around 200\,$\mu$m).
        (C) Scanning electron microscopy (SEM) showing representative morphology (window scaled to particle size).
        (D) Energy-dispersive X-ray spectroscopy (EDS) elemental composition data.}
        \label{fig:SI_4}
        \vspace{-1em}
    \end{figure}
    \begin{figure}[ht]
        \centering
        \includegraphics[width=0.7\linewidth]{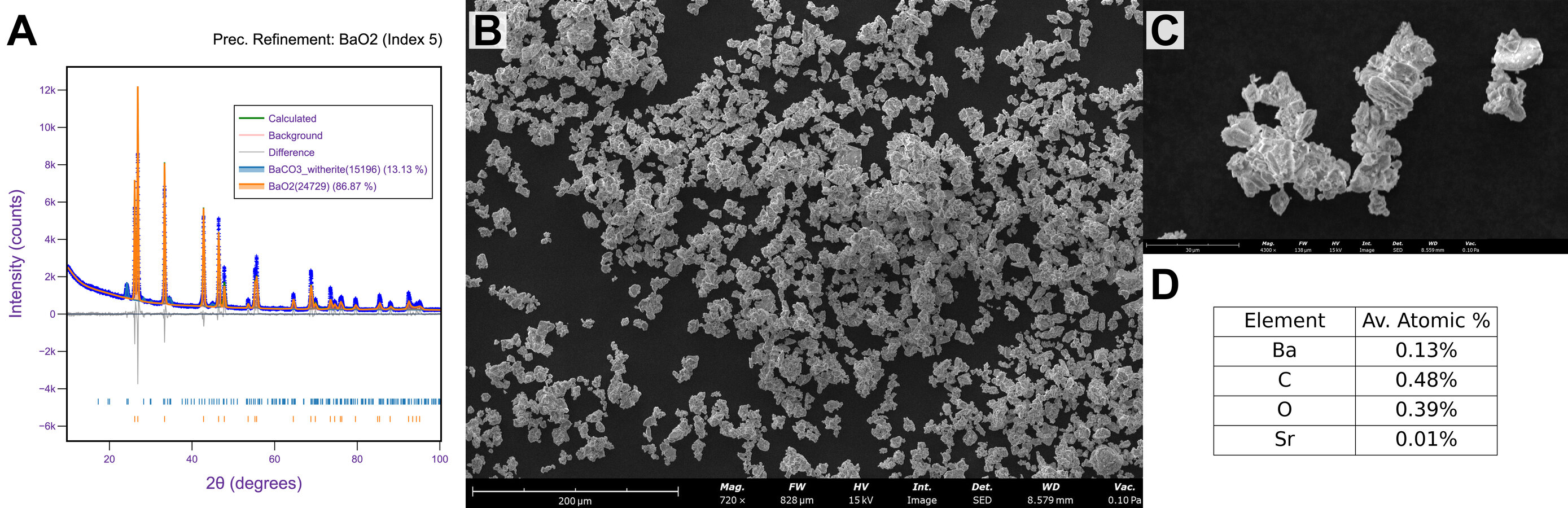}
        \caption{\textbf{Characterization of precursor BaO$_2$ (Project ID: 5).} (A) Powder X-ray diffraction (XRD) pattern with Rietveld refinement.
        (B) Scanning electron microscopy (SEM) showing representative morphology (scale bar around 200\,$\mu$m).
        (C) Scanning electron microscopy (SEM) showing representative morphology (window scaled to particle size).
        (D) Energy-dispersive X-ray spectroscopy (EDS) elemental composition data.}
        \label{fig:SI_5}
        \vspace{-1em}
    \end{figure}
    \begin{figure}[ht]
        \centering
        \includegraphics[width=0.7\linewidth]{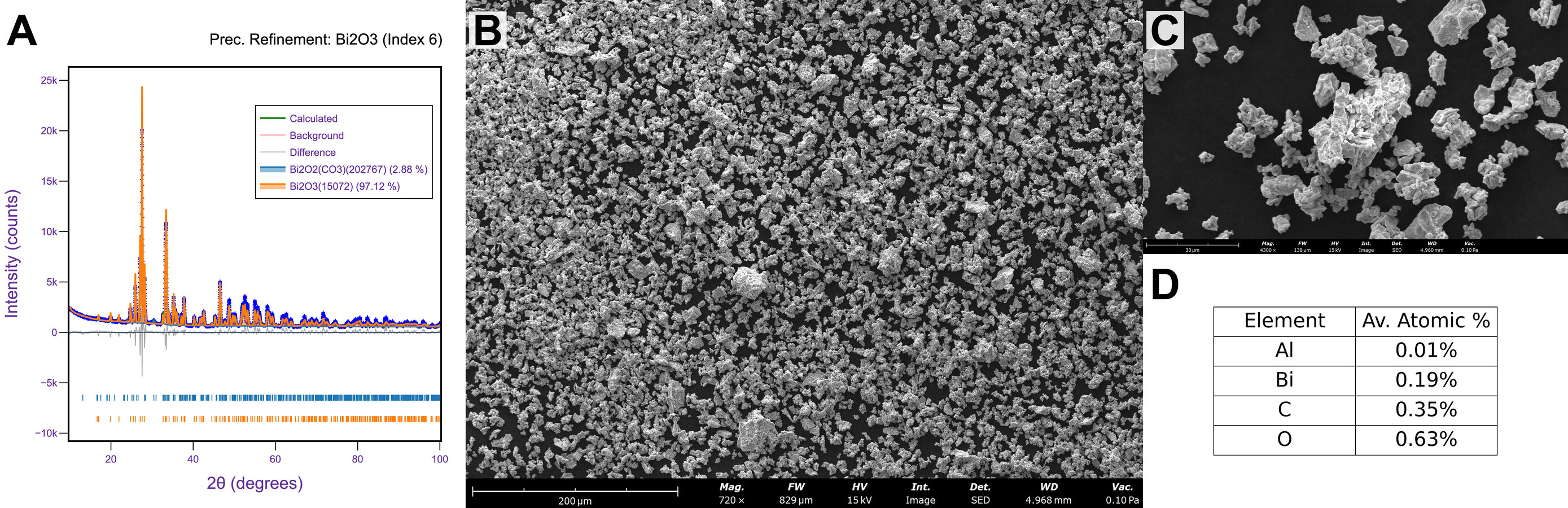}
        \caption{\textbf{Characterization of precursor Bi$_2$O$_3$ (Project ID: 6).} (A) Powder X-ray diffraction (XRD) pattern with Rietveld refinement.
        (B) Scanning electron microscopy (SEM) showing representative morphology (scale bar around 200\,$\mu$m).
        (C) Scanning electron microscopy (SEM) showing representative morphology (window scaled to particle size).
        (D) Energy-dispersive X-ray spectroscopy (EDS) elemental composition data.}
        \label{fig:SI_6}
        \vspace{-1em}
    \end{figure}
    \begin{figure}[ht]
        \centering
        \includegraphics[width=0.7\linewidth]{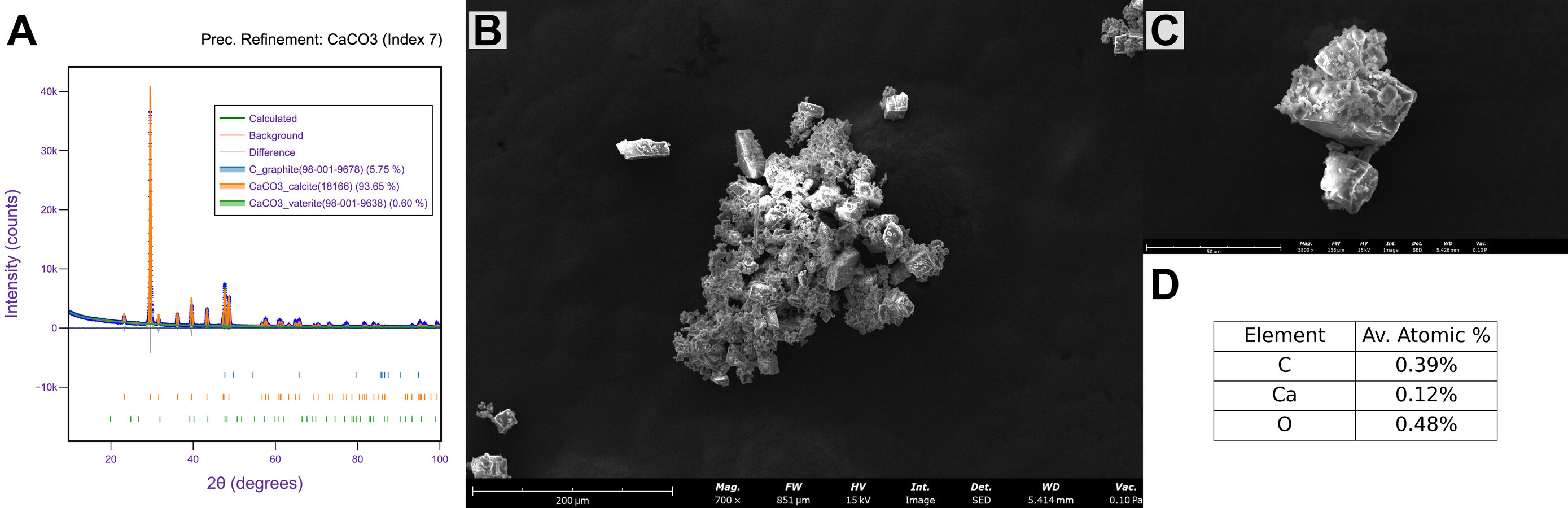}
        \caption{\textbf{Characterization of precursor CaCO$_3$ (Project ID: 7).} (A) Powder X-ray diffraction (XRD) pattern with Rietveld refinement.
        (B) Scanning electron microscopy (SEM) showing representative morphology (scale bar around 200\,$\mu$m).
        (C) Scanning electron microscopy (SEM) showing representative morphology (window scaled to particle size).
        (D) Energy-dispersive X-ray spectroscopy (EDS) elemental composition data.}
        \label{fig:SI_7}
        \vspace{-1em}
    \end{figure}
    \begin{figure}[ht]
        \centering
        \includegraphics[width=0.7\linewidth]{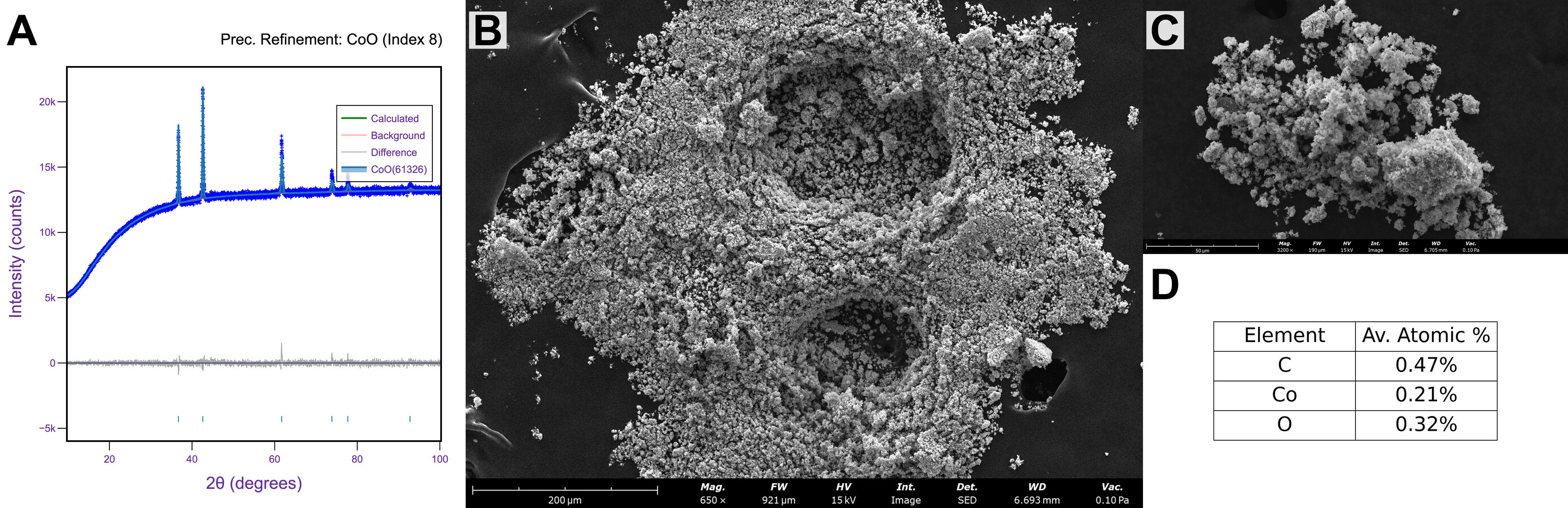}
        \caption{\textbf{Characterization of precursor CoO (Project ID: 8).} (A) Powder X-ray diffraction (XRD) pattern with Rietveld refinement.
        (B) Scanning electron microscopy (SEM) showing representative morphology (scale bar around 200\,$\mu$m).
        (C) Scanning electron microscopy (SEM) showing representative morphology (window scaled to particle size).
        (D) Energy-dispersive X-ray spectroscopy (EDS) elemental composition data.}
        \label{fig:SI_8}
        \vspace{-1em}
    \end{figure}
    \begin{figure}[ht]
        \centering
        \includegraphics[width=0.7\linewidth]{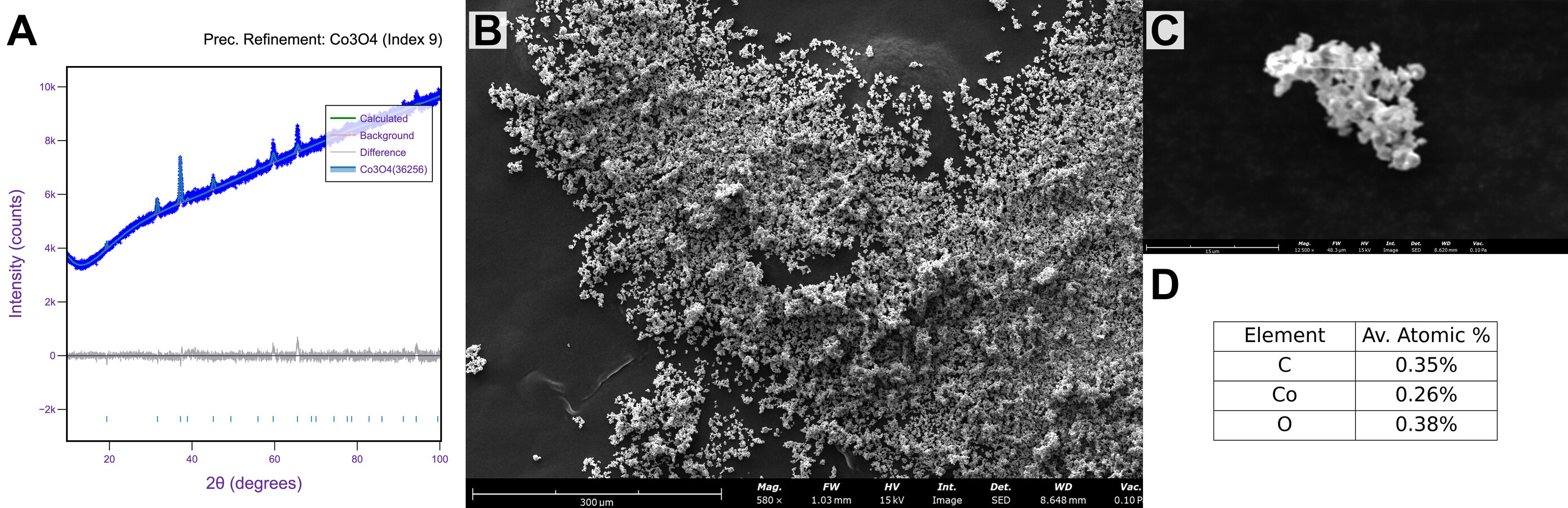}
        \caption{\textbf{Characterization of precursor Co$_3$O$_4$ (Project ID: 9).} (A) Powder X-ray diffraction (XRD) pattern with Rietveld refinement.
        (B) Scanning electron microscopy (SEM) showing representative morphology (scale bar around 200\,$\mu$m).
        (C) Scanning electron microscopy (SEM) showing representative morphology (window scaled to particle size).
        (D) Energy-dispersive X-ray spectroscopy (EDS) elemental composition data.}
        \label{fig:SI_9}
        \vspace{-1em}
    \end{figure}
    \begin{figure}[ht]
        \centering
        \includegraphics[width=0.7\linewidth]{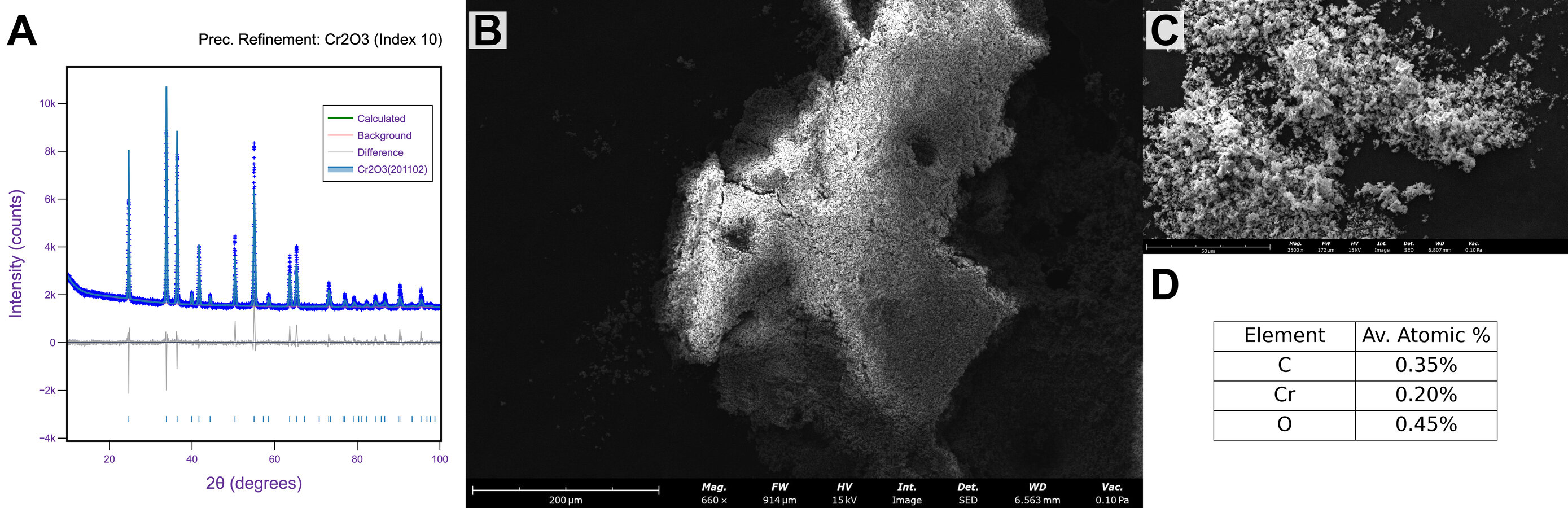}
        \caption{\textbf{Characterization of precursor Cr$_2$O$_3$ (Project ID: 10).} (A) Powder X-ray diffraction (XRD) pattern with Rietveld refinement.
        (B) Scanning electron microscopy (SEM) showing representative morphology (scale bar around 200\,$\mu$m).
        (C) Scanning electron microscopy (SEM) showing representative morphology (window scaled to particle size).
        (D) Energy-dispersive X-ray spectroscopy (EDS) elemental composition data.}
        \label{fig:SI_10}
        \vspace{-1em}
    \end{figure}
    \begin{figure}[ht]
        \centering
        \includegraphics[width=0.7\linewidth]{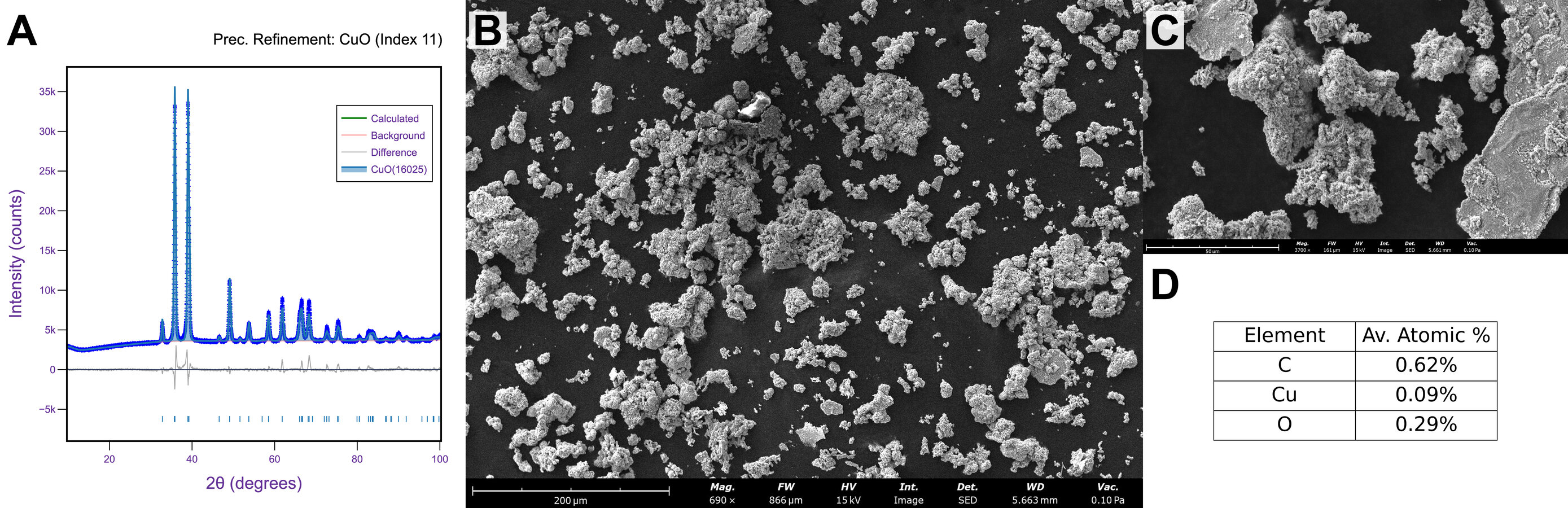}
        \caption{\textbf{Characterization of precursor CuO (Project ID: 11).} (A) Powder X-ray diffraction (XRD) pattern with Rietveld refinement.
        (B) Scanning electron microscopy (SEM) showing representative morphology (scale bar around 200\,$\mu$m).
        (C) Scanning electron microscopy (SEM) showing representative morphology (window scaled to particle size).
        (D) Energy-dispersive X-ray spectroscopy (EDS) elemental composition data.}
        \label{fig:SI_11}
        \vspace{-1em}
    \end{figure}
    \begin{figure}[ht]
        \centering
        \includegraphics[width=0.7\linewidth]{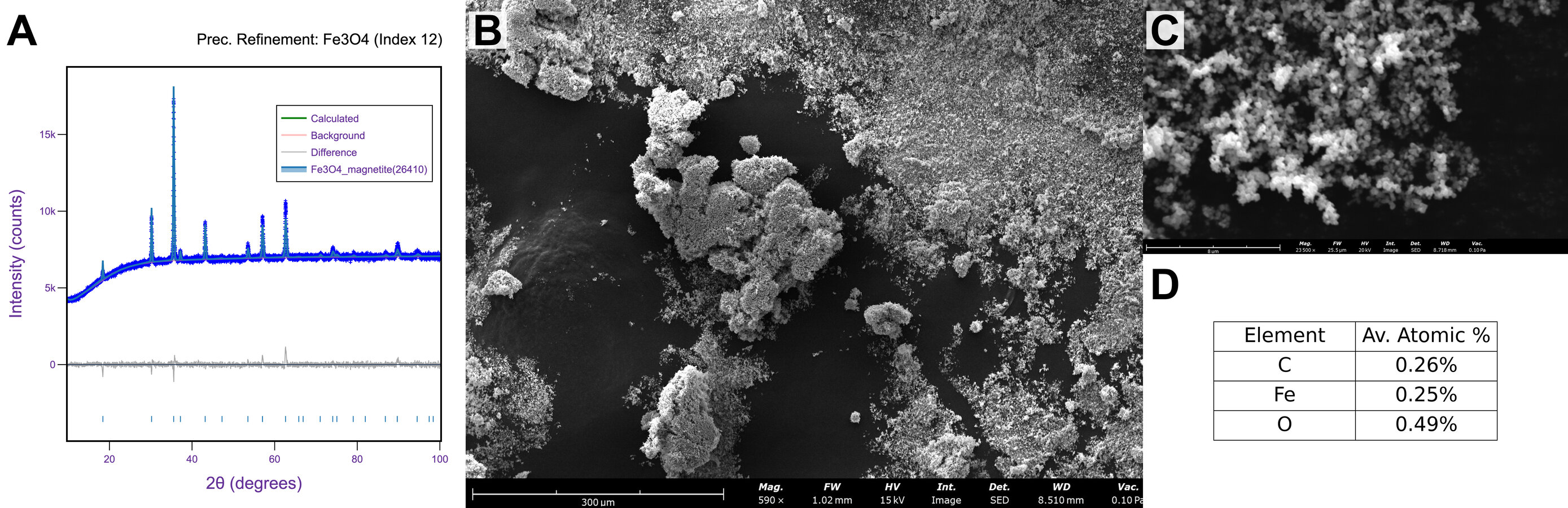}
        \caption{\textbf{Characterization of precursor Fe$_3$O$_4$ (Project ID: 12).} (A) Powder X-ray diffraction (XRD) pattern with Rietveld refinement.
        (B) Scanning electron microscopy (SEM) showing representative morphology (scale bar around 200\,$\mu$m).
        (C) Scanning electron microscopy (SEM) showing representative morphology (window scaled to particle size).
        (D) Energy-dispersive X-ray spectroscopy (EDS) elemental composition data.}
        \label{fig:SI_12}
        \vspace{-1em}
    \end{figure}
    \begin{figure}[ht]
        \centering
        \includegraphics[width=0.7\linewidth]{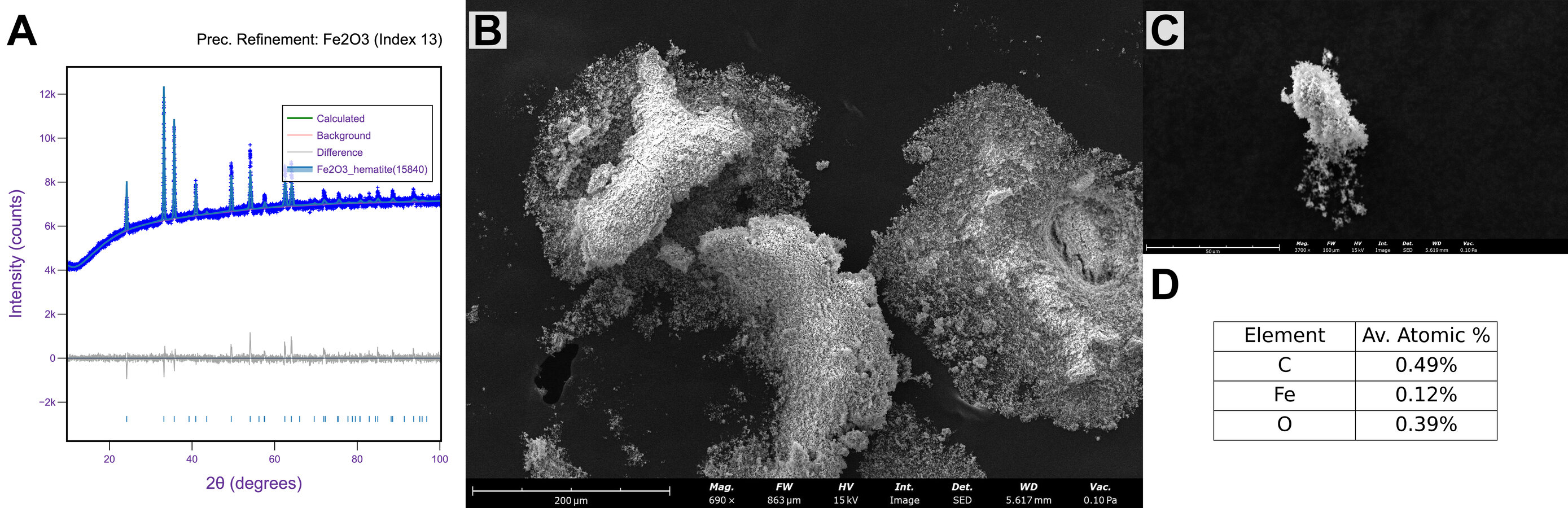}
        \caption{\textbf{Characterization of precursor Fe$_2$O$_3$ (Project ID: 13).} (A) Powder X-ray diffraction (XRD) pattern with Rietveld refinement.
        (B) Scanning electron microscopy (SEM) showing representative morphology (scale bar around 200\,$\mu$m).
        (C) Scanning electron microscopy (SEM) showing representative morphology (window scaled to particle size).
        (D) Energy-dispersive X-ray spectroscopy (EDS) elemental composition data.}
        \label{fig:SI_13}
        \vspace{-1em}
    \end{figure}
    \begin{figure}[ht]
        \centering
        \includegraphics[width=0.7\linewidth]{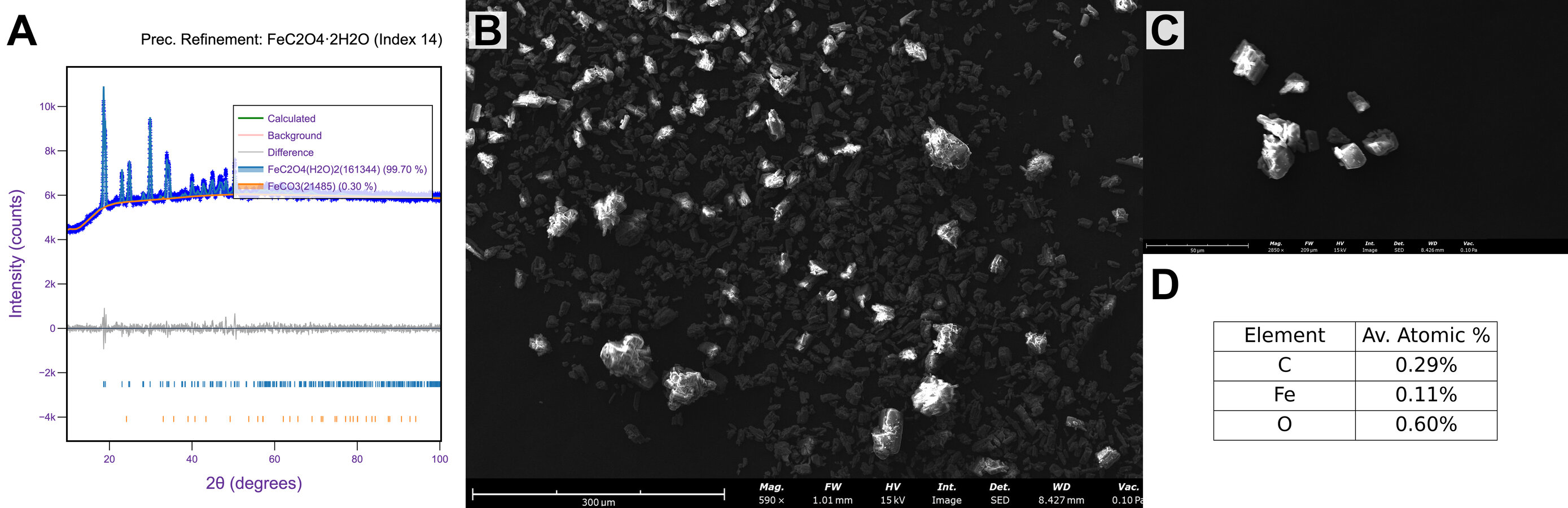}
        \caption{\textbf{Characterization of precursor FeC$_2$O$_4\cdot$2H$_2$O (Project ID: 14).} (A) Powder X-ray diffraction (XRD) pattern with Rietveld refinement.
        (B) Scanning electron microscopy (SEM) showing representative morphology (scale bar around 200\,$\mu$m).
        (C) Scanning electron microscopy (SEM) showing representative morphology (window scaled to particle size).
        (D) Energy-dispersive X-ray spectroscopy (EDS) elemental composition data.}
        \label{fig:SI_14}
        \vspace{-1em}
    \end{figure}
    \begin{figure}[ht]
        \centering
        \includegraphics[width=0.7\linewidth]{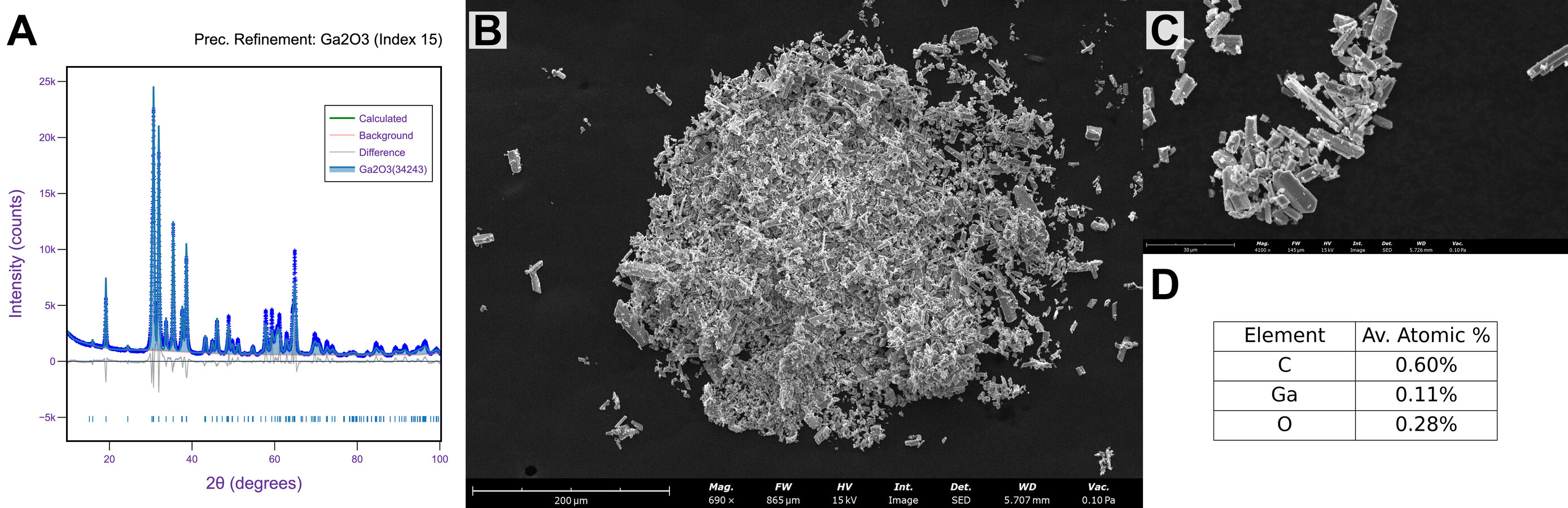}
        \caption{\textbf{Characterization of precursor Ga$_2$O$_3$ (Project ID: 15).} (A) Powder X-ray diffraction (XRD) pattern with Rietveld refinement.
        (B) Scanning electron microscopy (SEM) showing representative morphology (scale bar around 200\,$\mu$m).
        (C) Scanning electron microscopy (SEM) showing representative morphology (window scaled to particle size).
        (D) Energy-dispersive X-ray spectroscopy (EDS) elemental composition data.}
        \label{fig:SI_15}
        \vspace{-1em}
    \end{figure}
    \begin{figure}[ht]
        \centering
        \includegraphics[width=0.7\linewidth]{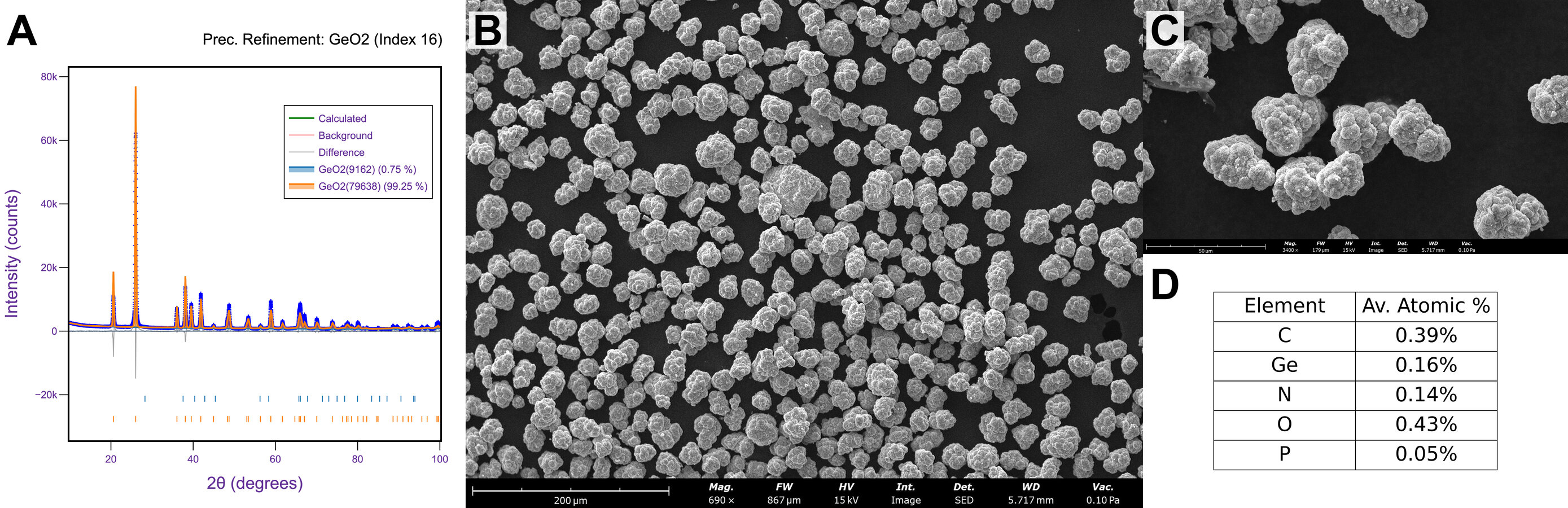}
        \caption{\textbf{Characterization of precursor GeO$_2$ (Project ID: 16).} (A) Powder X-ray diffraction (XRD) pattern with Rietveld refinement.
        (B) Scanning electron microscopy (SEM) showing representative morphology (scale bar around 200\,$\mu$m).
        (C) Scanning electron microscopy (SEM) showing representative morphology (window scaled to particle size).
        (D) Energy-dispersive X-ray spectroscopy (EDS) elemental composition data.}
        \label{fig:SI_16}
        \vspace{-1em}
    \end{figure}
    \begin{figure}[ht]
        \centering
        \includegraphics[width=0.7\linewidth]{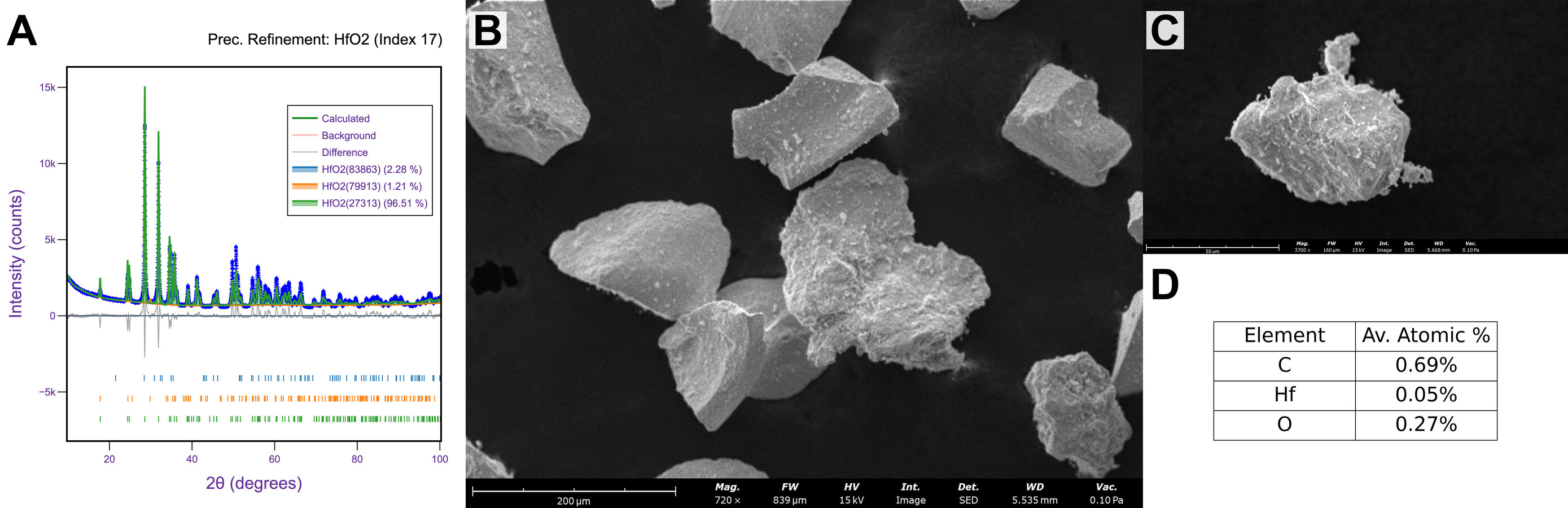}
        \caption{\textbf{Characterization of precursor HfO$_2$ (Project ID: 17).} (A) Powder X-ray diffraction (XRD) pattern with Rietveld refinement.
        (B) Scanning electron microscopy (SEM) showing representative morphology (scale bar around 200\,$\mu$m).
        (C) Scanning electron microscopy (SEM) showing representative morphology (window scaled to particle size).
        (D) Energy-dispersive X-ray spectroscopy (EDS) elemental composition data.}
        \label{fig:SI_17}
        \vspace{-1em}
    \end{figure}
    \begin{figure}[ht]
        \centering
        \includegraphics[width=0.7\linewidth]{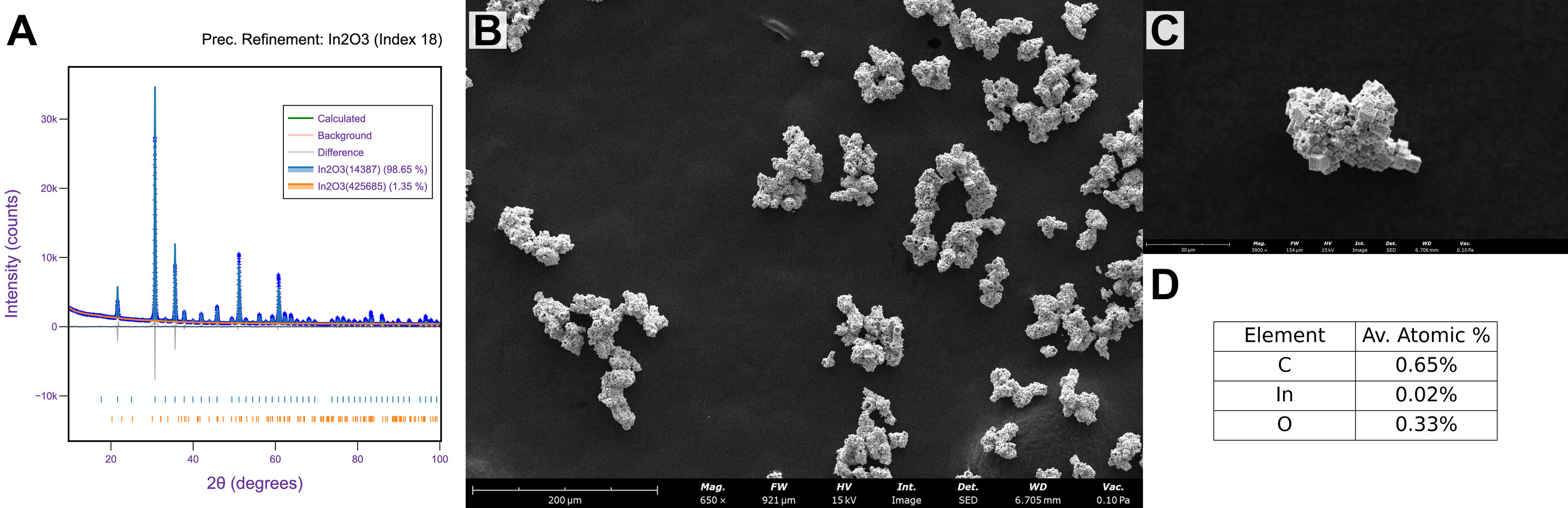}
        \caption{\textbf{Characterization of precursor In$_2$O$_3$ (Project ID: 18).} (A) Powder X-ray diffraction (XRD) pattern with Rietveld refinement.
        (B) Scanning electron microscopy (SEM) showing representative morphology (scale bar around 200\,$\mu$m).
        (C) Scanning electron microscopy (SEM) showing representative morphology (window scaled to particle size).
        (D) Energy-dispersive X-ray spectroscopy (EDS) elemental composition data.}
        \label{fig:SI_18}
        \vspace{-1em}
    \end{figure}
    \begin{figure}[ht]
        \centering
        \includegraphics[width=0.7\linewidth]{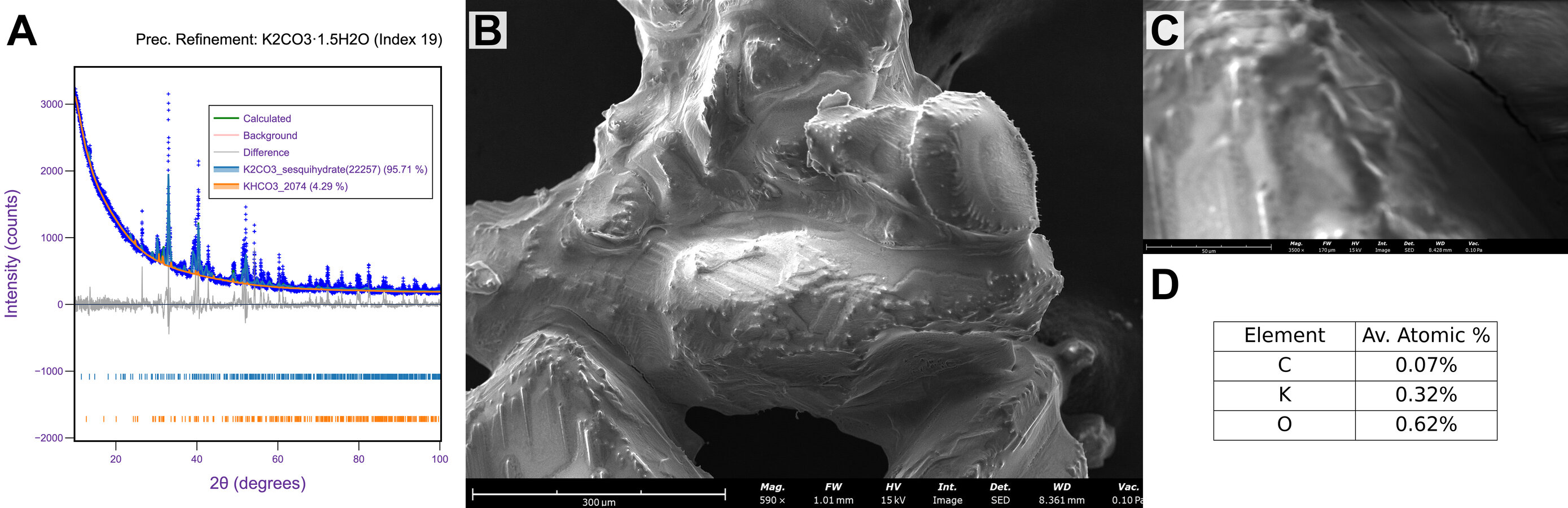}
        \caption{\textbf{Characterization of precursor K$_2$CO$_3\cdot$1.5H$_2$O (Project ID: 19).} (A) Powder X-ray diffraction (XRD) pattern with Rietveld refinement.
        (B) Scanning electron microscopy (SEM) showing representative morphology (scale bar around 200\,$\mu$m).
        (C) Scanning electron microscopy (SEM) showing representative morphology (window scaled to particle size).
        (D) Energy-dispersive X-ray spectroscopy (EDS) elemental composition data.}
        \label{fig:SI_19}
        \vspace{-1em}
    \end{figure}
    \begin{figure}[ht]
        \centering
        \includegraphics[width=0.7\linewidth]{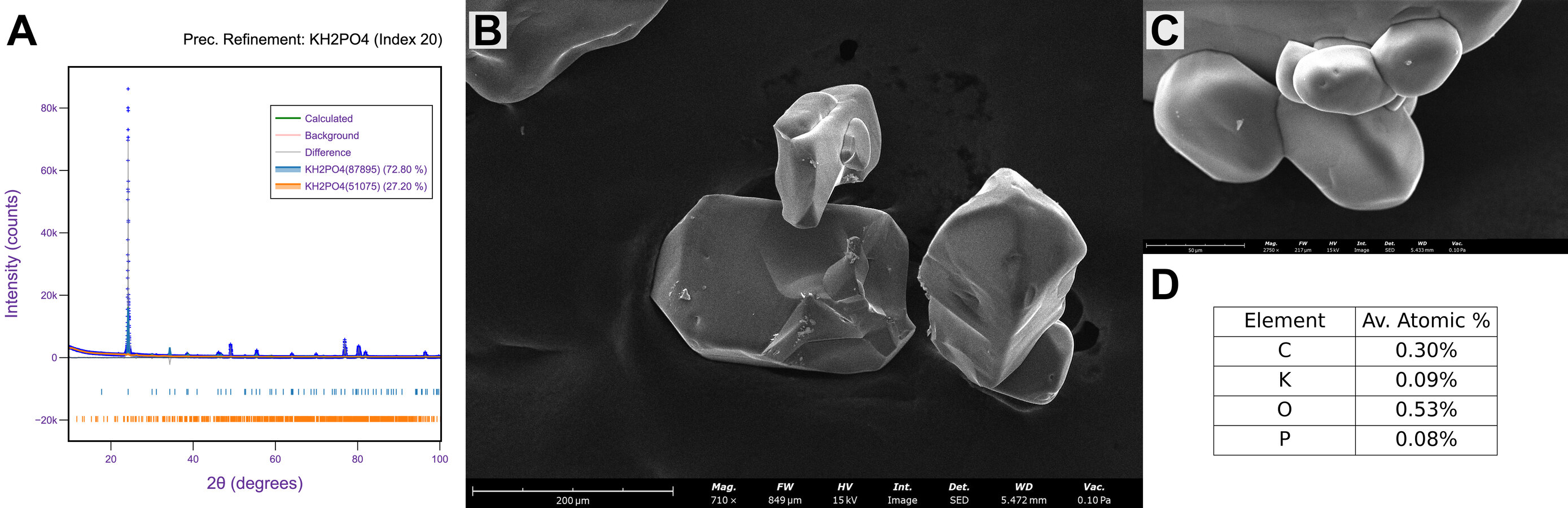}
        \caption{\textbf{Characterization of precursor KH$_2$PO$_4$ (Project ID: 20).} (A) Powder X-ray diffraction (XRD) pattern with Rietveld refinement.
        (B) Scanning electron microscopy (SEM) showing representative morphology (scale bar around 200\,$\mu$m).
        (C) Scanning electron microscopy (SEM) showing representative morphology (window scaled to particle size).
        (D) Energy-dispersive X-ray spectroscopy (EDS) elemental composition data.}
        \label{fig:SI_20}
        \vspace{-1em}
    \end{figure}
    \begin{figure}[ht]
        \centering
        \includegraphics[width=0.7\linewidth]{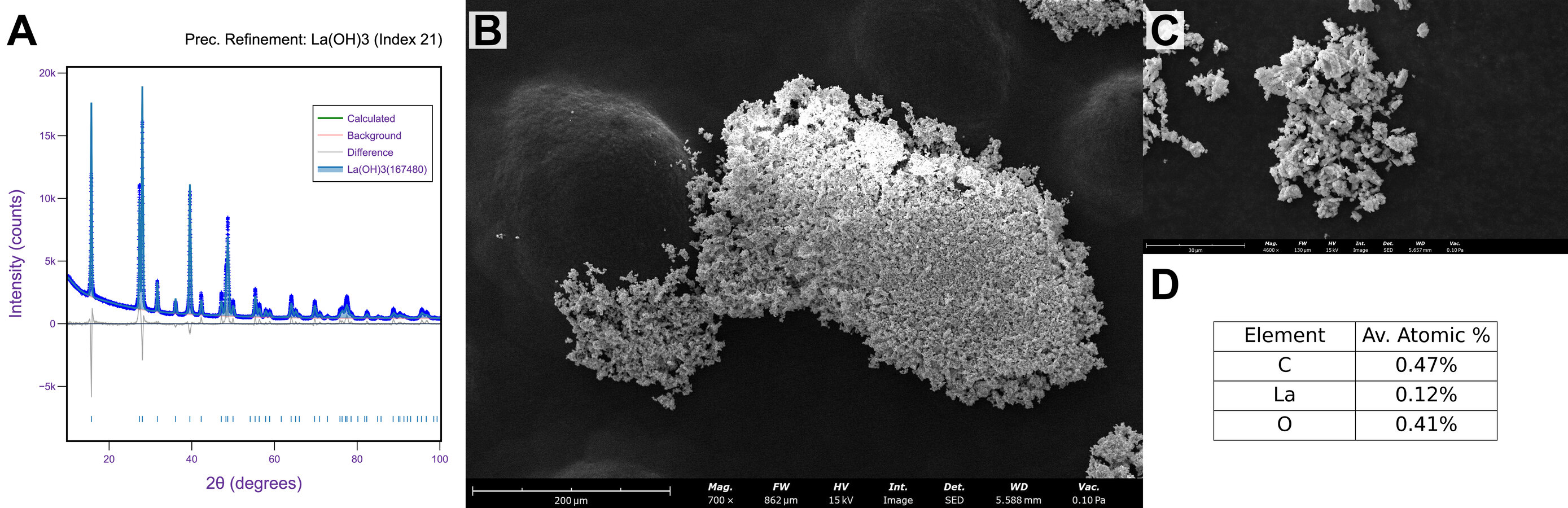}
        \caption{\textbf{Characterization of precursor La(OH)$_3$ (Project ID: 21).} (A) Powder X-ray diffraction (XRD) pattern with Rietveld refinement.
        (B) Scanning electron microscopy (SEM) showing representative morphology (scale bar around 200\,$\mu$m).
        (C) Scanning electron microscopy (SEM) showing representative morphology (window scaled to particle size).
        (D) Energy-dispersive X-ray spectroscopy (EDS) elemental composition data.}
        \label{fig:SI_21}
        \vspace{-1em}
    \end{figure}
    \begin{figure}[ht]
        \centering
        \includegraphics[width=0.7\linewidth]{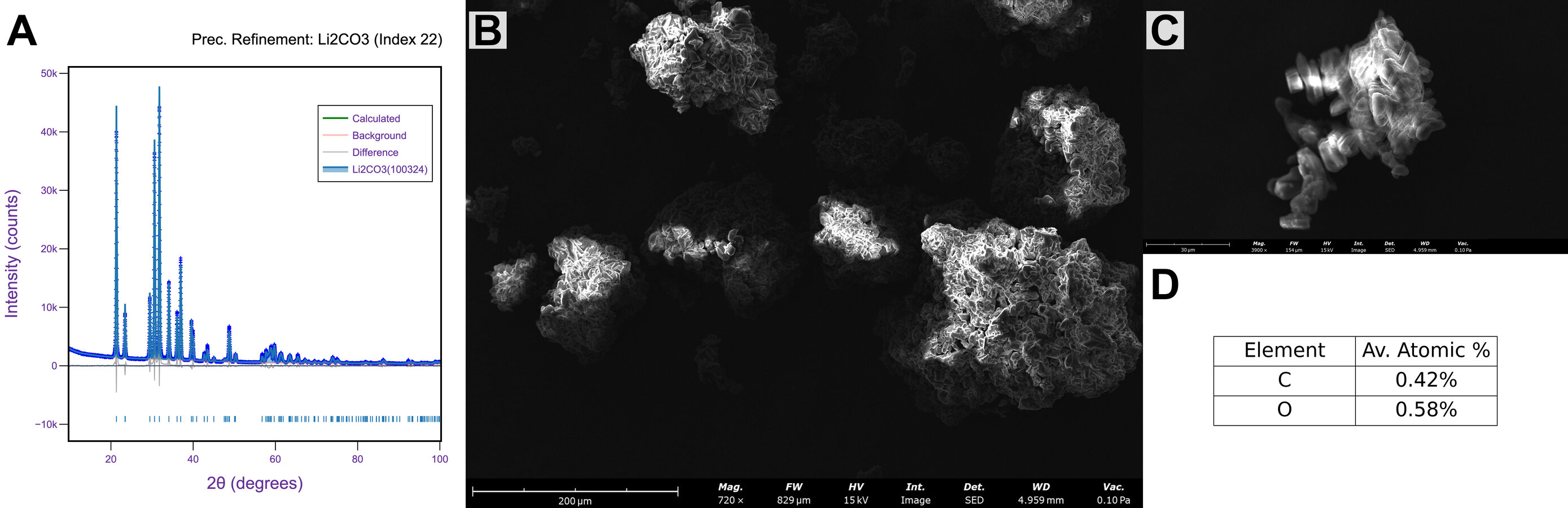}
        \caption{\textbf{Characterization of precursor Li$_2$CO$_3$ (Project ID: 22).} (A) Powder X-ray diffraction (XRD) pattern with Rietveld refinement.
        (B) Scanning electron microscopy (SEM) showing representative morphology (scale bar around 200\,$\mu$m).
        (C) Scanning electron microscopy (SEM) showing representative morphology (window scaled to particle size).
        (D) Energy-dispersive X-ray spectroscopy (EDS) elemental composition data.}
        \label{fig:SI_22}
        \vspace{-1em}
    \end{figure}
    \begin{figure}[ht]
        \centering
        \includegraphics[width=0.7\linewidth]{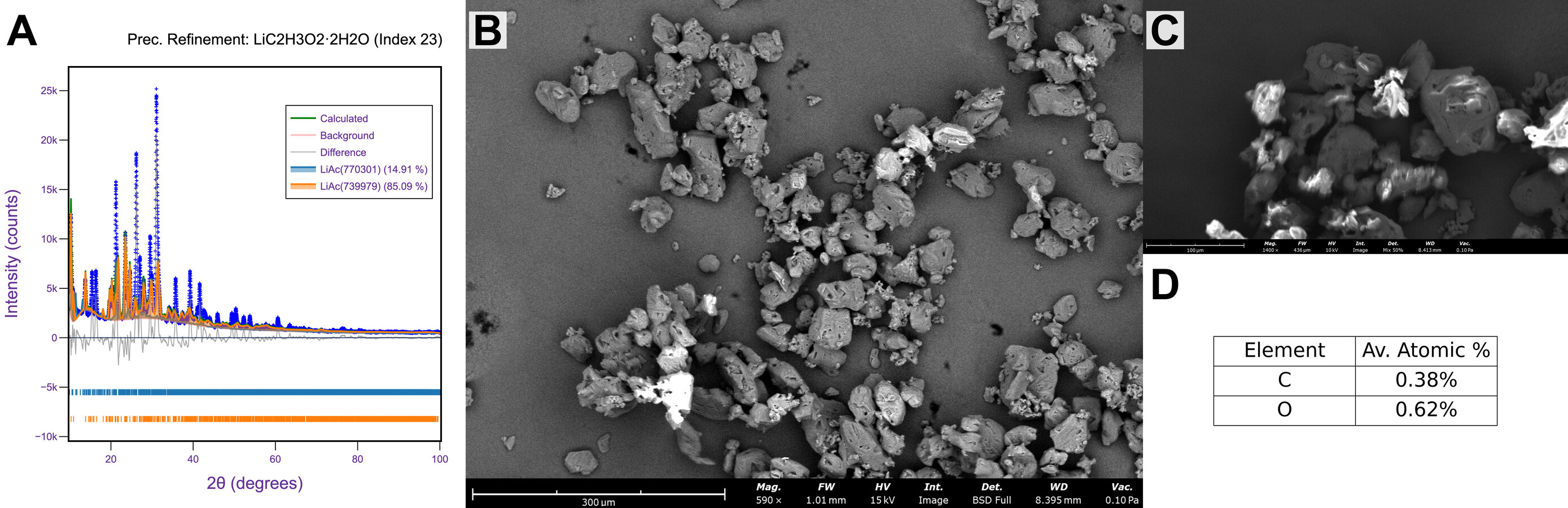}
        \caption{\textbf{Characterization of precursor LiC$_2$H$_3$O$_2\cdot$2H$_2$O (Project ID: 23).} (A) Powder X-ray diffraction (XRD) pattern with Rietveld refinement.
        (B) Scanning electron microscopy (SEM) showing representative morphology (scale bar around 200\,$\mu$m).
        (C) Scanning electron microscopy (SEM) showing representative morphology (window scaled to particle size).
        (D) Energy-dispersive X-ray spectroscopy (EDS) elemental composition data.}
        \label{fig:SI_23}
        \vspace{-1em}
    \end{figure}
    \begin{figure}[ht]
        \centering
        \includegraphics[width=0.7\linewidth]{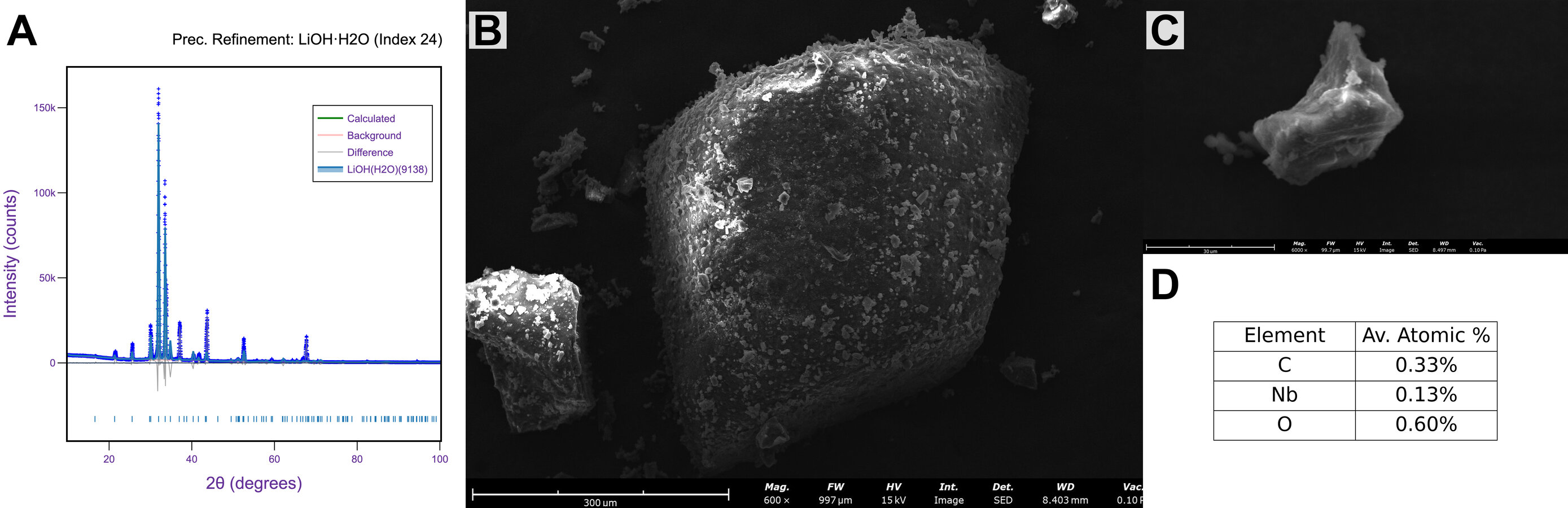}
        \caption{\textbf{Characterization of precursor LiOH$\cdot$H$_2$O (Project ID: 24).} (A) Powder X-ray diffraction (XRD) pattern with Rietveld refinement.
        (B) Scanning electron microscopy (SEM) showing representative morphology (scale bar around 200\,$\mu$m).
        (C) Scanning electron microscopy (SEM) showing representative morphology (window scaled to particle size).
        (D) Energy-dispersive X-ray spectroscopy (EDS) elemental composition data.}
        \label{fig:SI_24}
        \vspace{-1em}
    \end{figure}
    \begin{figure}[ht]
        \centering
        \includegraphics[width=0.7\linewidth]{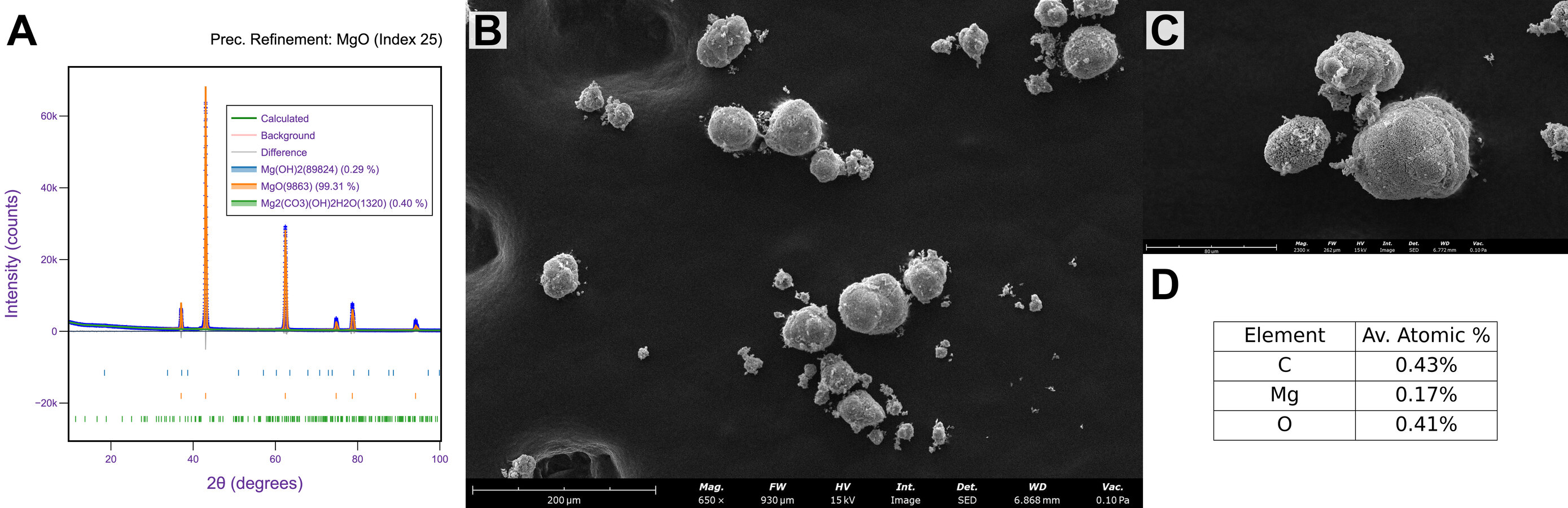}
        \caption{\textbf{Characterization of precursor MgO (Project ID: 25).} (A) Powder X-ray diffraction (XRD) pattern with Rietveld refinement.
        (B) Scanning electron microscopy (SEM) showing representative morphology (scale bar around 200\,$\mu$m).
        (C) Scanning electron microscopy (SEM) showing representative morphology (window scaled to particle size).
        (D) Energy-dispersive X-ray spectroscopy (EDS) elemental composition data.}
        \label{fig:SI_25}
        \vspace{-1em}
    \end{figure}
    \begin{figure}[ht]
        \centering
        \includegraphics[width=0.7\linewidth]{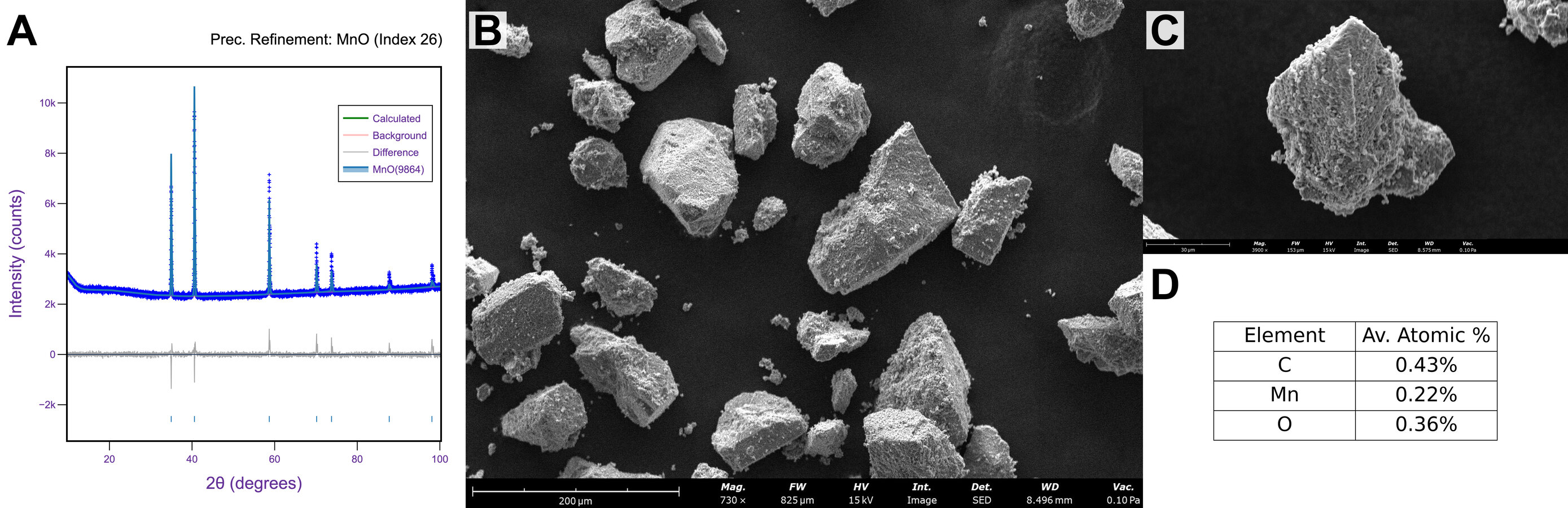}
        \caption{\textbf{Characterization of precursor MnO (Project ID: 26).} (A) Powder X-ray diffraction (XRD) pattern with Rietveld refinement.
        (B) Scanning electron microscopy (SEM) showing representative morphology (scale bar around 200\,$\mu$m).
        (C) Scanning electron microscopy (SEM) showing representative morphology (window scaled to particle size).
        (D) Energy-dispersive X-ray spectroscopy (EDS) elemental composition data.}
        \label{fig:SI_26}
        \vspace{-1em}
    \end{figure}
    \begin{figure}[ht]
        \centering
        \includegraphics[width=0.7\linewidth]{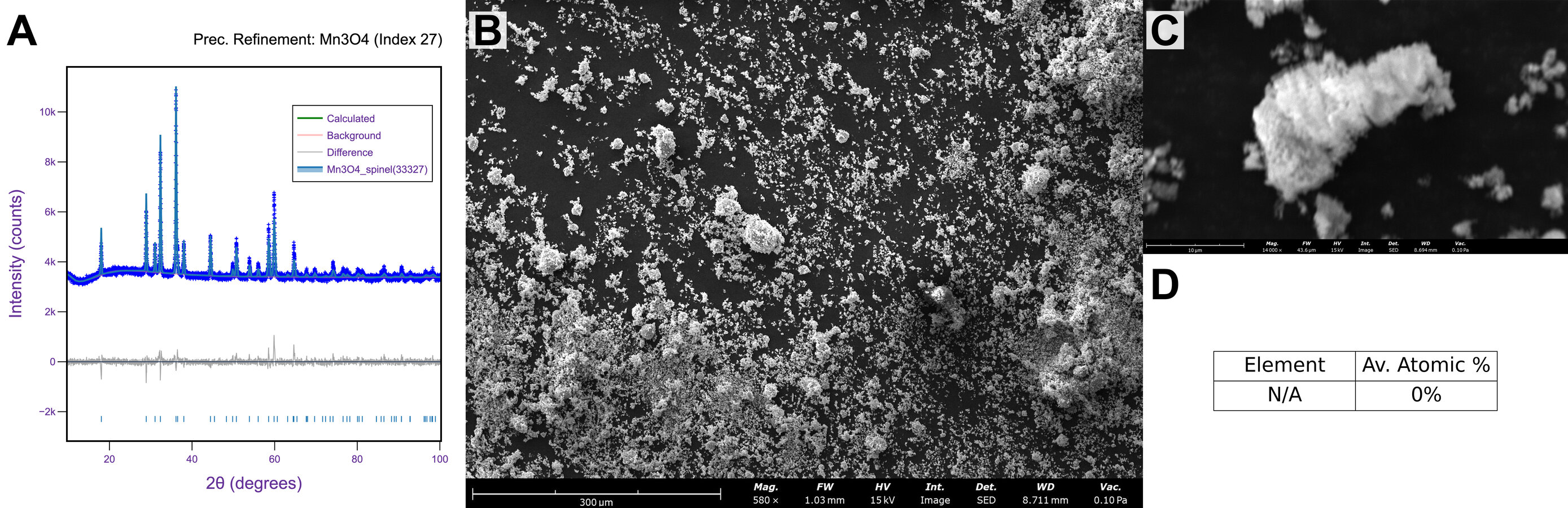}
        \caption{\textbf{Characterization of precursor Mn$_3$O$_4$ (Project ID: 27).} (A) Powder X-ray diffraction (XRD) pattern with Rietveld refinement.
        (B) Scanning electron microscopy (SEM) showing representative morphology (scale bar around 200\,$\mu$m).
        (C) Scanning electron microscopy (SEM) showing representative morphology (window scaled to particle size).
        (D) Energy-dispersive X-ray spectroscopy (EDS) elemental composition data.}
        \label{fig:SI_27}
        \vspace{-1em}
    \end{figure}
    \begin{figure}[ht]
        \centering
        \includegraphics[width=0.7\linewidth]{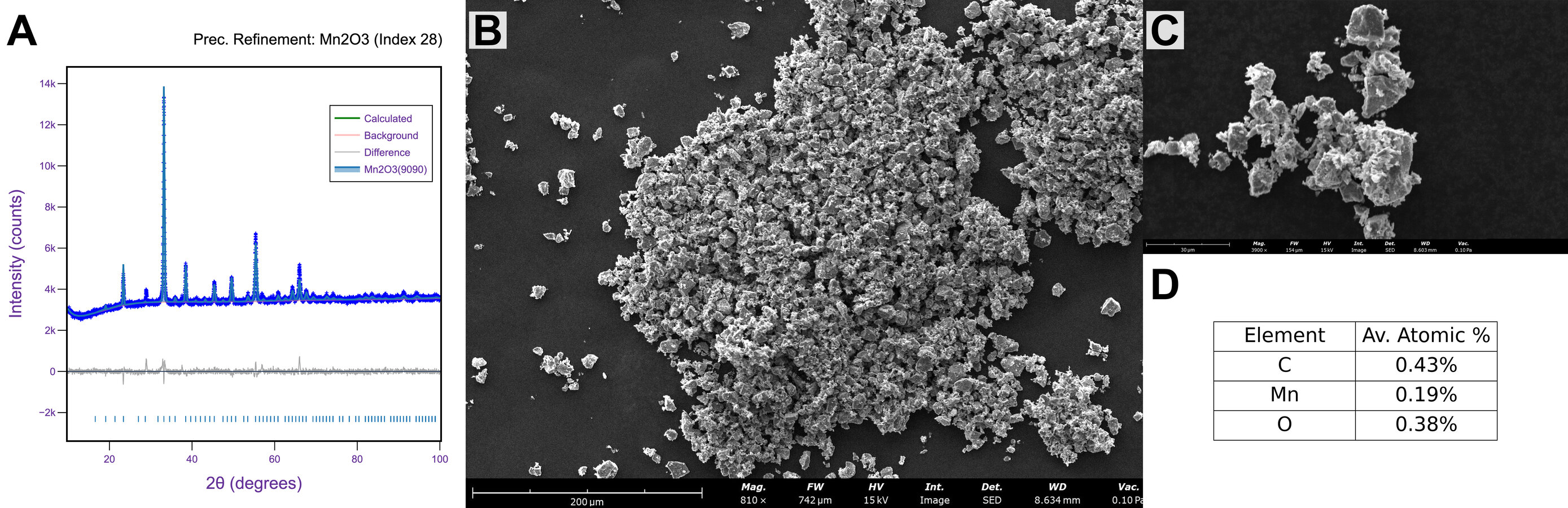}
        \caption{\textbf{Characterization of precursor Mn$_2$O$_3$ (Project ID: 28).} (A) Powder X-ray diffraction (XRD) pattern with Rietveld refinement.
        (B) Scanning electron microscopy (SEM) showing representative morphology (scale bar around 200\,$\mu$m).
        (C) Scanning electron microscopy (SEM) showing representative morphology (window scaled to particle size).
        (D) Energy-dispersive X-ray spectroscopy (EDS) elemental composition data.}
        \label{fig:SI_28}
        \vspace{-1em}
    \end{figure}
    \begin{figure}[ht]
        \centering
        \includegraphics[width=0.7\linewidth]{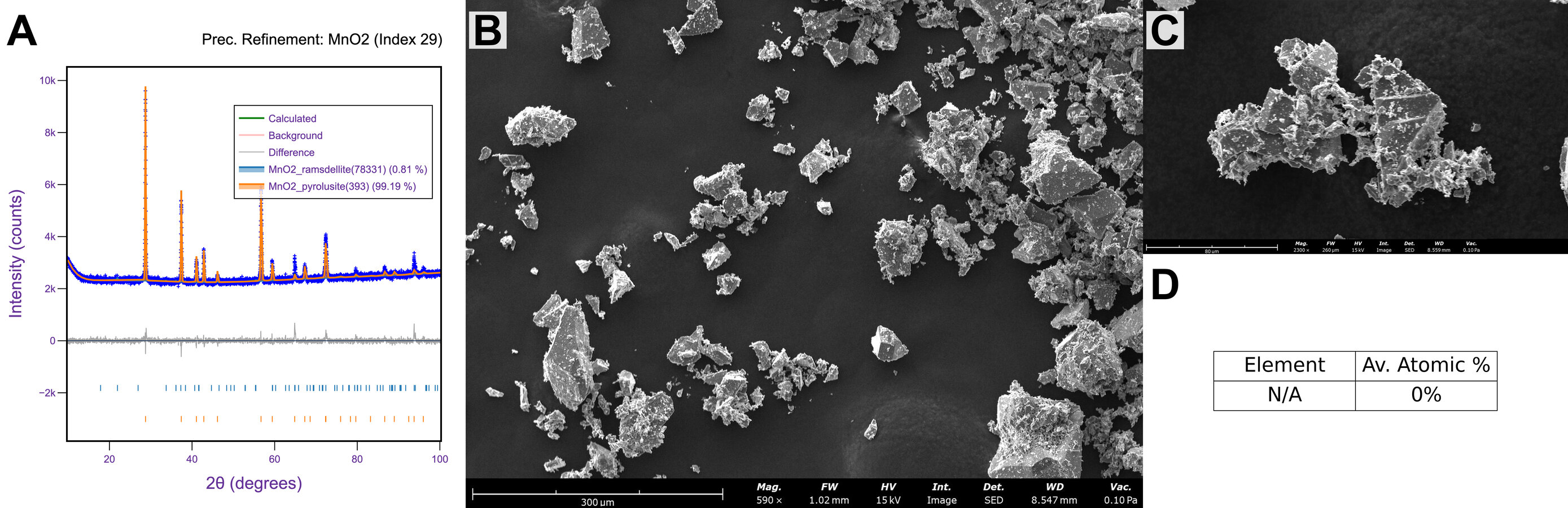}
        \caption{\textbf{Characterization of precursor MnO$_2$ (Project ID: 29).} (A) Powder X-ray diffraction (XRD) pattern with Rietveld refinement.
        (B) Scanning electron microscopy (SEM) showing representative morphology (scale bar around 200\,$\mu$m).
        (C) Scanning electron microscopy (SEM) showing representative morphology (window scaled to particle size).
        (D) Energy-dispersive X-ray spectroscopy (EDS) elemental composition data.}
        \label{fig:SI_29}
        \vspace{-1em}
    \end{figure}
    \begin{figure}[ht]
        \centering
        \includegraphics[width=0.7\linewidth]{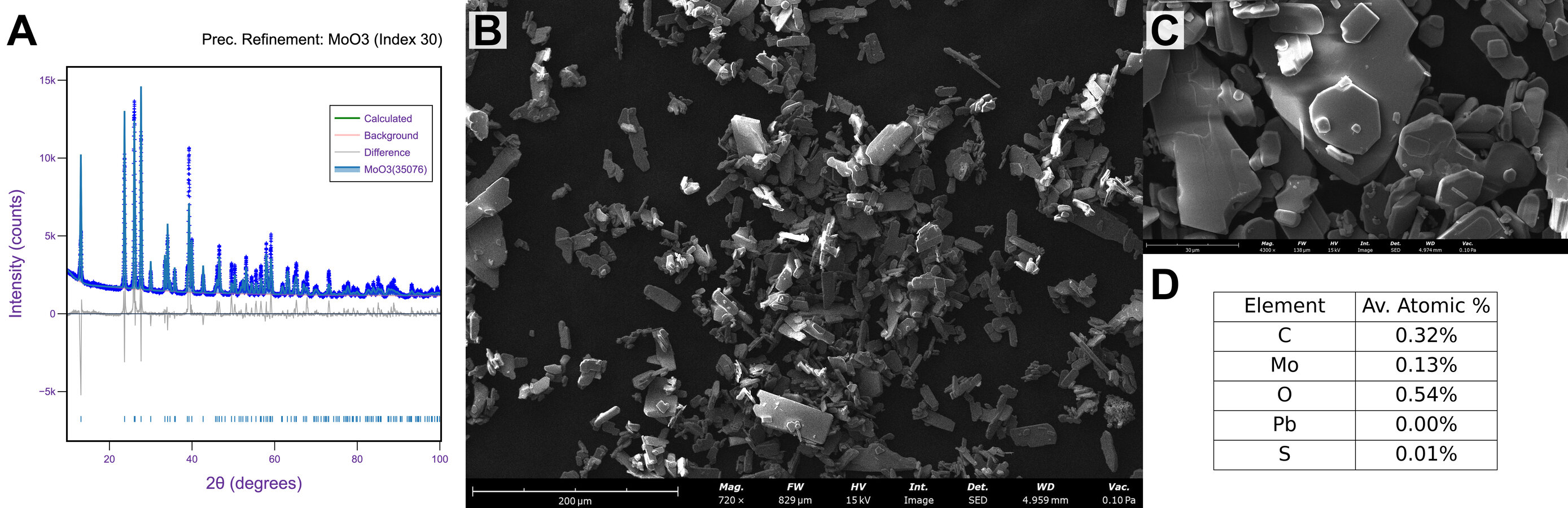}
        \caption{\textbf{Characterization of precursor MoO$_3$ (Project ID: 30).} (A) Powder X-ray diffraction (XRD) pattern with Rietveld refinement.
        (B) Scanning electron microscopy (SEM) showing representative morphology (scale bar around 200\,$\mu$m).
        (C) Scanning electron microscopy (SEM) showing representative morphology (window scaled to particle size).
        (D) Energy-dispersive X-ray spectroscopy (EDS) elemental composition data.}
        \label{fig:SI_30}
        \vspace{-1em}
    \end{figure}
    \begin{figure}[ht]
        \centering
        \includegraphics[width=0.7\linewidth]{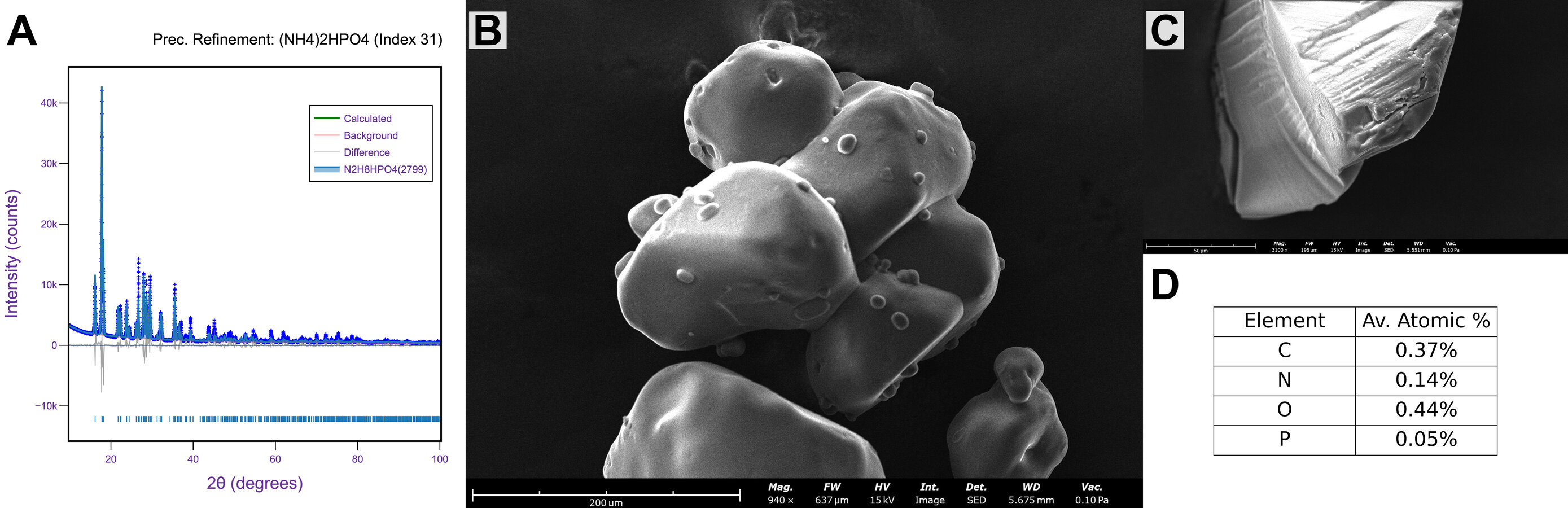}
        \caption{\textbf{Characterization of precursor (NH$_4$)$_2$HPO$_4$ (Project ID: 31).} (A) Powder X-ray diffraction (XRD) pattern with Rietveld refinement.
        (B) Scanning electron microscopy (SEM) showing representative morphology (scale bar around 200\,$\mu$m).
        (C) Scanning electron microscopy (SEM) showing representative morphology (window scaled to particle size).
        (D) Energy-dispersive X-ray spectroscopy (EDS) elemental composition data.}
        \label{fig:SI_31}
        \vspace{-1em}
    \end{figure}
    \begin{figure}[ht]
        \centering
        \includegraphics[width=0.7\linewidth]{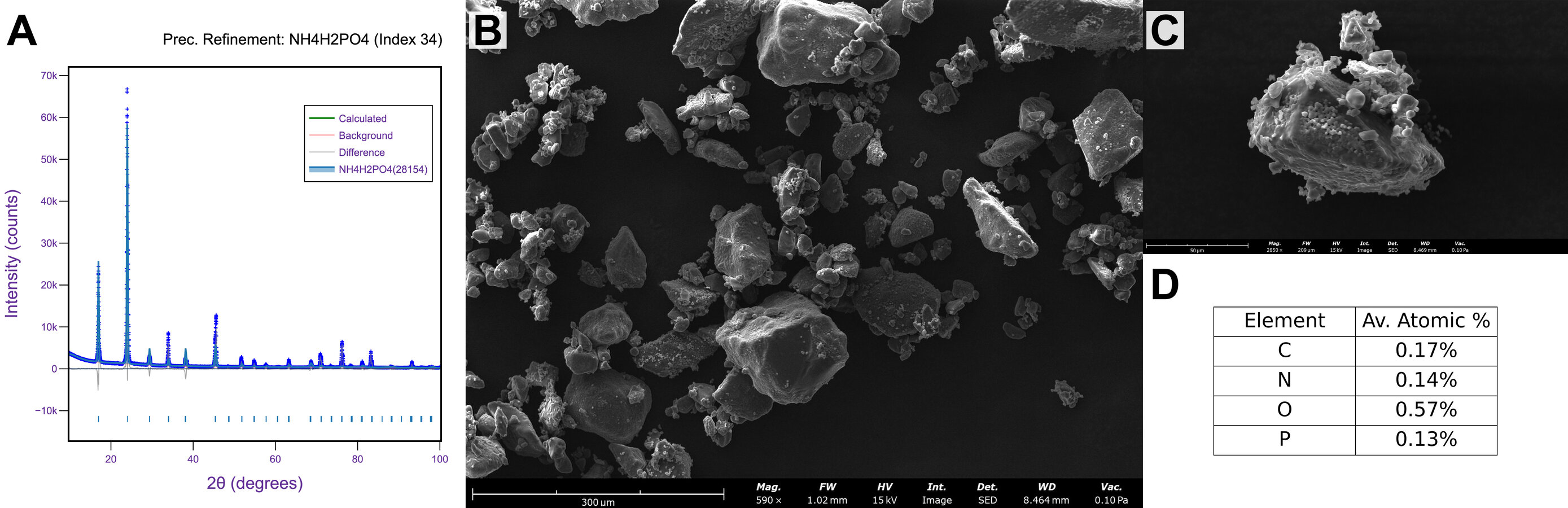}
        \caption{\textbf{Characterization of precursor NH$_4$H$_2$PO$_4$ (Project ID: 34).} (A) Powder X-ray diffraction (XRD) pattern with Rietveld refinement.
        (B) Scanning electron microscopy (SEM) showing representative morphology (scale bar around 200\,$\mu$m).
        (C) Scanning electron microscopy (SEM) showing representative morphology (window scaled to particle size).
        (D) Energy-dispersive X-ray spectroscopy (EDS) elemental composition data.}
        \label{fig:SI_34}
        \vspace{-1em}
    \end{figure}
    \begin{figure}[ht]
        \centering
        \includegraphics[width=0.7\linewidth]{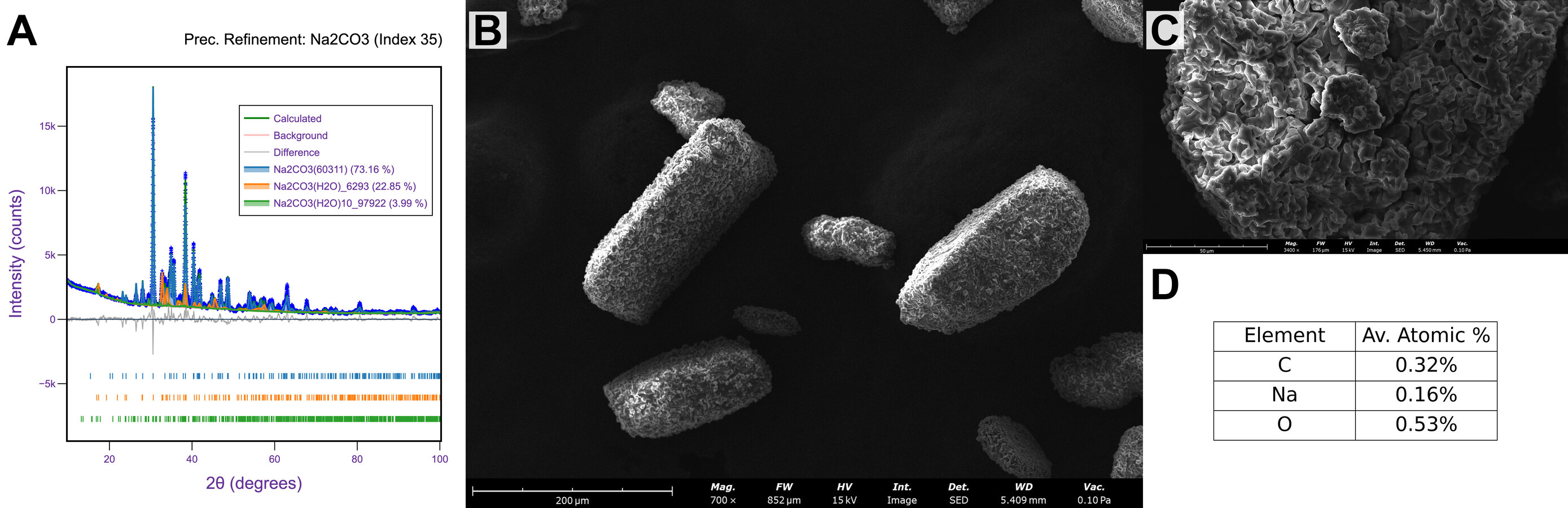}
        \caption{\textbf{Characterization of precursor Na$_2$CO$_3$ (Project ID: 35).} (A) Powder X-ray diffraction (XRD) pattern with Rietveld refinement.
        (B) Scanning electron microscopy (SEM) showing representative morphology (scale bar around 200\,$\mu$m).
        (C) Scanning electron microscopy (SEM) showing representative morphology (window scaled to particle size).
        (D) Energy-dispersive X-ray spectroscopy (EDS) elemental composition data.}
        \label{fig:SI_35}
        \vspace{-1em}
    \end{figure}
    \begin{figure}[ht]
        \centering
        \includegraphics[width=0.7\linewidth]{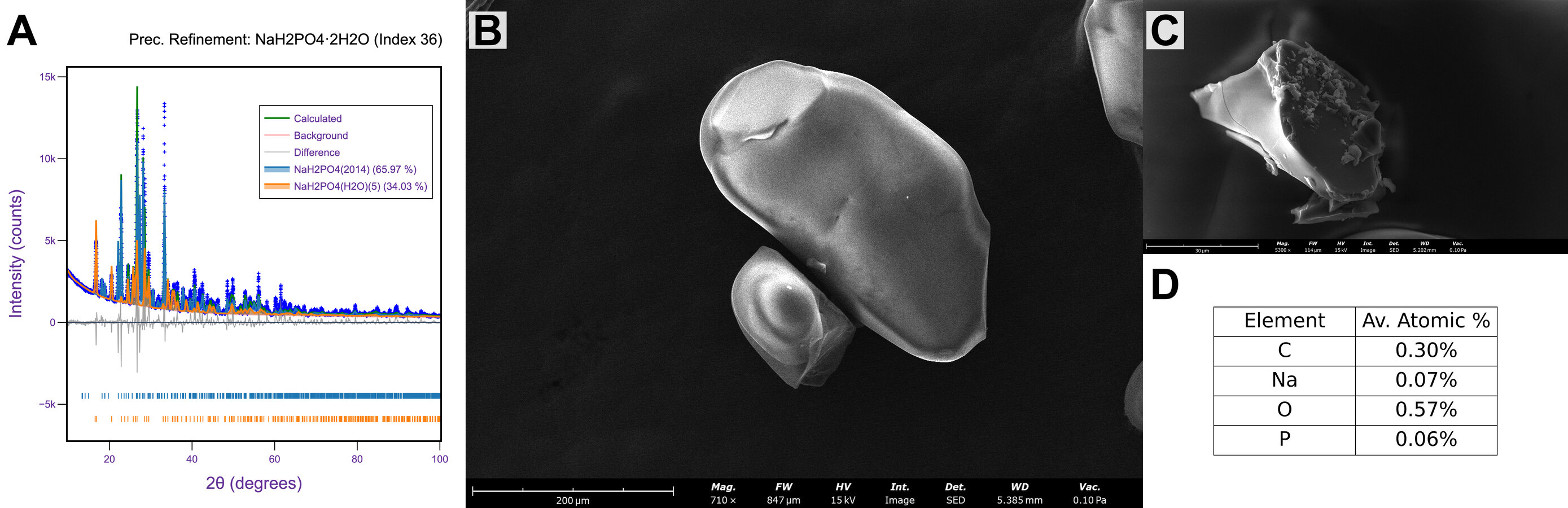}
        \caption{\textbf{Characterization of precursor NaH$_2$PO$_4\cdot$2H$_2$O (Project ID: 36).} (A) Powder X-ray diffraction (XRD) pattern with Rietveld refinement.
        (B) Scanning electron microscopy (SEM) showing representative morphology (scale bar around 200\,$\mu$m).
        (C) Scanning electron microscopy (SEM) showing representative morphology (window scaled to particle size).
        (D) Energy-dispersive X-ray spectroscopy (EDS) elemental composition data.}
        \label{fig:SI_36}
        \vspace{-1em}
    \end{figure}
    \begin{figure}[ht]
        \centering
        \includegraphics[width=0.7\linewidth]{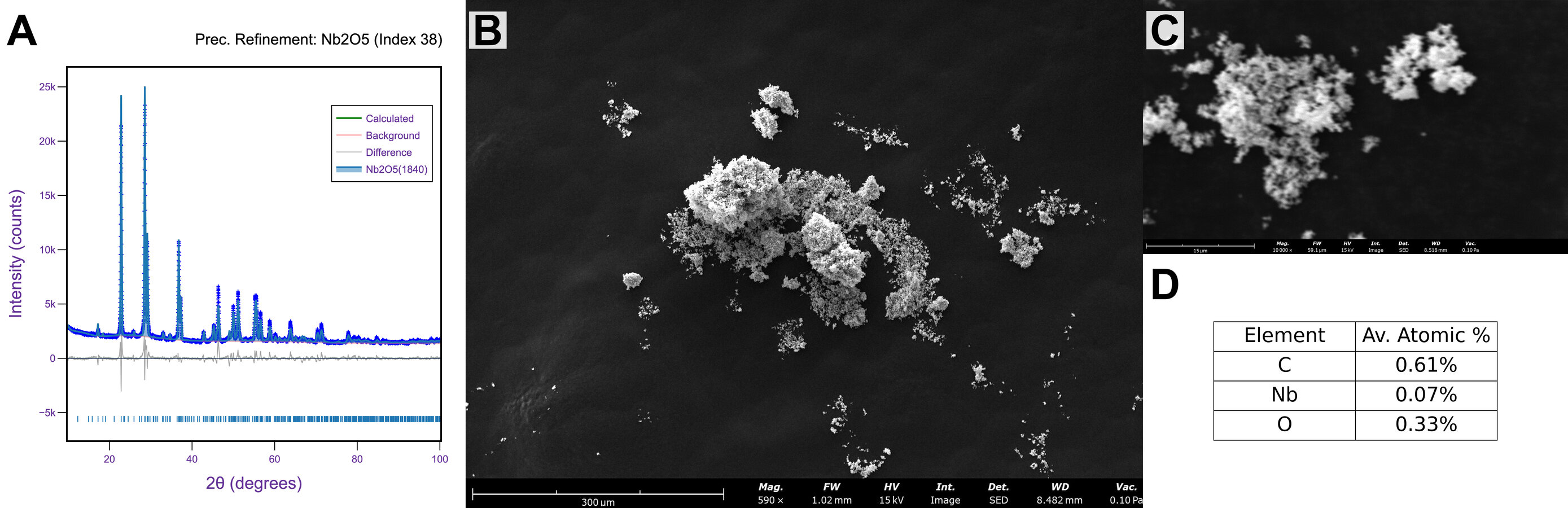}
        \caption{\textbf{Characterization of precursor Nb$_2$O$_5$ (Project ID: 38).} (A) Powder X-ray diffraction (XRD) pattern with Rietveld refinement.
        (B) Scanning electron microscopy (SEM) showing representative morphology (scale bar around 200\,$\mu$m).
        (C) Scanning electron microscopy (SEM) showing representative morphology (window scaled to particle size).
        (D) Energy-dispersive X-ray spectroscopy (EDS) elemental composition data.}
        \label{fig:SI_38}
        \vspace{-1em}
    \end{figure}
    \begin{figure}[ht]
        \centering
        \includegraphics[width=0.7\linewidth]{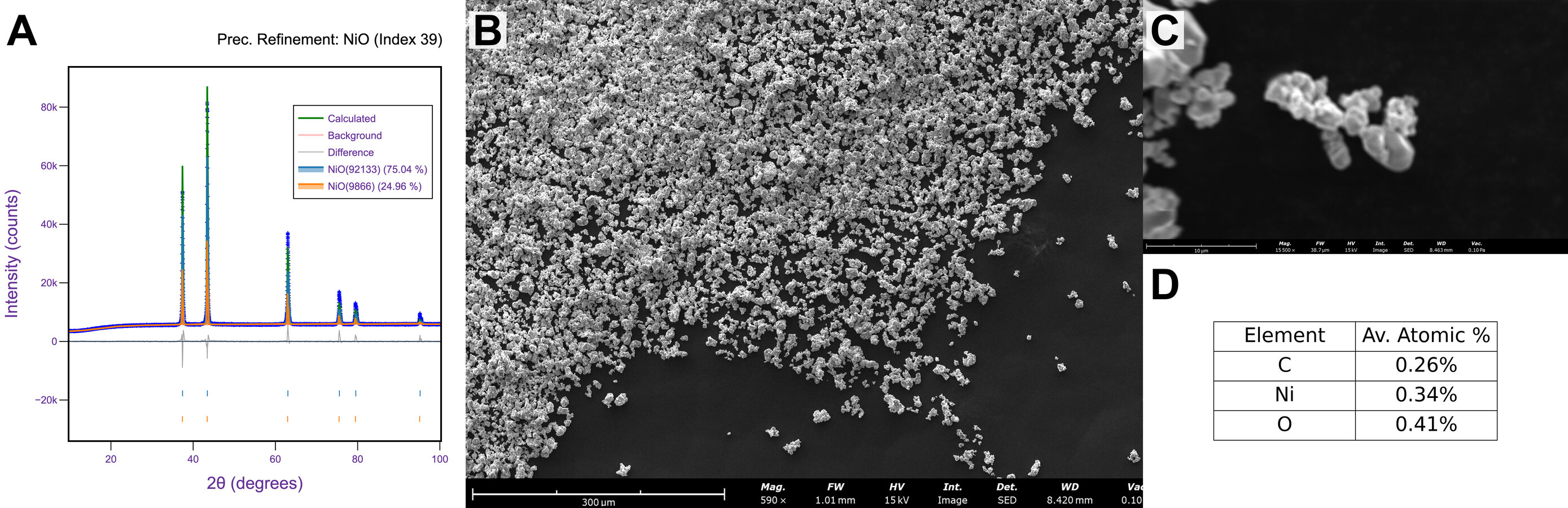}
        \caption{\textbf{Characterization of precursor NiO (Project ID: 39).} (A) Powder X-ray diffraction (XRD) pattern with Rietveld refinement.
        (B) Scanning electron microscopy (SEM) showing representative morphology (scale bar around 200\,$\mu$m).
        (C) Scanning electron microscopy (SEM) showing representative morphology (window scaled to particle size).
        (D) Energy-dispersive X-ray spectroscopy (EDS) elemental composition data.}
        \label{fig:SI_39}
        \vspace{-1em}
    \end{figure}
    \begin{figure}[ht]
        \centering
        \includegraphics[width=0.7\linewidth]{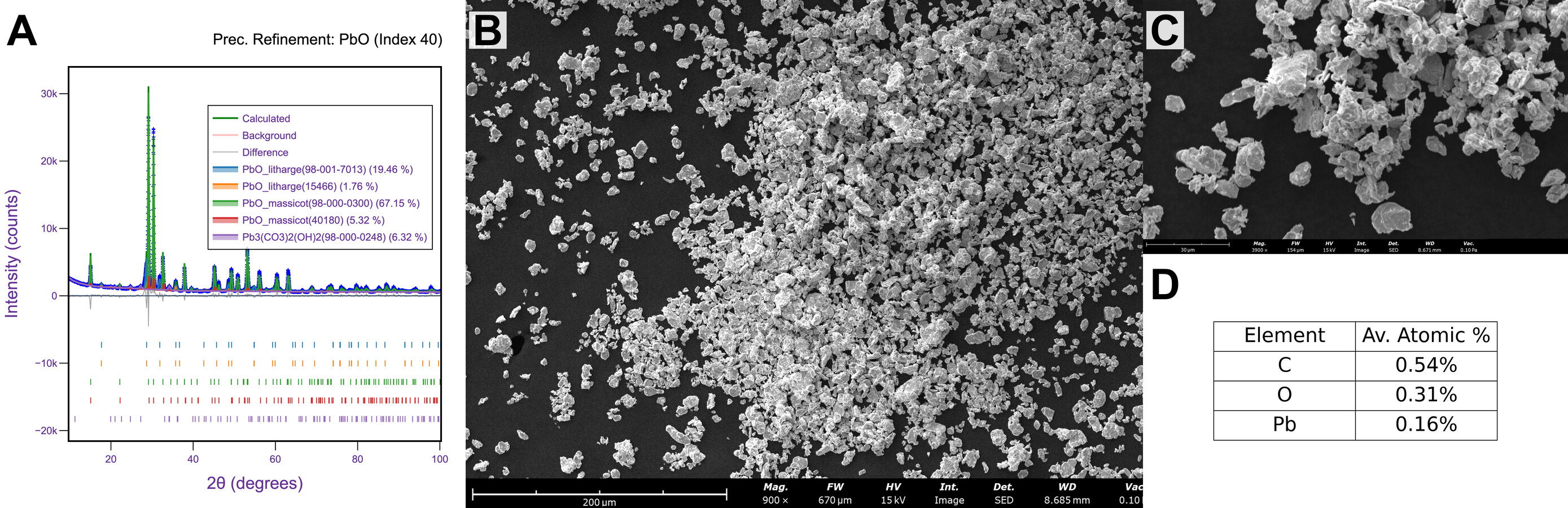}
        \caption{\textbf{Characterization of precursor PbO (Project ID: 40).} (A) Powder X-ray diffraction (XRD) pattern with Rietveld refinement.
        (B) Scanning electron microscopy (SEM) showing representative morphology (scale bar around 200\,$\mu$m).
        (C) Scanning electron microscopy (SEM) showing representative morphology (window scaled to particle size).
        (D) Energy-dispersive X-ray spectroscopy (EDS) elemental composition data.}
        \label{fig:SI_40}
        \vspace{-1em}
    \end{figure}
    \begin{figure}[ht]
        \centering
        \includegraphics[width=0.7\linewidth]{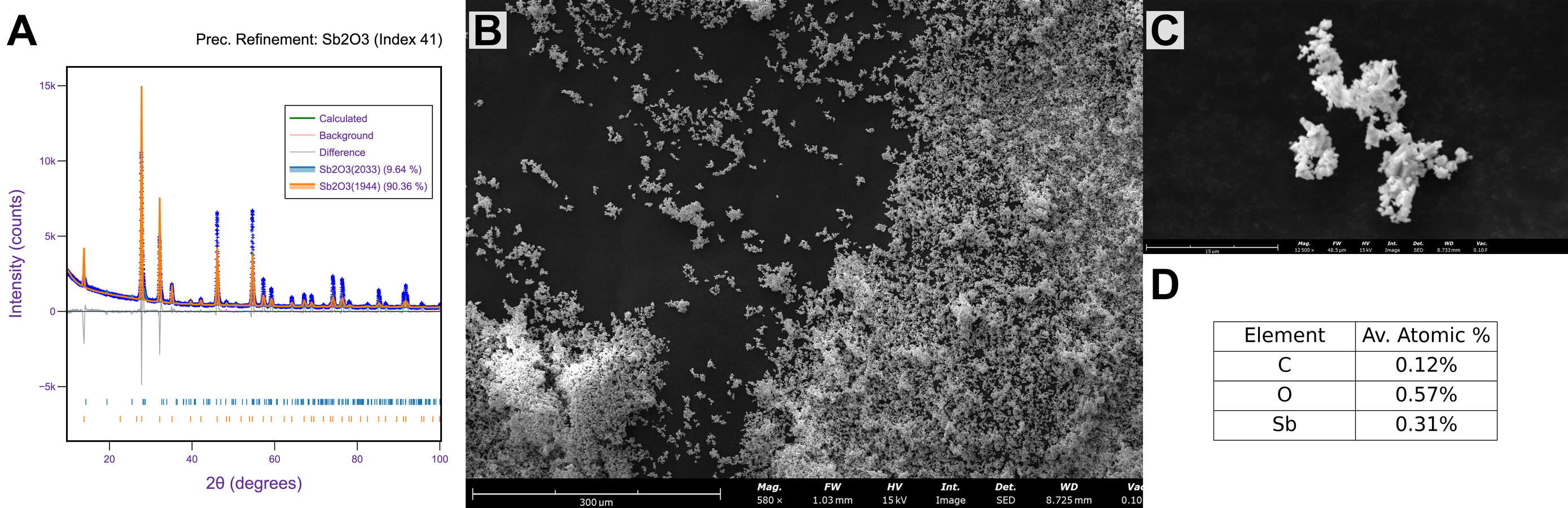}
        \caption{\textbf{Characterization of precursor Sb$_2$O$_3$ (Project ID: 41).} (A) Powder X-ray diffraction (XRD) pattern with Rietveld refinement.
        (B) Scanning electron microscopy (SEM) showing representative morphology (scale bar around 200\,$\mu$m).
        (C) Scanning electron microscopy (SEM) showing representative morphology (window scaled to particle size).
        (D) Energy-dispersive X-ray spectroscopy (EDS) elemental composition data.}
        \label{fig:SI_41}
        \vspace{-1em}
    \end{figure}
    \begin{figure}[ht]
        \centering
        \includegraphics[width=0.7\linewidth]{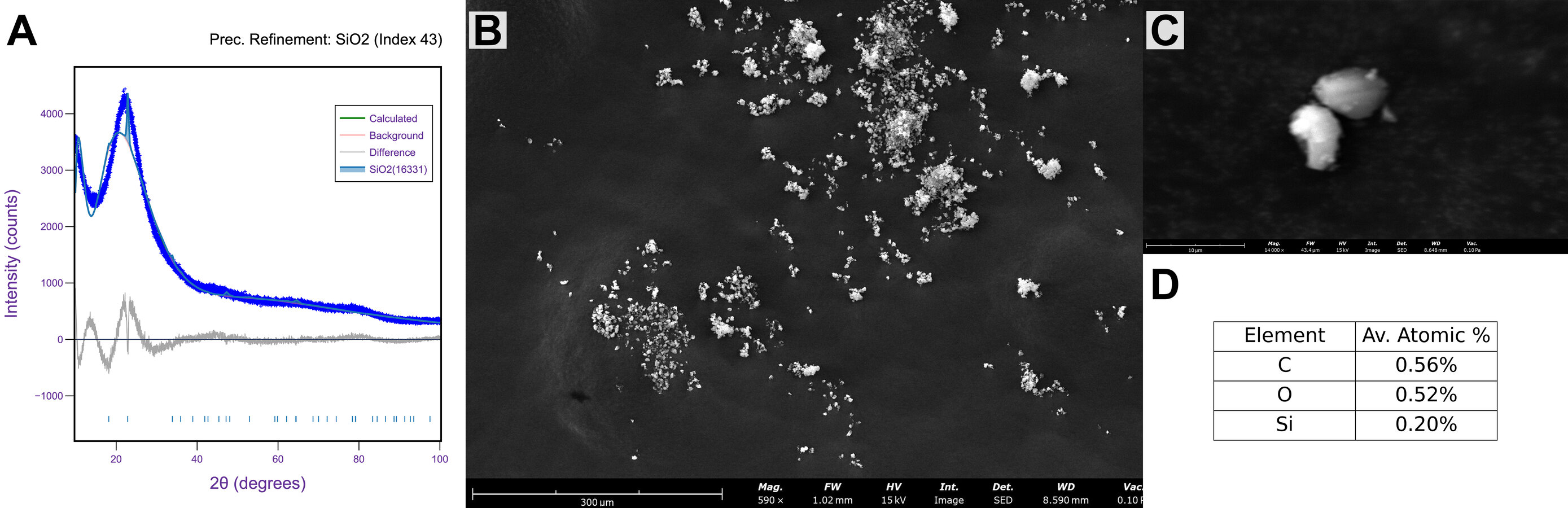}
        \caption{\textbf{Characterization of precursor SiO$_2$ (Project ID: 43).} (A) Powder X-ray diffraction (XRD) pattern with Rietveld refinement.
        (B) Scanning electron microscopy (SEM) showing representative morphology (scale bar around 200\,$\mu$m).
        (C) Scanning electron microscopy (SEM) showing representative morphology (window scaled to particle size).
        (D) Energy-dispersive X-ray spectroscopy (EDS) elemental composition data.}
        \label{fig:SI_43}
        \vspace{-1em}
    \end{figure}
    \begin{figure}[ht]
        \centering
        \includegraphics[width=0.7\linewidth]{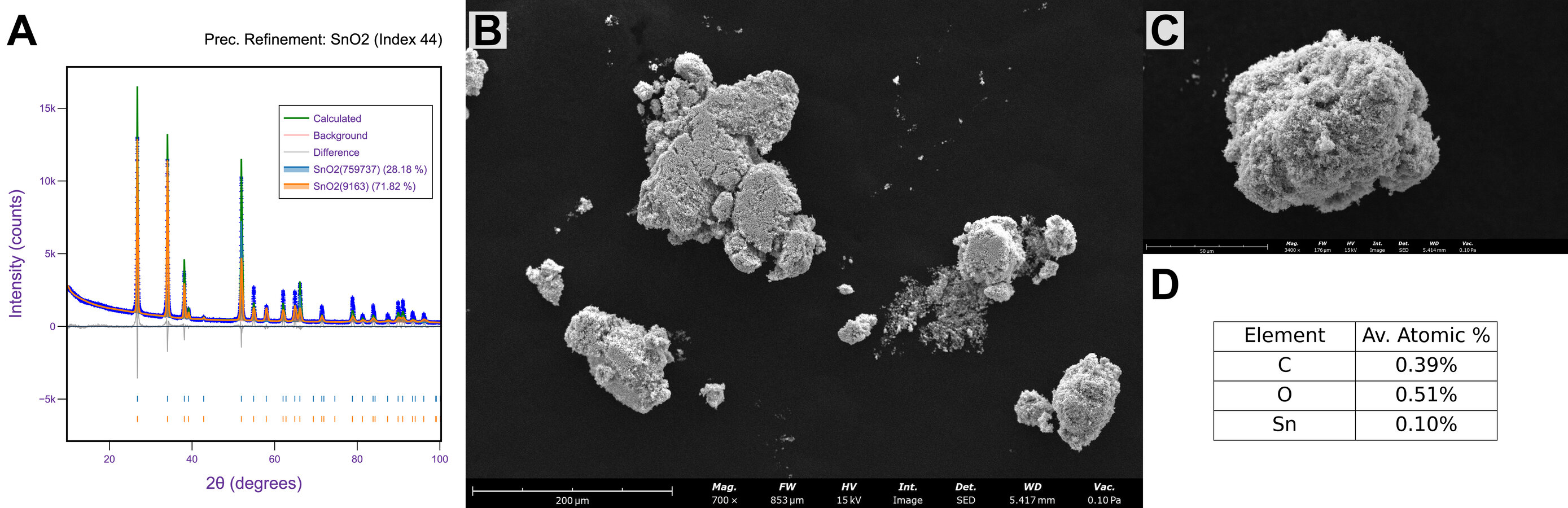}
        \caption{\textbf{Characterization of precursor SnO$_2$ (Project ID: 44).} (A) Powder X-ray diffraction (XRD) pattern with Rietveld refinement.
        (B) Scanning electron microscopy (SEM) showing representative morphology (scale bar around 200\,$\mu$m).
        (C) Scanning electron microscopy (SEM) showing representative morphology (window scaled to particle size).
        (D) Energy-dispersive X-ray spectroscopy (EDS) elemental composition data.}
        \label{fig:SI_44}
        \vspace{-1em}
    \end{figure}
    \begin{figure}[ht]
        \centering
        \includegraphics[width=0.7\linewidth]{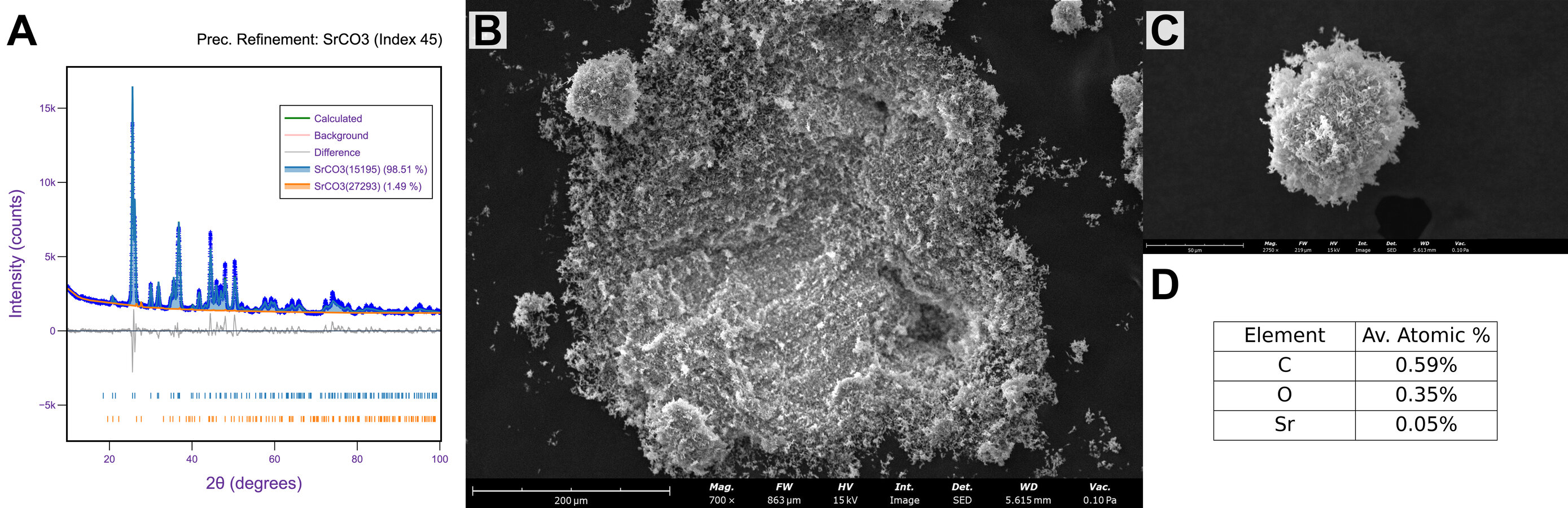}
        \caption{\textbf{Characterization of precursor SrCO$_3$ (Project ID: 45).} (A) Powder X-ray diffraction (XRD) pattern with Rietveld refinement.
        (B) Scanning electron microscopy (SEM) showing representative morphology (scale bar around 200\,$\mu$m).
        (C) Scanning electron microscopy (SEM) showing representative morphology (window scaled to particle size).
        (D) Energy-dispersive X-ray spectroscopy (EDS) elemental composition data.}
        \label{fig:SI_45}
        \vspace{-1em}
    \end{figure}
    \begin{figure}[ht]
        \centering
        \includegraphics[width=0.7\linewidth]{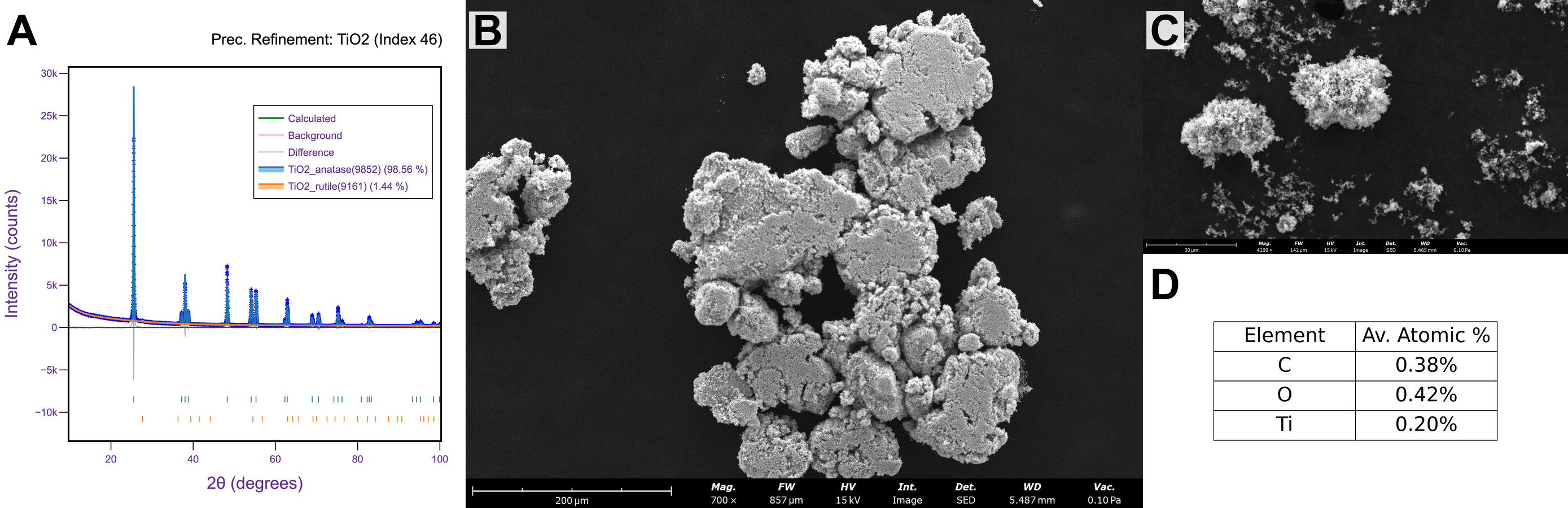}
        \caption{\textbf{Characterization of precursor TiO$_2$ (Project ID: 46).} (A) Powder X-ray diffraction (XRD) pattern with Rietveld refinement.
        (B) Scanning electron microscopy (SEM) showing representative morphology (scale bar around 200\,$\mu$m).
        (C) Scanning electron microscopy (SEM) showing representative morphology (window scaled to particle size).
        (D) Energy-dispersive X-ray spectroscopy (EDS) elemental composition data.}
        \label{fig:SI_46}
        \vspace{-1em}
    \end{figure}
    \begin{figure}[ht]
        \centering
        \includegraphics[width=0.7\linewidth]{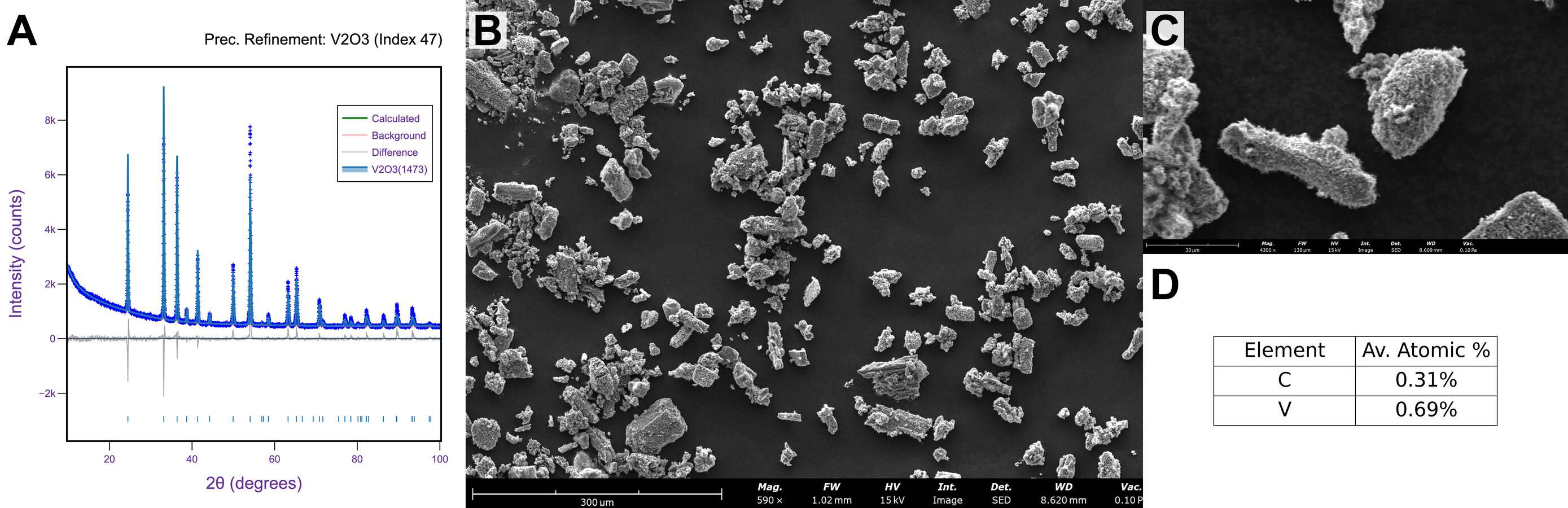}
        \caption{\textbf{Characterization of precursor V$_2$O$_3$ (Project ID: 47).} (A) Powder X-ray diffraction (XRD) pattern with Rietveld refinement.
        (B) Scanning electron microscopy (SEM) showing representative morphology (scale bar around 200\,$\mu$m).
        (C) Scanning electron microscopy (SEM) showing representative morphology (window scaled to particle size).
        (D) Energy-dispersive X-ray spectroscopy (EDS) elemental composition data.}
        \label{fig:SI_47}
        \vspace{-1em}
    \end{figure}
    \begin{figure}[ht]
        \centering
        \includegraphics[width=0.7\linewidth]{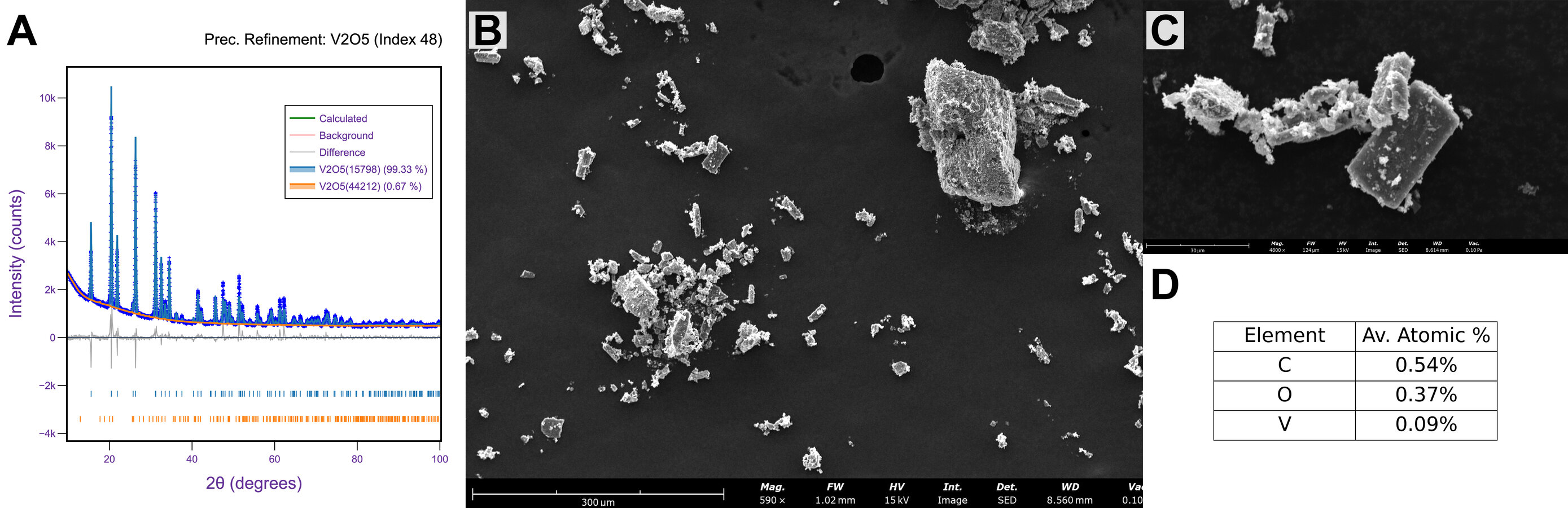}
        \caption{\textbf{Characterization of precursor V$_2$O$_5$ (Project ID: 48).} (A) Powder X-ray diffraction (XRD) pattern with Rietveld refinement.
        (B) Scanning electron microscopy (SEM) showing representative morphology (scale bar around 200\,$\mu$m).
        (C) Scanning electron microscopy (SEM) showing representative morphology (window scaled to particle size).
        (D) Energy-dispersive X-ray spectroscopy (EDS) elemental composition data.}
        \label{fig:SI_48}
        \vspace{-1em}
    \end{figure}
    \begin{figure}[ht]
        \centering
        \includegraphics[width=0.7\linewidth]{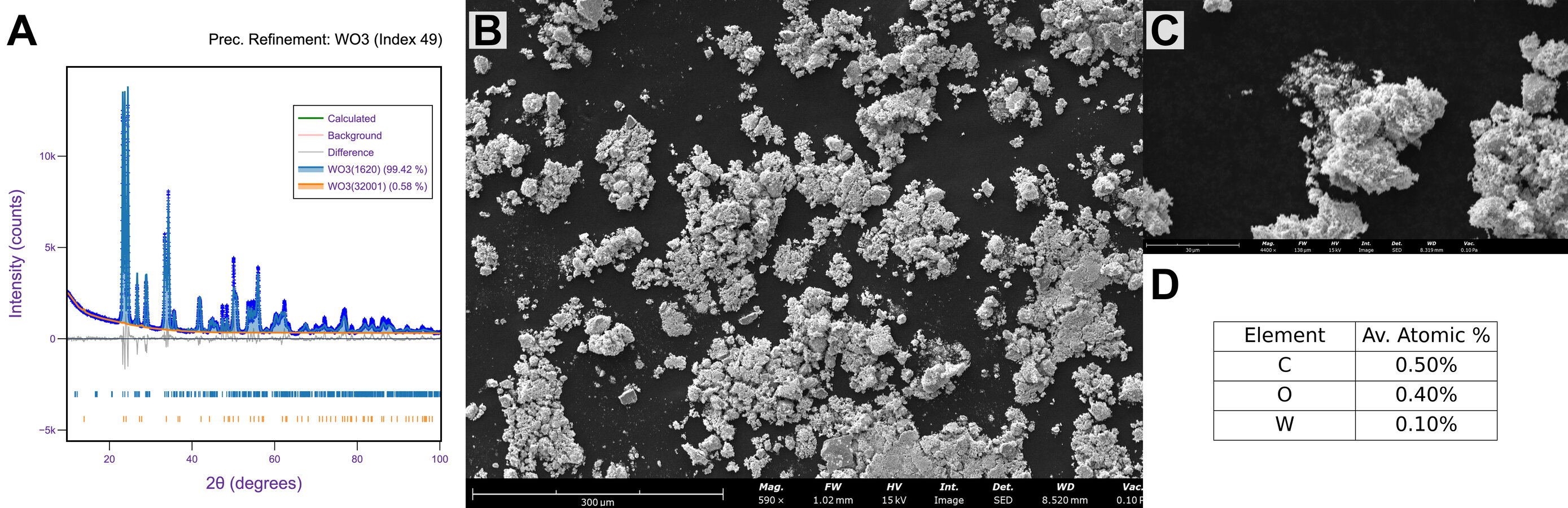}
        \caption{\textbf{Characterization of precursor WO$_3$ (Project ID: 49).} (A) Powder X-ray diffraction (XRD) pattern with Rietveld refinement.
        (B) Scanning electron microscopy (SEM) showing representative morphology (scale bar around 200\,$\mu$m).
        (C) Scanning electron microscopy (SEM) showing representative morphology (window scaled to particle size).
        (D) Energy-dispersive X-ray spectroscopy (EDS) elemental composition data.}
        \label{fig:SI_49}
        \vspace{-1em}
    \end{figure}
    \begin{figure}[ht]
        \centering
        \includegraphics[width=0.7\linewidth]{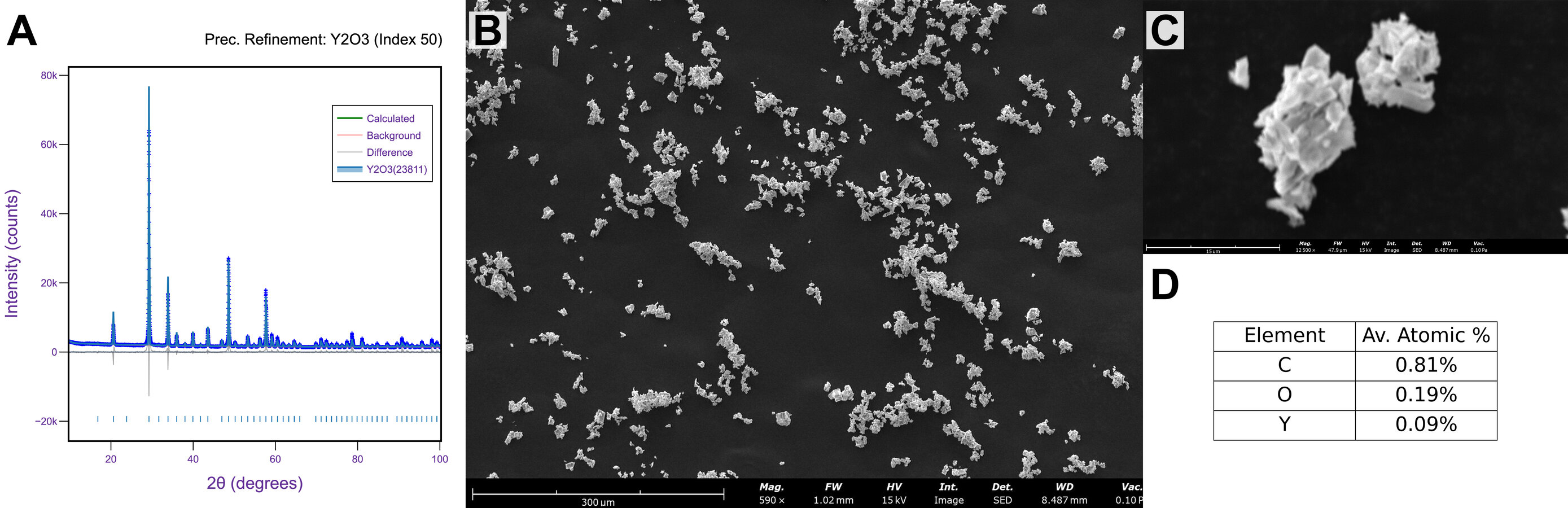}
        \caption{\textbf{Characterization of precursor Y$_2$O$_3$ (Project ID: 50).} (A) Powder X-ray diffraction (XRD) pattern with Rietveld refinement.
        (B) Scanning electron microscopy (SEM) showing representative morphology (scale bar around 200\,$\mu$m).
        (C) Scanning electron microscopy (SEM) showing representative morphology (window scaled to particle size).
        (D) Energy-dispersive X-ray spectroscopy (EDS) elemental composition data.}
        \label{fig:SI_50}
        \vspace{-1em}
    \end{figure}
    \begin{figure}[ht]
        \centering
        \includegraphics[width=0.7\linewidth]{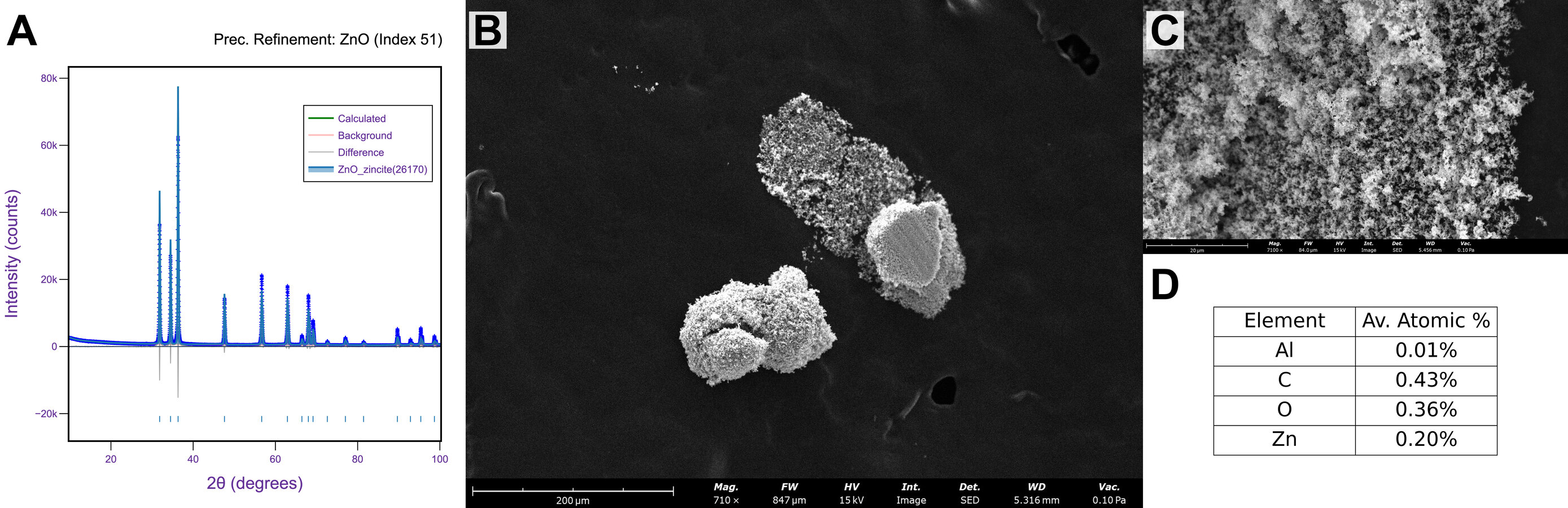}
        \caption{\textbf{Characterization of precursor ZnO (Project ID: 51).} (A) Powder X-ray diffraction (XRD) pattern with Rietveld refinement.
        (B) Scanning electron microscopy (SEM) showing representative morphology (scale bar around 200\,$\mu$m).
        (C) Scanning electron microscopy (SEM) showing representative morphology (window scaled to particle size).
        (D) Energy-dispersive X-ray spectroscopy (EDS) elemental composition data.}
        \label{fig:SI_51}
        \vspace{-1em}
    \end{figure}

\clearpage
    \begin{figure}[ht]
        \centering
        \includegraphics[width=0.7\linewidth]{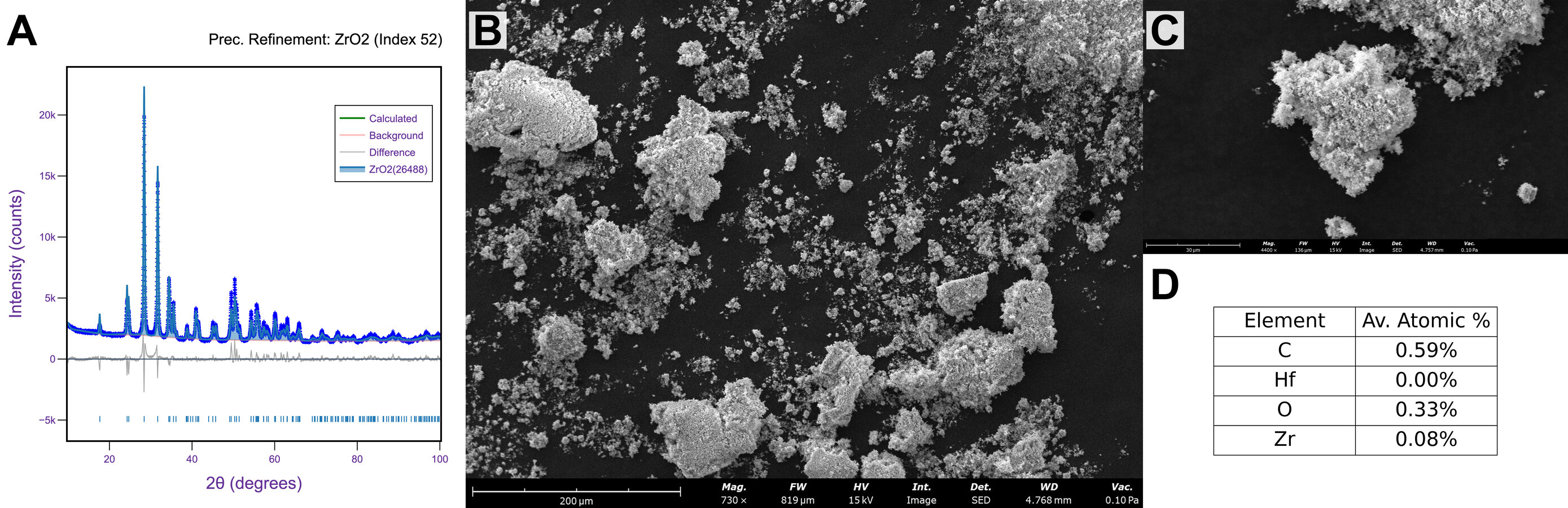}
        \caption{\textbf{Characterization of precursor ZrO$_2$ (Project ID: 52).} (A) Powder X-ray diffraction (XRD) pattern with Rietveld refinement.
        (B) Scanning electron microscopy (SEM) showing representative morphology (scale bar around 200\,$\mu$m).
        (C) Scanning electron microscopy (SEM) showing representative morphology (window scaled to particle size).
        (D) Energy-dispersive X-ray spectroscopy (EDS) elemental composition data.}
        \label{fig:SI_52}
        \vspace{-1em}
    \end{figure}

\clearpage

\subsubsection*{Section S3. Reaction Outcome Analysis}

\begin{figure}[!htbp]
    \centering
    \includegraphics[width=0.5\linewidth]{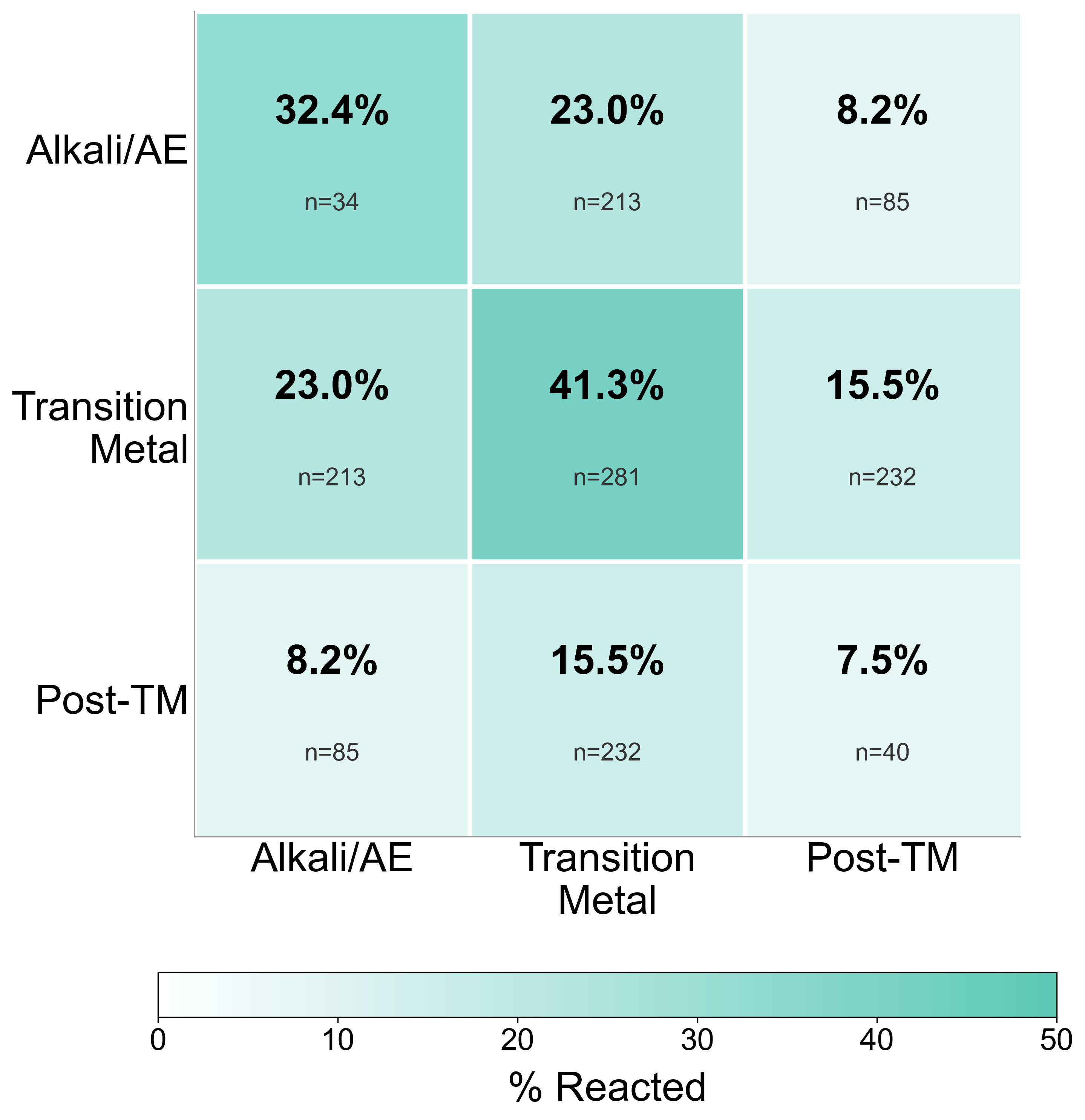}
    \caption{\textbf{Heatmap of the percentage of precursor pairs that reacted, categorized by cation metal type (alkali/alkaline earth, transition metal, or post-transition metal).}}
    \label{fig:heatmap-element-type-only}
\end{figure}

\clearpage 
\subsubsection*{S3.1. Single Feature Analysis}

\begin{figure}[!htbp]
    \centering
    \includegraphics[width=1.0\linewidth]{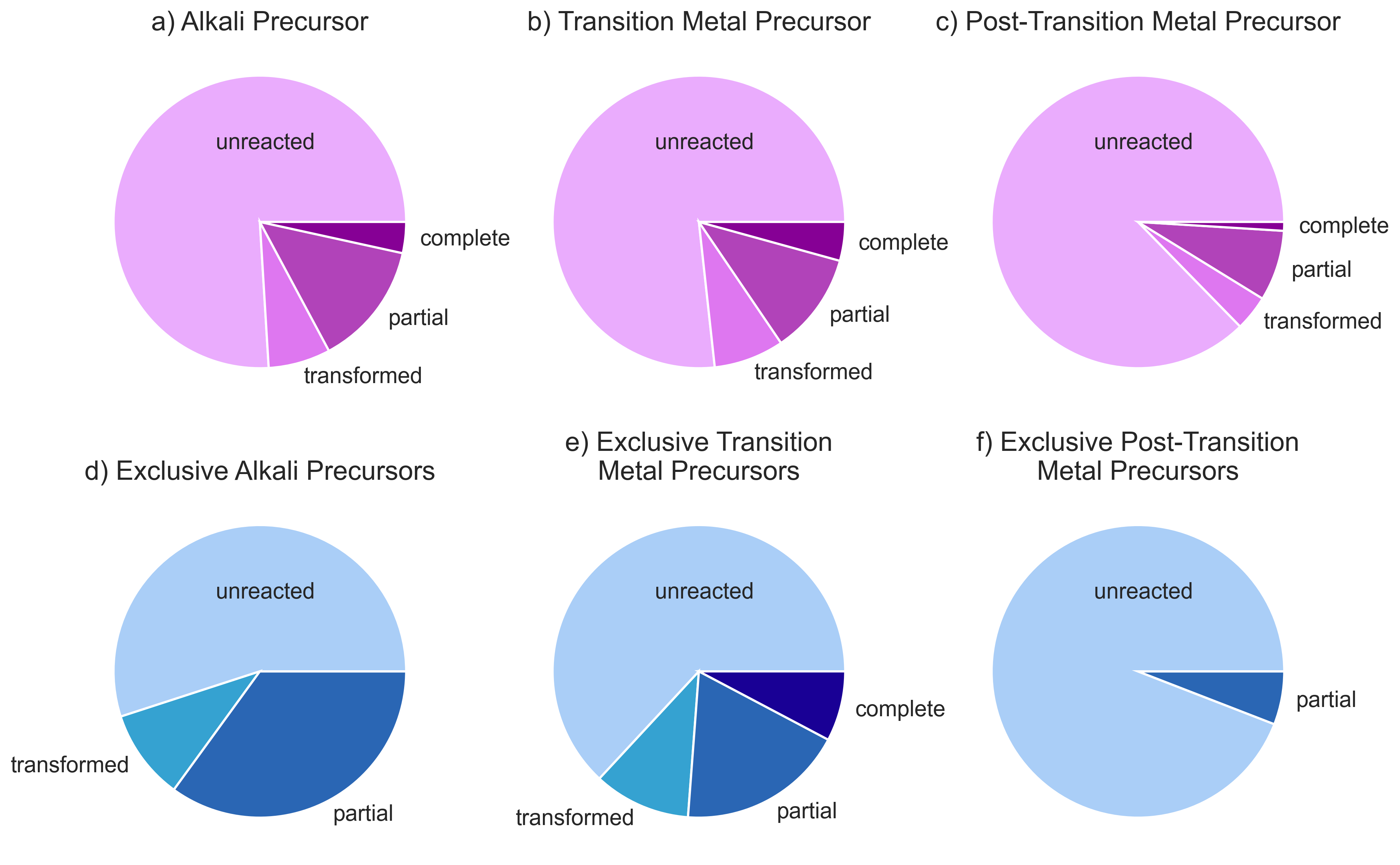}
    \caption{\textbf{Pie charts showing reaction outcome (complete reaction, partial reaction, transformed precursors, or unreacted precursors) recovered.}
    In the top row, reactions in each pie chart are grouped by the presence of having a single precursor of (\textbf{a}) an alkali or alkaline earth metal, (\textbf{b}) a transition metal, or (\textbf{c}) a post-transition metal.
    In the bottom row, reactions in each pie chart are grouped by the presence of having both precursors that are either (\textbf{d}) an alkali or alkaline earth metal, (\textbf{e}) a transition metal, or (\textbf{f}) a post-transition metal.}
    \label{fig:prec-chem-pie}
\end{figure}

\begin{figure}[!htbp]
    \centering
    \includegraphics[width=1.0\linewidth]{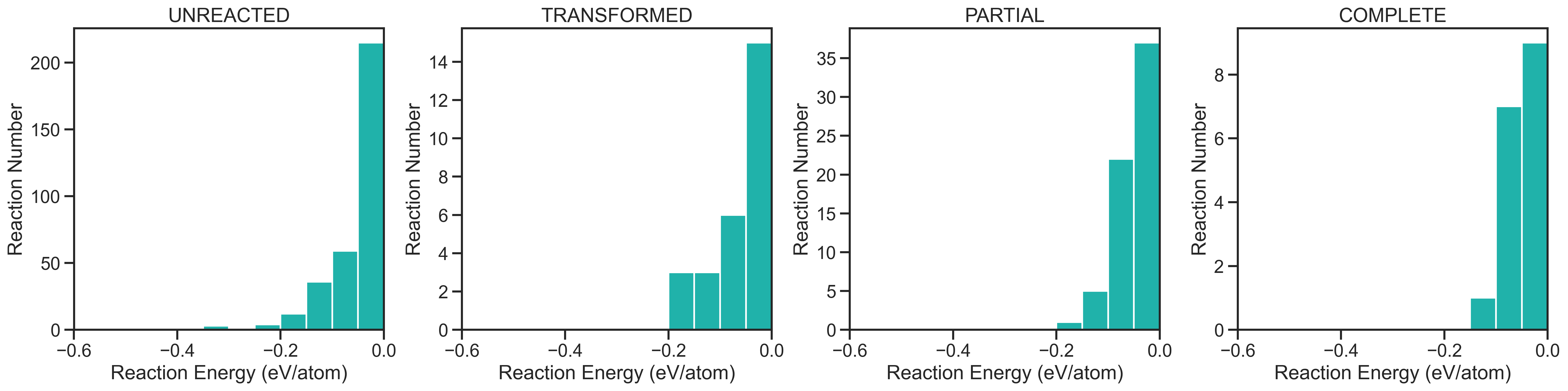}
    \caption{\textbf{Distribution of reaction energies across different synthesis outcomes.} Histograms (left to right) illustrate the frequency of reactions categorized as unreacted precursors, transformed precursors, partial reactions, and complete reactions.}
    \label{fig:hist-rxn-e}
\end{figure}

\begin{figure}[!htbp]
    \centering
    \includegraphics[width=1.0\linewidth]{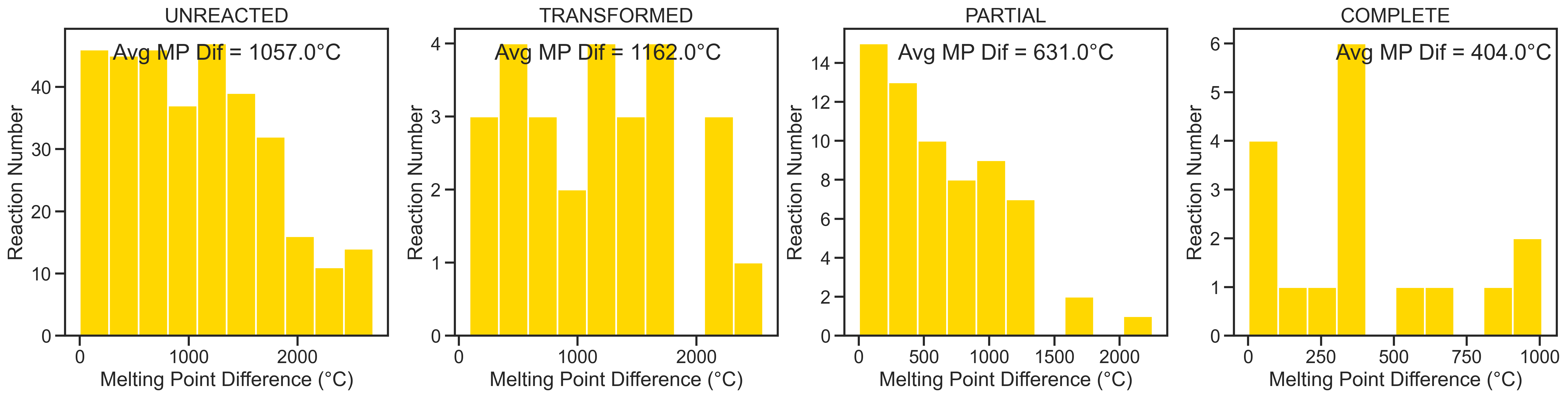}
    \caption{\textbf{Distribution of melting point difference between precursor pairs across different synthesis outcomes.} Histograms (left to right) illustrate the frequency of reactions categorized as unreacted precursors, transformed precursors, partial reactions, and complete reactions.}
    \label{fig:hist-mp-dif}
\end{figure}

\begin{figure}[!htbp]
    \centering
    \includegraphics[width=1.0\linewidth]{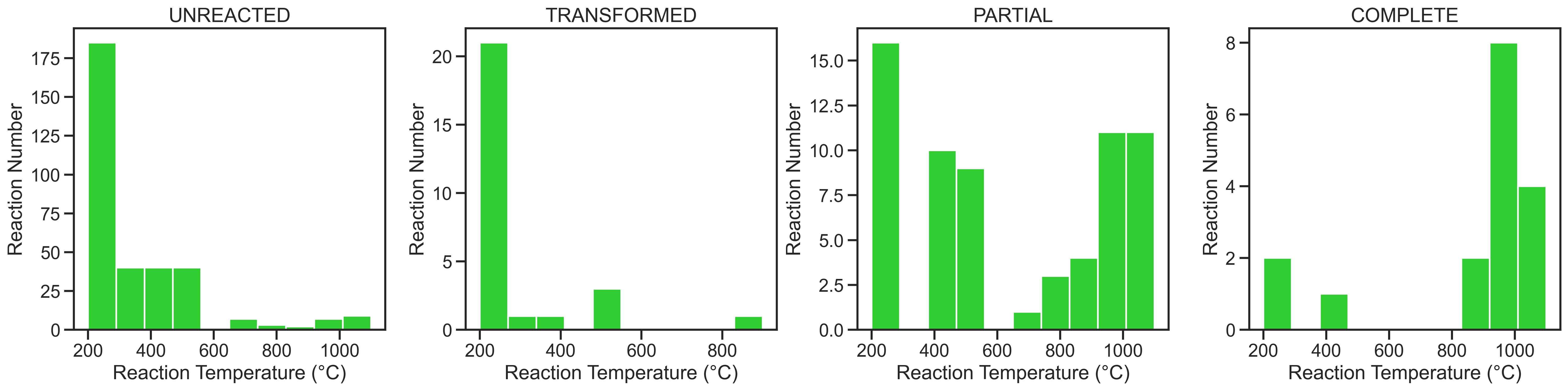}
    \caption{\textbf{Distribution of reaction temperature utilized in this study across different synthesis outcomes.} Histograms (left to right) illustrate the frequency of reactions categorized as unreacted precursors, transformed precursors, partial reactions, and complete reactions.}
    \label{fig:hist-rxn-temp}
\end{figure}

\begin{table}[!htbp]
\linespread{1}\selectfont 
\centering
\tiny
\caption{\textbf{Summary of reaction outcomes categorized by precursor, including raw counts for unreacted, transformed, partially reacted, and completely reacted states.} Reactions that failed physically (e.g. melted, vitrified, or fused to the crucible) could not be assigned a reaction category and are tabulated separately as errored; \% reacted is computed over the categorized reactions only, i.e. (partially + completely reacted) / (total $-$ errored).}
\label{tab:precursor-reaction-outcomes-table}
\begin{tabular}{lccccccc}
\hline \hline
Precursor & Unreacted & Transformed & Partially & Completely & Errored & Total & \% Reacted \\
 & & & Reacted & Reacted & (Phys. Incomp.) & & \\
\hline
V$_2$O$_3$ & 6 & 14 & 6 & 15 & 4 & 45 & 51.2 \\
ZnO & 15 & 7 & 8 & 14 & 1 & 45 & 50.0 \\
SrCO$_3$ & 12 & 11 & 11 & 11 & 0 & 45 & 48.9 \\
WO$_3$ & 18 & 5 & 12 & 9 & 1 & 45 & 47.7 \\
Mn$_3$O$_4$ & 10 & 14 & 9 & 12 & 0 & 45 & 46.7 \\
Nb$_2$O$_5$ & 14 & 10 & 12 & 9 & 0 & 45 & 46.7 \\
Cr$_2$O$_3$ & 15 & 9 & 6 & 14 & 1 & 45 & 45.5 \\
Fe$_3$O$_4$ & 6 & 20 & 7 & 12 & 0 & 45 & 42.2 \\
MnO & 7 & 19 & 7 & 12 & 0 & 45 & 42.2 \\
CoO & 5 & 21 & 2 & 16 & 1 & 45 & 40.9 \\
K$_2$CO$_3$ & 1 & 25 & 7 & 8 & 4 & 45 & 36.6 \\
TiO$_2$ & 13 & 14 & 2 & 13 & 3 & 45 & 35.7 \\
MgO & 16 & 13 & 4 & 12 & 0 & 45 & 35.6 \\
Fe$_2$O$_3$ & 17 & 11 & 7 & 8 & 2 & 45 & 34.9 \\
Na$_2$CO$_3$ & 18 & 10 & 7 & 8 & 2 & 45 & 34.9 \\
(NH$_4$)$_2$HPO$_4$ & 0 & 26 & 6 & 7 & 6 & 45 & 33.3 \\
Li$_2$CO$_3$ & 15 & 15 & 8 & 7 & 0 & 45 & 33.3 \\
CuO & 21 & 10 & 7 & 7 & 0 & 45 & 31.1 \\
V$_2$O$_5$ & 23 & 8 & 7 & 7 & 0 & 45 & 31.1 \\
Y$_2$O$_3$ & 21 & 10 & 10 & 4 & 0 & 45 & 31.1 \\
NiO & 17 & 14 & 3 & 10 & 1 & 45 & 29.5 \\
MoO$_3$ & 20 & 11 & 9 & 4 & 1 & 45 & 29.5 \\
PbO & 6 & 26 & 8 & 5 & 0 & 45 & 28.9 \\
Bi$_2$O$_3$ & 21 & 12 & 7 & 5 & 0 & 45 & 26.7 \\
NH$_4$H$_2$PO$_4$ & 6 & 21 & 2 & 6 & 10 & 45 & 22.9 \\
Ag$_2$O & 1 & 33 & 5 & 5 & 1 & 45 & 22.7 \\
Ga$_2$O$_3$ & 26 & 9 & 6 & 4 & 0 & 45 & 22.2 \\
LiOH & 0 & 36 & 4 & 4 & 1 & 45 & 18.2 \\
Mn$_2$O$_3$ & 20 & 17 & 6 & 2 & 0 & 45 & 17.8 \\
In$_2$O$_3$ & 23 & 15 & 3 & 4 & 0 & 45 & 15.6 \\
KH$_2$PO$_4$ & 26 & 10 & 4 & 2 & 3 & 45 & 14.3 \\
SiO$_2$ & 22 & 15 & 1 & 5 & 2 & 45 & 14.0 \\
SnO$_2$ & 13 & 24 & 1 & 5 & 2 & 45 & 14.0 \\
Co$_3$O$_4$ & 7 & 31 & 1 & 5 & 1 & 45 & 13.6 \\
GeO$_2$ & 13 & 26 & 4 & 2 & 0 & 45 & 13.3 \\
CaCO$_3$ & 27 & 13 & 3 & 2 & 0 & 45 & 11.1 \\
FeC$_2$O$_4$ & 1 & 39 & 3 & 2 & 0 & 45 & 11.1 \\
ZrO$_2$ & 24 & 16 & 2 & 1 & 2 & 45 & 7.0 \\
Sb$_2$O$_3$ & 25 & 17 & 2 & 1 & 0 & 45 & 6.7 \\
BaCO$_3$ & 27 & 15 & 3 & 0 & 0 & 45 & 6.7 \\
B(OH)$_3$ & 33 & 8 & 2 & 0 & 2 & 45 & 4.7 \\
HfO$_2$ & 26 & 16 & 0 & 2 & 1 & 45 & 4.5 \\
La(OH)$_3$ & 31 & 12 & 0 & 2 & 0 & 45 & 4.4 \\
MnO$_2$ & 27 & 16 & 0 & 2 & 0 & 45 & 4.4 \\
BaO$_2$ & 7 & 36 & 1 & 1 & 0 & 45 & 4.4 \\
Al(OH)$_3$ & 36 & 8 & 1 & 0 & 0 & 45 & 2.2 \\
\hline \hline
\end{tabular}
\end{table}

\clearpage 
\subsubsection*{Section S4. Samples Incompatible with Automated Sample Handling}

The samples listed in Table~\ref{tab:errored-samples} were identified as mechanically incompatible with the automated recovery and characterization system employed by A-Lab. Instead of powder, manual inspection revealed that these samples were primarily viscous liquids, very hard solids, or had clay-like textures. Consequently, powder X-ray diffraction (XRD) characterization was omitted for these samples.

\begin{table}[!htbp]
\linespread{1}\selectfont 
\caption{\label{tab:errored-samples} \textbf{Samples that were executed in A-Lab, but reaction outcome led to mechanical incompatibility and prevented automated sample handling.} The observations associated with inability to recover and characterize samples automatically is given in the right column.}
\begin{tabular}{lcc}
\hline \hline
\textbf{Precursors} &\textbf{Physical State Observation} \\
\hline
Ag$_2$O + KH$_2$PO$_4$	& liquid, tough \\ 
Co$_3$O$_4$ + KH$_2$PO$_4$	& tough solid \\ 
 NiO  + NH$_4$H$_2$PO$_4$ & very sticky solid \\ 
  Fe$_2$O$_3$   +  K$_2$CO$_3$(H$_2$O)$_{1.5}$     &  sticky, hard liquid      \\ 
    HfO$_2$   +  NH$_4$H$_2$PO$_4$     &  clay-like, liquid     \\ 
    NH$_4$H$_2$PO$_4$   +  SnO$_2$     &  clay-like, liquid     \\ 
(NH$_4$)$_2$HPO$_4$   +  SnO$_2$     &  clay-like, liquid     \\ 
    NH$_4$H$_2$PO$_4$   +  WO$_3$     &    clay-like, liquid   \\ 
    (NH$_4$)$_2$HPO$_4$   +  TiO$_2$     &   clay-like, liquid     \\ 
    NH$_4$H$_2$PO$_4$  +  ZrO$_2$     &   clay-like with powder    \\ 
    V$_2$O$_3$   +  ZnO     &   hard solid, very little powder    \\ 
    TiO$_2$   +  V$_2$O$_3$     &   hard solid, very little powder    \\ 
    NH$_4$H$_2$PO$_4$   +  TiO$_2$     &   clay-like, liquid, sticky    \\ 
    NH$_4$H$_2$PO$_4$   +  (NH$_4$)$_2$HPO$_4$     &    viscous liquid   \\ 
    V$_2$O$_3$   +  ZrO$_2$     &  hard solid, very little powder    \\ 
    Fe$_2$O$_3$   +  V$_2$O$_3$     &    hard solid, very little powder, significant reaction with crucible   \\ 
    K$_2$CO$_3$(H$_2$O)$_{1.5}$   +  KH$_2$PO$_4$     &    liquid   \\ 
    K$_2$CO$_3$(H$_2$O)$_{1.5}$   +  NH$_4$H$_2$PO$_4$     &    liquid, volatilized   \\ 
    K$_2$CO$_3$(H$_2$O)$_{1.5}$   +  Na$_2$CO$_3$     &   volatilized, limited number of crystals    \\ 
    B(OH)$_3$   +  NH$_4$H$_2$PO$_4$     &   viscous liquid or soft clay    \\ 
    B(OH)$_3$   +  (NH$_4$)$_2$HPO$_4$     &   clay    \\ 
    MoO$_3$   +  NH$_4$H$_2$PO$_4$     &    clay   \\ 
    LiOH(H$_2$O)   +  (NH$_4$)$_2$HPO$_4$     &    clay   \\ 
    Cr$_2$O$_3$   +  (NH$_4$)$_2$HPO$_4$     &   clay    \\ 
    Na$_2$CO$_3$   +  SiO$_2$     &   clay    \\ 
    CoO   +  SiO$_2$     &   hard solid, very little powder    \\ 

\hline \hline
\end{tabular}
\end{table}

\clearpage 
\subsubsection*{Section S5. Machine Learning Model Output}

\subsubsection*{S5.1. Additional Analysis to Understand Feature Inclusion}

\begin{figure}[!htbp]
    \centering
    \includegraphics[width=0.5\linewidth]{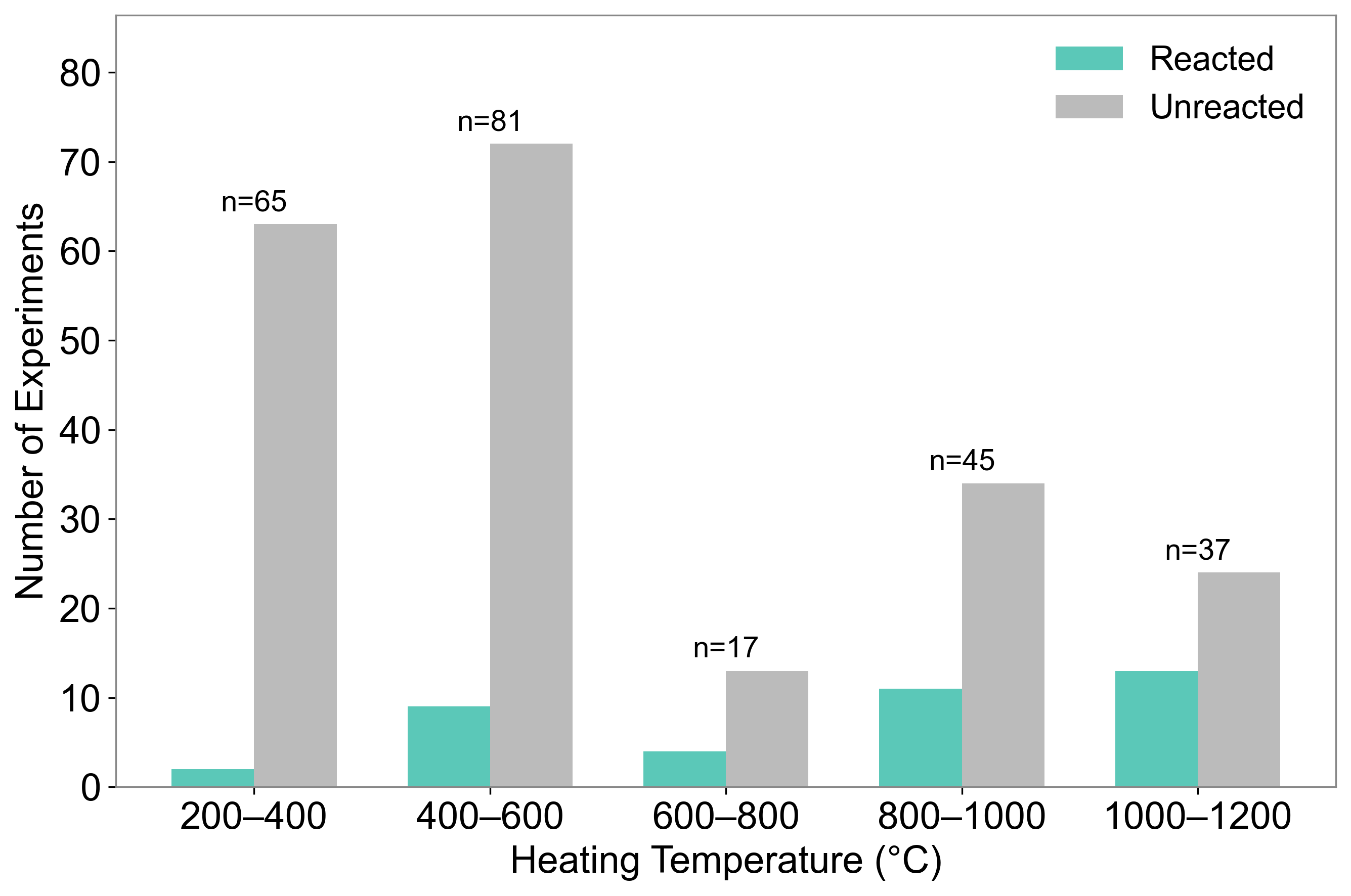}
    \caption{\textbf{Reaction outcomes across temperature bins for experiments predicted to be thermodynamically unfavorable ($\Delta G_{rxn} \geq 0$).} Teal bars indicate reacted samples, while gray bars indicate unreacted samples. The total number of experiments in each bin is labeled above the corresponding bar.
    }
    \label{fig:supp_positive_energy_bins}
\end{figure}

\begin{figure}[!htbp]
    \centering
    \includegraphics[width=1.0\linewidth]{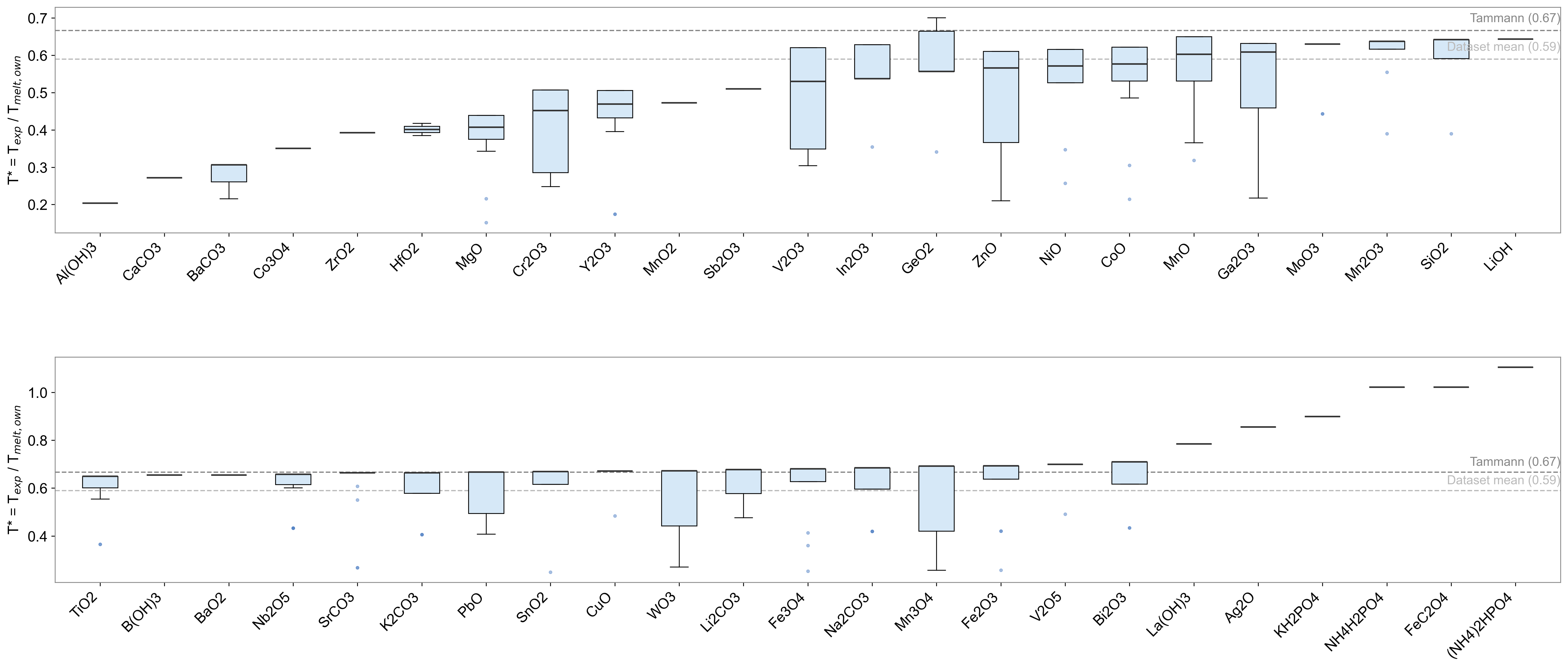}
    \caption{\textbf{Box-and-whisker plots of recorded dwell temperatures for reacted experiments, categorized by precursor.} Temperatures are normalized by the reference temperature ($T_{ref}$) of each respective precursor. Precursors are ordered along the x-axis by ascending mean normalized dwell temperature.}
    \label{fig:supp_precursor_onset_boxwhisker}
\end{figure}

\clearpage 
\subsubsection*{S5.2. Additional Models}

\begin{figure}[!htbp]
    \centering
    \includegraphics[width=0.5\linewidth]{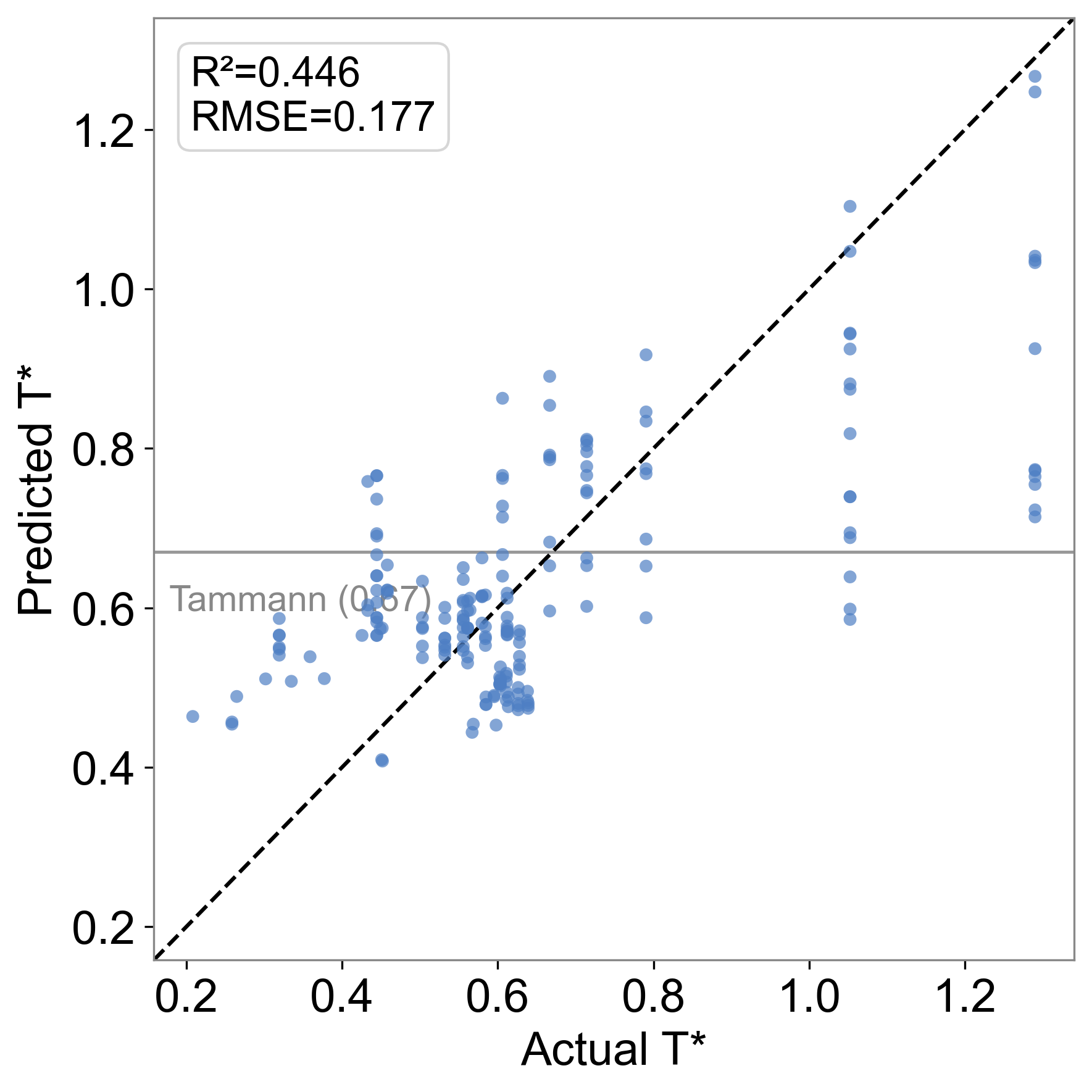}
    \caption{\textbf{Linear regression model for $T^*$ fit using the minimum melting point of the precursor pairs.} The plot compares experimentally recorded homologous ($T_{ref}$-normalized) $T^*$ against predicted $T^*$. The experimental Tammann rule ($T^* = 0.67$) is indicated by a gray horizontal line. Model performance is characterized by $R^2 = 0.136$ and $RMSE = 0.142$.}
    \label{fig:supp_meltingpoints_only}
\end{figure}

\begin{figure}[!htbp]
    \centering
    \includegraphics[width=0.5\linewidth]{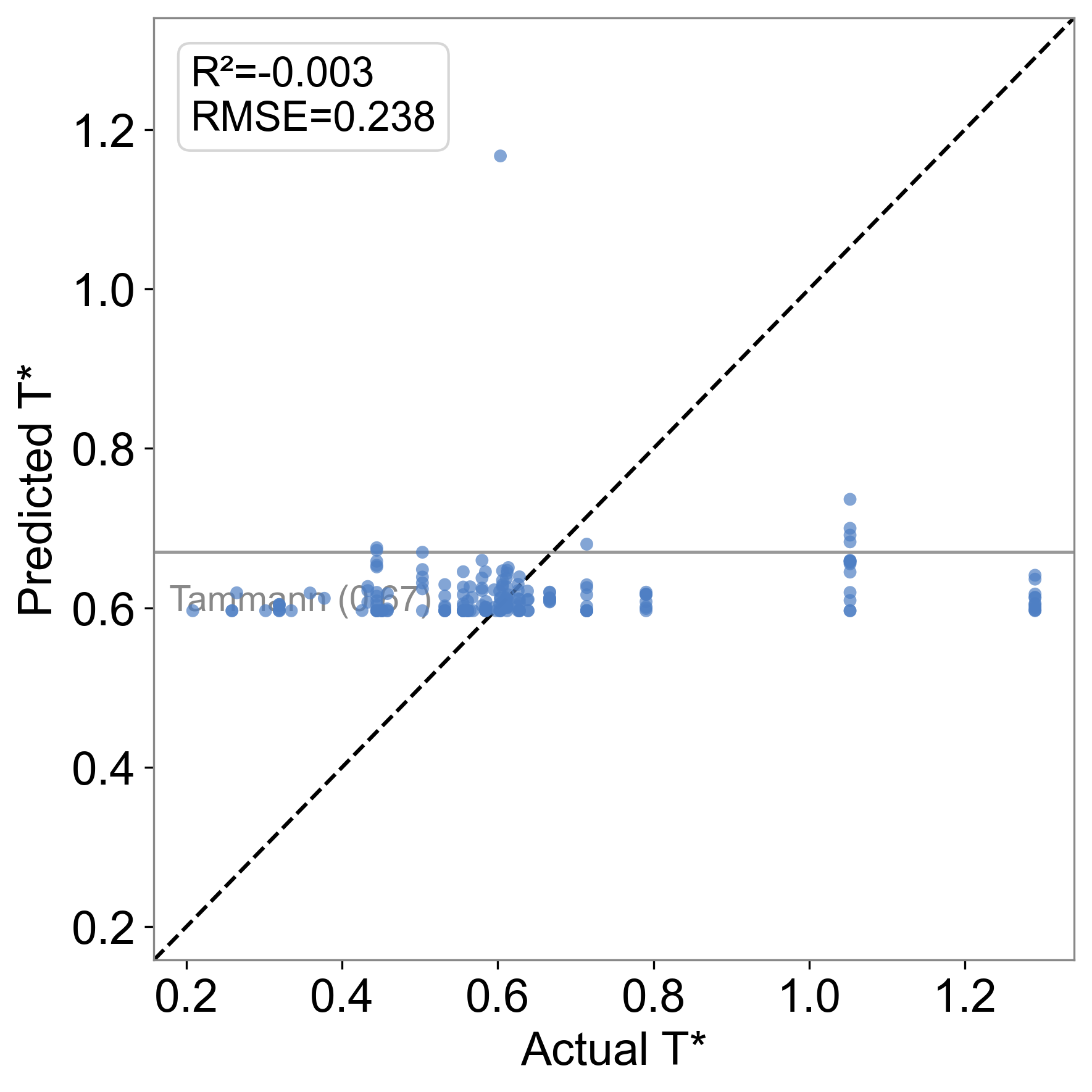}
    \caption{\textbf{Linear regression model for $T^*$ fit using the reaction energy ($G_{rxn}$) of the precursor pairs.} The plot compares experimentally recorded homologous ($T_m$-normalized) $T^*$ against predicted $T^*$. The experimental Tammann rule ($T^* = 0.67$) is indicated by a gray horizontal line. Model performance is characterized by $R^2 = -0.001$ and $RMSE = 0.153$.}
    \label{fig:supp_energy_only}
\end{figure}

\begin{figure}[!htbp]
    \centering
    \includegraphics[width=0.5\linewidth]{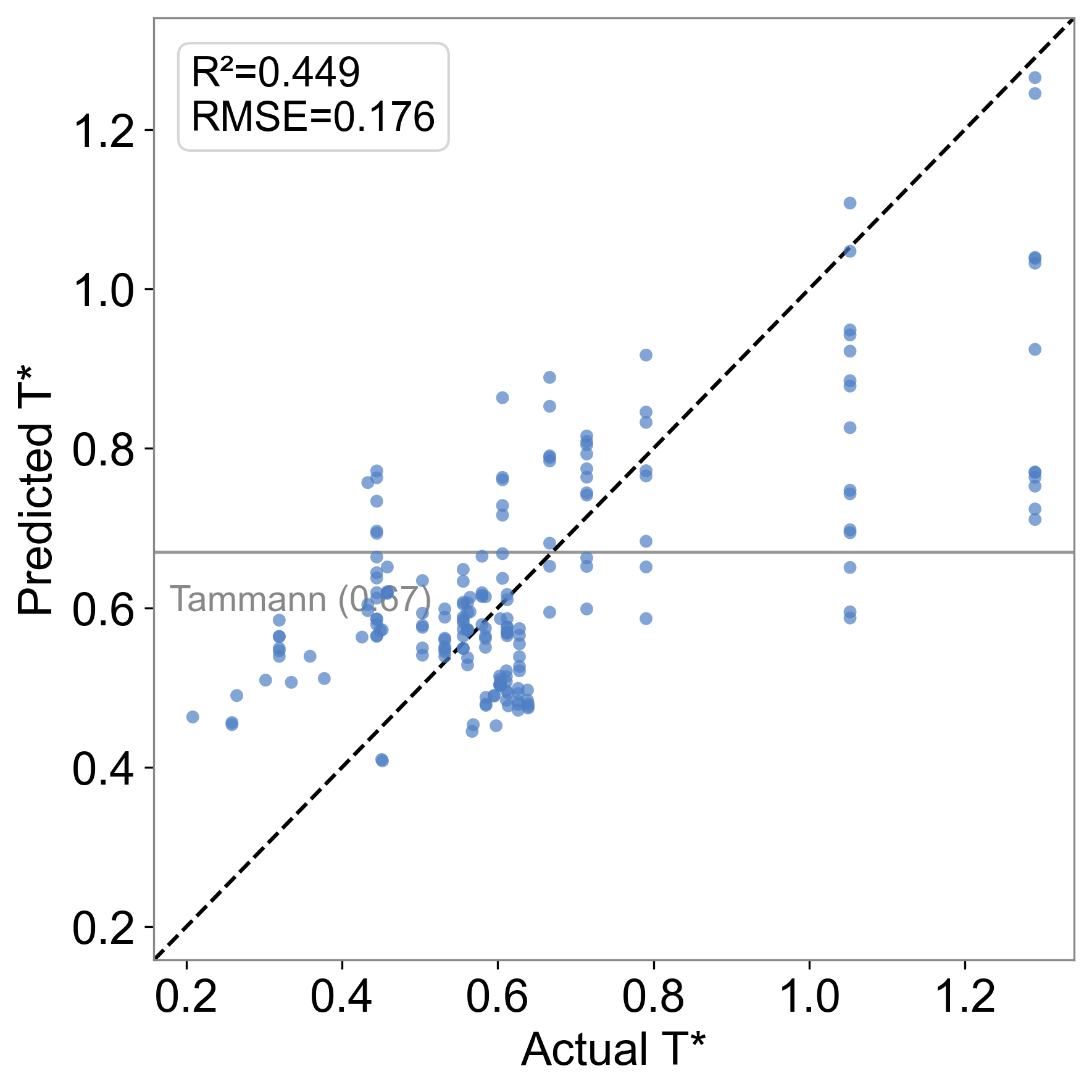}
    \caption{\textbf{Linear regression model for $T^*$ fit using the minimum melting point of the precursor pairs and the reaction energy ($G_{rxn}$) of the precursor pairs.} The plot compares experimentally recorded homologous ($T_m$-normalized) $T^*$ against predicted $T^*$. The experimental Tammann rule ($T^* = 0.67$) is indicated by a gray horizontal line. Model performance is characterized by $R^2 = 0.130$ and $RMSE = 0.143$.}
    \label{fig:supp_meltingpoints_energy}
\end{figure}

\begin{figure}[!htbp]
    \centering
    \includegraphics[width=0.5\linewidth]{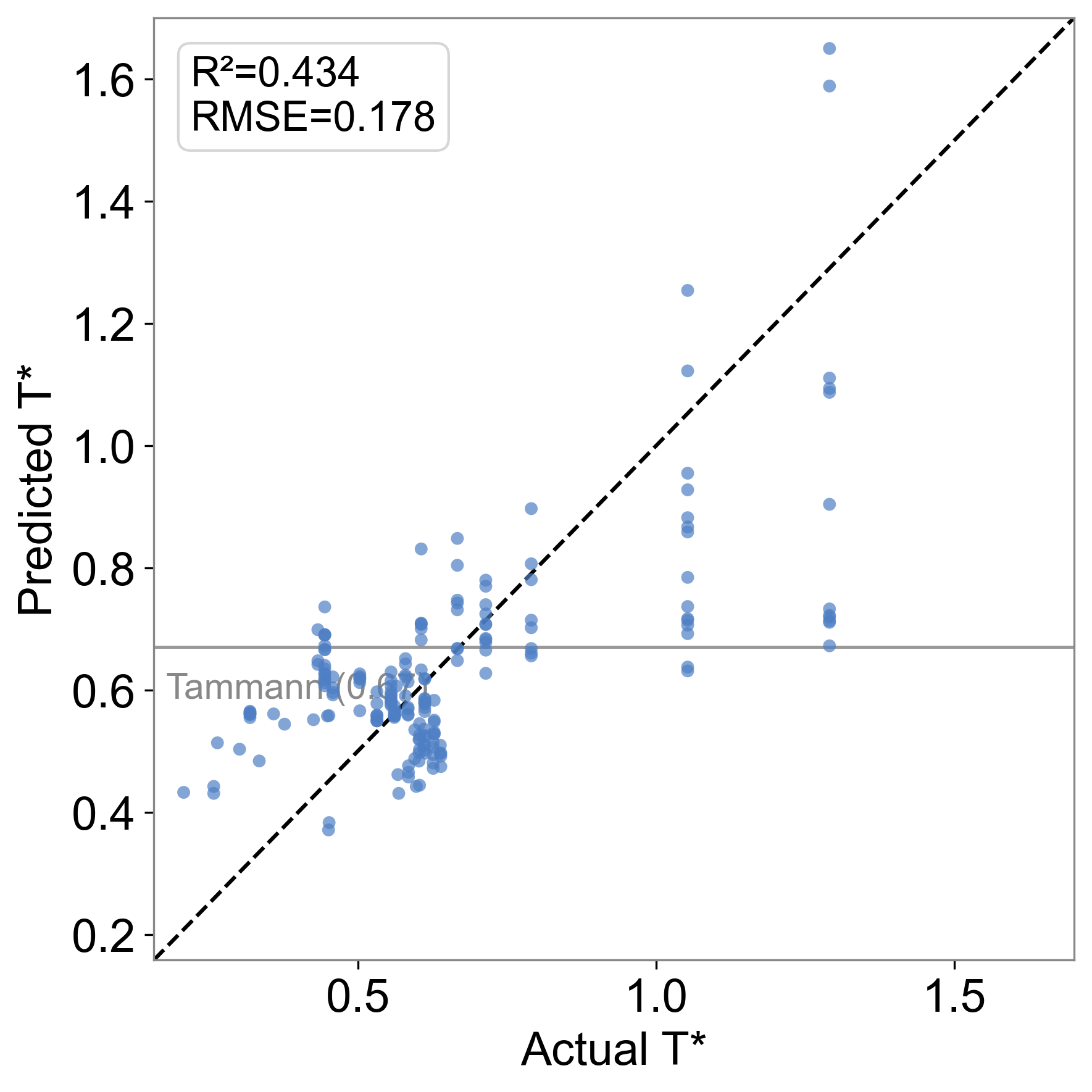}
    \caption{\textbf{Linear regression model for $T^*$ fit using only aggregate features plotted in full linear regression model.} The plot compares experimentally recorded homologous ($T_m$-normalized) $T^*$ against predicted $T^*$. The experimental Tammann rule ($T^* = 0.67$) is indicated by a gray horizontal line. Model performance is characterized by $R^2 = 0.127$ and $RMSE = 0.143$.}
    \label{fig:supp_nonlinear_only}
\end{figure}

\clearpage 
\subsubsection*{S5.3. Additional Training and Models}

\begin{figure}[!htbp]
    \centering
    \includegraphics[width=1.0\linewidth]{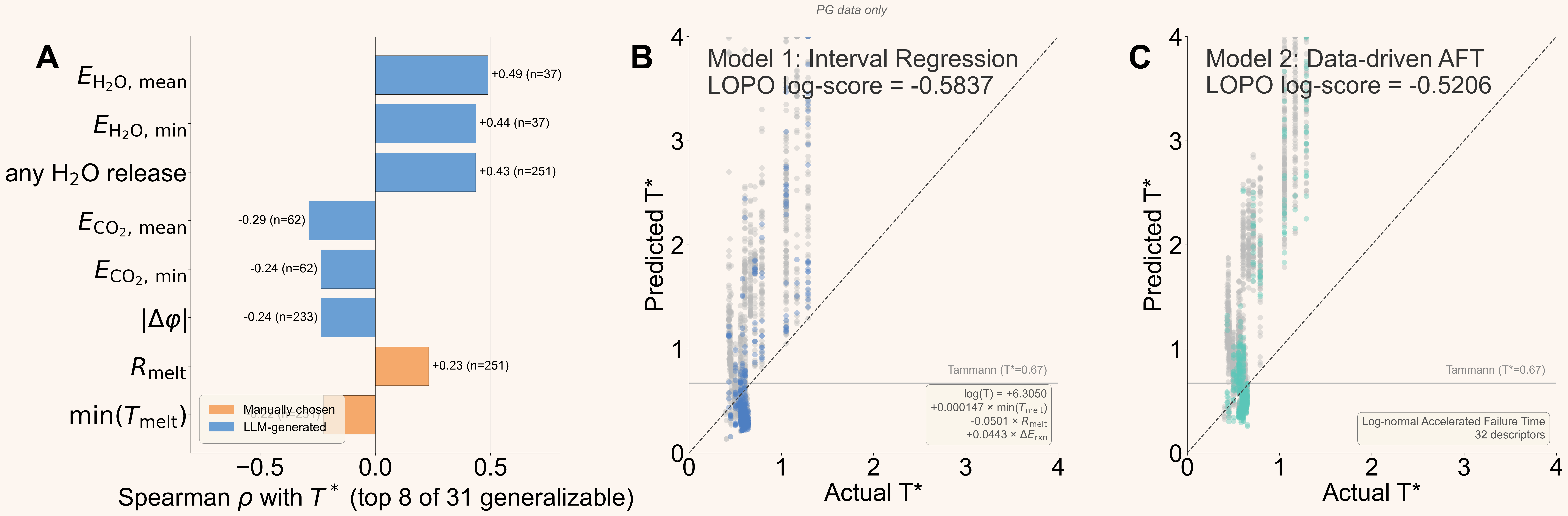}
    \caption{\textbf{Machine learning models to predict reaction onset temperature trained on only data from this work (precursor genome).}
    (\textbf{A}) Spearman correlation coefficients \cite{Spearman1904_CorrelationCoefficients} between physicochemical features and normalized reaction onset temperature (T$^*$= T$_{dwell}$ / T$_{melt,min}$). Melting point properties dominate, alongside LLM generated parameters such as energy of water released ($E_{\mathrm{H_2O}}$).
    (\textbf{B}-\textbf{C}) Predicted versus actual $T^*$ for two machine learning models, with reacted observations in color and unreacted in gray. LOPO log-scores are: (\textbf{B}) Model 1 (3-feature AFT), $-0.492$; (\textbf{C}) accelerated failure time: Model 2 (32-feature AFT), $-0.484$. The dashed line is drawn at 1:1, where reacted, colored points are expected at or below the parity line. The horizontal line marks the Tammann reference ($T^*=0.67$). 
    }
    \label{fig:models_comparison_PGonly}
\end{figure}

\begin{figure}[!htbp]
    \centering
    \includegraphics[width=0.5\linewidth]{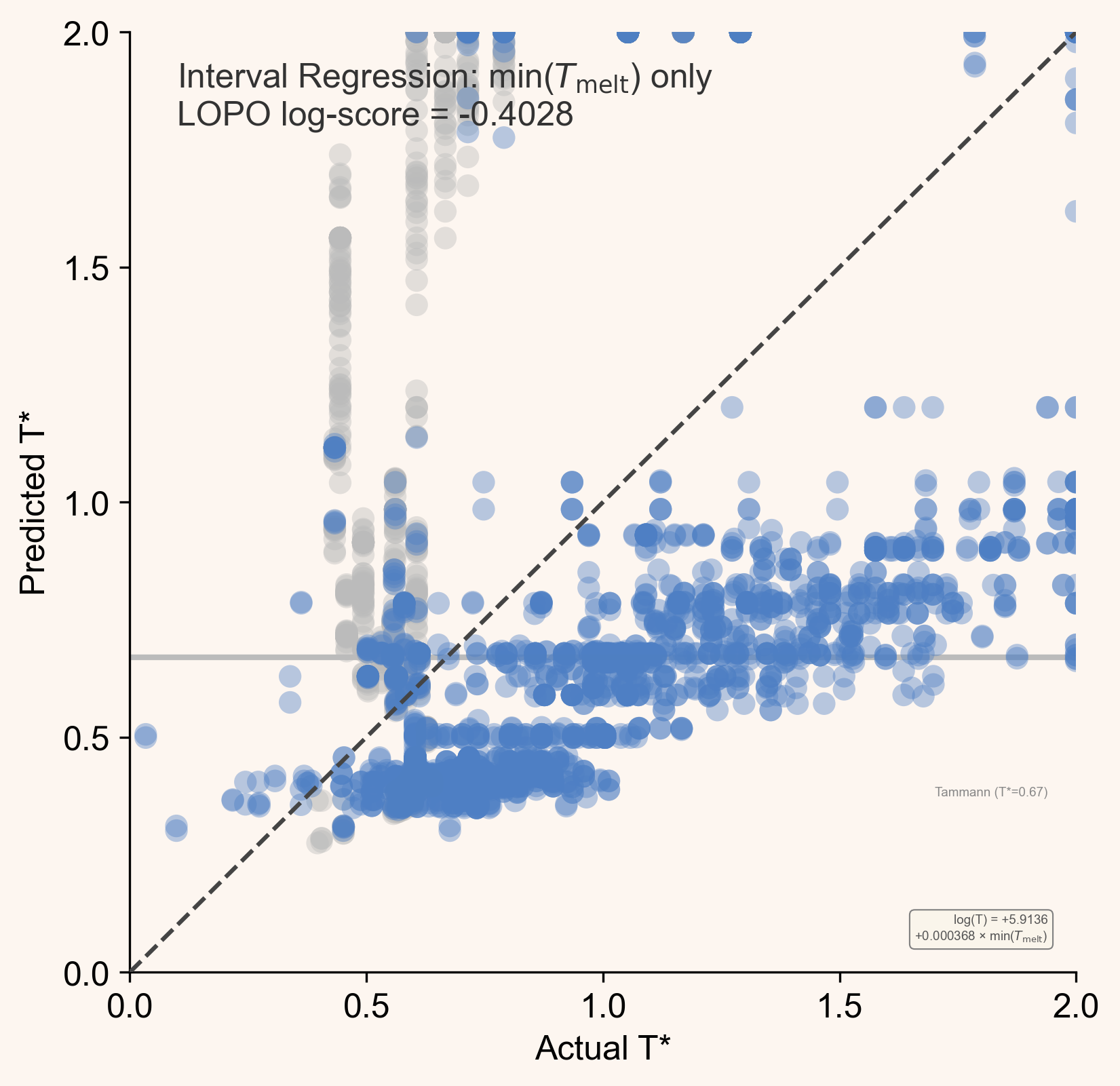}
    \caption{\textbf{Predicted versus actual $T^*$ for an interval regression machine learning model with only minimum melting point as a feature.} The model is trained on precursor genome data and on the data set from Ref. \cite{Kononova2019_OlgaDataSet}, with reacted observations in blue and unreacted in gray. The dashed line is 1:1; the horizontal line marks the Tammann reference ($T^*=0.67$). The LOPO log-score is $-0.4028$.}
    \label{fig:models_meltingpoint_PGonly}
\end{figure}

\begin{figure}[!htbp]
    \centering
    \includegraphics[width=1.0\linewidth]{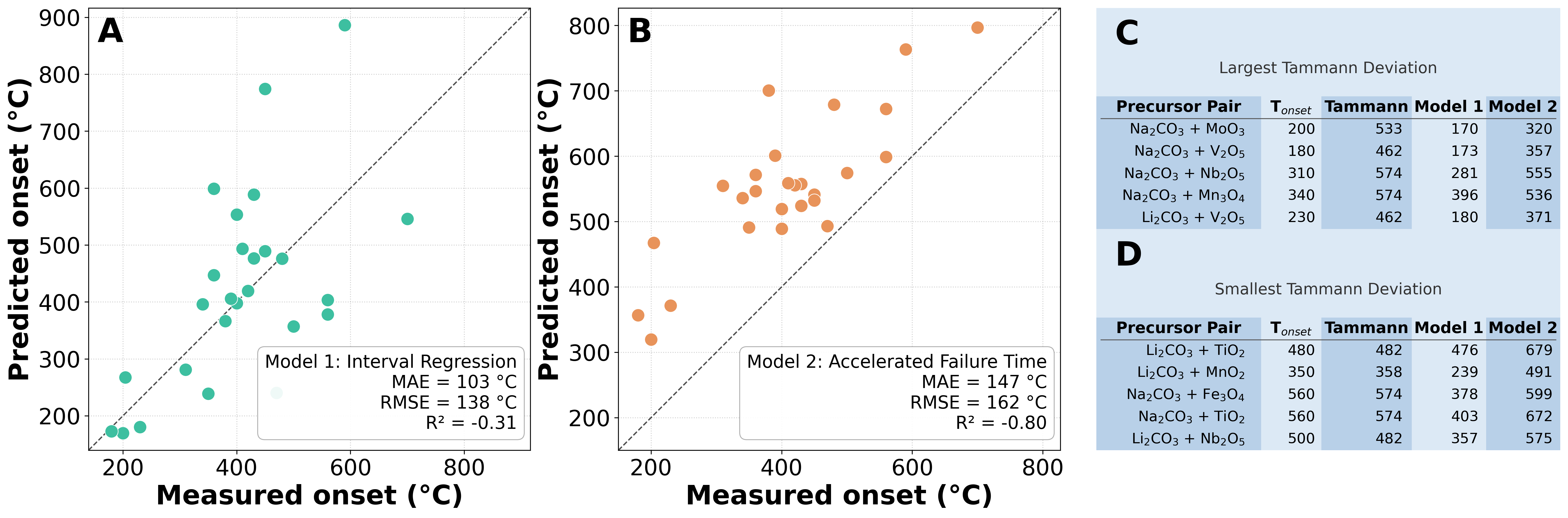}
    \caption{Comparison of predictive models for measured solid-state reaction onset temperature trained only on precursor genome data.
    (\textbf{A}) Measured versus predicted onset temperatures ($T_{onset}$) for the three-feature AFT model (Model 1), with a calculated MAE of 180$^\circ$C, RMSE of 201$^\circ$C, and an $R^2$ of -1.78.
    (\textbf{B}) Measured versus predicted onset temperatures ($T_{onset}$) for the accelerated failure time model (Model 2), with a calculated MAE of 189$^\circ$C, RMSE of 256$^\circ$C, and an $R^2$ of 3.51.
    (\textbf{C}) Tabulated subset of reactions exhibiting the smallest deviation from the Tammann-rule temperature.
    (\textbf{D}) Tabulated subset of reactions exhibiting the largest deviation from the Tammann-rule temperature.
    Tables in C and D provide the precursor pairs, measured experimental onset temperatures, Tammann-rule temperatures, and predicted onset temperatures for both Model 1 and Model 2.}
    \label{fig:insitu_validation_PGonly}
\end{figure}

\clearpage 
\subsubsection*{S5.4. Spearman Correlations}

\begin{table}[!htbp]
\linespread{1}\selectfont 
\centering
\small
\caption{Spearman rank correlation between each of the 32 model descriptors and the binary reaction outcome (reacted${}=1$ vs unreacted${}=0$), across the full PG\,+\,PG-LIT dataset (1{,}890 events; 26 \textit{in situ} pairs excluded). Positive $\rho$ indicates the feature tends to be larger for reacting pairs. $n$ is the number of events with a finite feature value. Sorted by $|\rho|$; the top eight appear in the main text.}
\label{tab:spearman-full-reacted}
\begin{tabular}{lccc}
\hline \hline
\textbf{Descriptor} & \textbf{Source} & \textbf{Spearman $\rho$} & \textbf{$n$} \\
\hline
any H$_2$O release & LLM & -0.431 & 1890 \\
$\min(T_{melt})$ & manual & +0.416 & 1890 \\
$R_{melt}=\max/\min(T_{melt})$ & manual & -0.390 & 1890 \\
AE$\,X$ dense 3D & LLM & +0.229 & 1890 \\
$|\Delta\varphi|$ (ionic potential) & LLM & +0.212 & 1804 \\
$T_{decomp}$ (precursor 2) & LLM & -0.183 & 1261 \\
$T_{decomp}$ (precursor 1) & LLM & +0.180 & 1020 \\
log residual oxide $T_{melt}$ & LLM & +0.166 & 1890 \\
acid-base strength product & LLM & +0.162 & 1890 \\
alkali$\,X$ max $E_{M-O}$ & LLM & -0.154 & 1889 \\
$E_{O,mean}$ & LLM & +0.134 & 1890 \\
$E_{O,min}$ & LLM & +0.123 & 1890 \\
any CO$_2$ release & LLM & +0.114 & 1890 \\
$|\Delta CN|$ (coordination) & LLM & -0.111 & 1890 \\
is dense 3D & LLM & +0.107 & 1890 \\
$\Delta E_{rxn}$ & manual & -0.100 & 1890 \\
alkali$\,X$ dense 3D & LLM & -0.074 & 1890 \\
is tunnel & LLM & +0.060 & 1890 \\
$E_{M-O,min}$ & LLM & -0.057 & 1889 \\
any precursor decomposes $<600^\circ$C & LLM & -0.042 & 1890 \\
$E_{H_2O,mean}$ & LLM & -0.040 & 326 \\
$E_{CO_2,min}$ & LLM & +0.037 & 643 \\
strong acid-base pair & LLM & +0.036 & 1890 \\
alkali$\,X$ tunnel & LLM & -0.033 & 1890 \\
alkali$\,X$ layered & LLM & +0.033 & 1890 \\
is open 3D & LLM & -0.031 & 1890 \\
$E_{H_2O,min}$ & LLM & -0.027 & 326 \\
$E_{M-O,mean}$ & LLM & -0.017 & 1889 \\
$E_{CO_2,mean}$ & LLM & +0.017 & 643 \\
is layered & LLM & -0.014 & 1890 \\
$E_{red,max}$ & LLM & -0.013 & 1890 \\
$E_{M-O,max}$ & LLM & +0.003 & 1889 \\
\hline \hline
\end{tabular}
\end{table}

\clearpage 
\subsubsection*{Section S6. In Situ Validation}

\begin{table}[!htbp]
\linespread{1}\selectfont 
\tiny
\caption{\label{tab:reaction-energies-onset} \textbf{Reaction energies and experimental onset temperatures for precursor pairs for the 26 pairs whose onset was measured with \textit{in situ} XRD.}}
\begin{tabular}{llccl}
\hline \hline
\textbf{Precursor 1} & \textbf{Precursor 2} & \textbf{Rxn. Energy (eV/atom)} & \textbf{Onset Temp. ($^\circ$C)} & \textbf{Source} \\
\hline
Li2CO3 & Nb2O5 & -0.039 & 500 & Szymanski, et al. 2024 \cite{Szymanski2024_thermoregime} \\
Li2CO3 & MnO & -0.015 & 430 & Szymanski, et al. 2024 \cite{Szymanski2024_thermoregime} \\
Li2CO3 & Mn3O4 & 0.0 & 420 & Szymanski, et al. 2024 \cite{Szymanski2024_thermoregime} \\
Li2CO3 & Mn2O3 & -0.016 & 400 & Szymanski, et al. 2024 \cite{Szymanski2024_thermoregime} \\
Li2CO3 & MnO2 & -0.045 & 350 & Szymanski, et al. 2024 \cite{Szymanski2024_thermoregime} \\
Li2CO3 & Fe3O4 & -0.026 & 390 & Szymanski, et al. 2024 \cite{Szymanski2024_thermoregime} \\
Li2CO3 & V2O5 & -0.071 & 230 & Szymanski, et al. 2024 \cite{Szymanski2024_thermoregime} \\
Li2CO3 & Co3O4 & -0.011 & 410 & Szymanski, et al. 2024 \cite{Szymanski2024_thermoregime} \\
Li2CO3 & ZrO2 & -0.016 & 590 & Szymanski, et al. 2024 \cite{Szymanski2024_thermoregime} \\
Li2CO3 & TiO2 & -0.037 & 480 & Szymanski, et al. 2024 \cite{Szymanski2024_thermoregime} \\
\hline
LiOH & Nb2O5 & -0.069 & 450 & Szymanski, et al. 2024 \cite{Szymanski2024_thermoregime} \\
LiOH & MnO & 0.0 & 450 & Szymanski, et al. 2024 \cite{Szymanski2024_thermoregime} \\
LiOH & Mn3O4 & -0.028 & 430 & Szymanski, et al. 2024 \cite{Szymanski2024_thermoregime} \\
LiOH & Mn2O3 & -0.063 & 400 & Szymanski, et al. 2024 \cite{Szymanski2024_thermoregime} \\
LiOH & MnO2 & -0.115 & 204 & Szymanski, et al. 2024 \cite{Szymanski2024_thermoregime} \\
LiOH & Fe3O4 & -0.043 & 360 & Szymanski, et al. 2024 \cite{Szymanski2024_thermoregime} \\
\hline
Na2CO3 & MnO & 0.0 & 360 & Szymanski, et al. 2024 \cite{Szymanski2024_thermoregime} \\
Na2CO3 & Mn3O4 & 0.0 & 340 & Szymanski, et al. 2024 \cite{Szymanski2024_thermoregime} \\
Na2CO3 & MnO2 & -0.042 & 470 & Szymanski, et al. 2024 \cite{Szymanski2024_thermoregime} \\
Na2CO3 & Fe3O4 & -0.015 & 560 & Szymanski, et al. 2024 \cite{Szymanski2024_thermoregime} \\
Na2CO3 & Nb2O5 & -0.054 & 310 & Szymanski, et al. 2024 \cite{Szymanski2024_thermoregime} \\
Na2CO3 & MoO3 & -0.090 & 200 & Szymanski, et al. 2024 \cite{Szymanski2024_thermoregime} \\
Na2CO3 & V2O5 & -0.093 & 180 & Szymanski, et al. 2024 \cite{Szymanski2024_thermoregime} \\
Na2CO3 & TiO2 & -0.043 & 560 & Szymanski, et al. 2024 \cite{Szymanski2024_thermoregime} \\
\hline
BaCO3 & TiO2 & -0.087 & 700 & McDermott, et al. 2023 \cite{McDermott2023_rxnnetwork2} \\
BaO2 & TiO2 & -0.149 & 380 & McDermott, et al. 2023 \cite{McDermott2023_rxnnetwork2} \\
\hline \hline
\end{tabular}
\end{table}


\end{document}